\documentclass[16pt]{article}
\pdfoutput=1
\usepackage{float}
\usepackage{amsthm}
\usepackage{amscd}
\usepackage{bbm}
\usepackage{mathrsfs}
\usepackage{amssymb}
\usepackage{graphics}
\usepackage{graphicx}
\usepackage{amsfonts}
\usepackage{amsmath}
\usepackage{array}
\usepackage{multirow}
\usepackage{booktabs}
\usepackage{subfigure}
\DeclareMathOperator{\sech}{sech}
\usepackage[numbers,sort&compress]{natbib}
\usepackage{sectsty}
\sectionfont{\fontsize{12}{16}\selectfont}
\subsectionfont{\fontsize{11}{8}\selectfont}
\subsubsectionfont{\fontsize{10}{0}\selectfont}
\paragraphfont{\fontsize{10}{0}\selectfont}

\begin{document}
\date{}
\title{The semi-discrete complex modified Korteweg-de Vries equation with zero and non-zero boundary conditions: Riemann-Hilbert approach and $N$-soliton solutions}%
\author{Bo-Jie Deng, Rui Guo$\thanks{Corresponding author:
gr81@sina.com}$, Jian-Wen Zhang\
\\
\\{\em
School of Mathematics, Taiyuan University  of} \\
{\em Technology, Taiyuan 030024, China}
} \maketitle

\begin{abstract}
We focus on the semi-discrete complex modified Korteweg-de Vries (DcmKdV) equation in this paper. The direct and inverse scattering theory is developed with zero and non-zero boundary conditions (BCs) of the potential. For direct problem, the properties of the eigenfunctions and the scattering matrix, including analyticity, asymptotics and symmetries, are investigated, which facilitates the establishment of the Riemann-Hilbert (RH) problems. By solving the RH problems in the inverse problem part, the reconstruction potential formulas are obtained, which allows us to derive the $N$-soliton solutions in the reflectionless case. Meanwhile, the trace formulas are derived by means of studying the corresponding RH problems. Furthermore, the dynamic characteristics of the $1$-soliton and $2$-soliton with zero and non-zero boundary are demonstrated by graphical simulation.

\vspace{3mm}\noindent\emph{Keywords}: Semi-discrete complex modified Korteweg-de Vries equation; Inverse scattering transformation; $N$-soliton solutions; Riemann-Hilbert problem
\end{abstract}

\section{ Introduction}
\renewcommand{\theequation}{\arabic{section}.\arabic{equation}}\setcounter{equation}{0}

\hspace{1.5em}As one of the most powerful methods in soliton theory and integrable systems,  the  inverse scattering transformation (IST) method is widely adopted to derive soliton solutions in messes of nonlinear integrable models. The method was  first proposed for studying the initial-value problems of the famous KdV equation in 1967~\cite{ck1}. Since then, plentiful researchers have focused on the application of the IST in diverse integrable models, among which the Zakharov Shabat-Ablowitz Kaup Newell Segur (ZS-AKNS) system including the nonlinear Schr\"{o}dinger (NLS) equation was successfully investigated by the IST~\cite{ck2,ck3,ck4,ck5,ck6}.

In 1975, a Riemann-Hilbert (RH) problem in the inverse scattering problem of a first-order differential equations was established~\cite{ck7}, and after that the generalized IST method named the RH approach whose inverse problem is formulated as a RH problem has been gradually formed with the unremitting efforts of numerous researchers~\cite{ck8,ck9}. As the modern version of the IST, the RH approach made up for the deficiencies of the traditional IST and immensely promoted the development of inverse scattering theory. As a matter of fact, extensive integrable models were investigated by the RH approach with zero boundary conditions (BCs)~\cite{ck10,ck11,ck12,ck13,ck14,ck27} and non-zero BCs~\cite{ck14,ck15,ck16,ck17,ck18,ck19}.

Over the past several decades, the integrable discrete models, which can describe abounding scientific phenomena in the field of condensed matter physics, molecular biology, nonlinear optics, and so on~\cite{ck20,ck21,ck22},  have aroused the intense interest of scholars and have become one of the hot spots in modern mathematical physics~\cite{ck23,ck24,ck25,ck26} . On account of the fact that the discrete models are as vital as continuous ones, the RH approach has been generalized for studying such models. It can be seen that under the zero BCs, some integrable discrete systems including discrete NLS equation, discrete matrix NLS equation and discrete nonlocal NLS equation have been solved by the RH approach~\cite{ck27,ck28}. And under the non-zero BCs,  the RH approach have been applied on discrete focusing and defocusing NLS equation~\cite{ck29,ck30,ck31}. Moreover, discrete sine-Gordon equation and discrete modified Korteweg-de Vries (mKdV) equation have been studied by constructing the RH problems with simple and double poles.~\cite{ck32,ck33}. These results have significantly enriched the research of discrete IST theory, integrability theory and soliton theory in nonlinear science, and have contributed to close the gap between the development of discrete systems and their continuous counterpart.

In the present paper, we deal with the DcmKdV equation as follows~\cite{ck34}, which is discrete in space and continuous in time:
\begin{equation}\label{1.1}
\begin{aligned}
\frac{\text{d}q_n}{\text{d}t}&=( 1-\sigma|q_n|^2 ) \big[ q_{n+2}-q_{n-2}+2q_{n-1}-2q_{n+1}+\sigma q_n( q_{n-1}\bar{q}_{n+1}-q_{n+1}\bar{q}_{n-1} )\\
&+\sigma \bar{q}_n( q_{n-1}^{2}-\sigma q_{n+1}^{2} )+q_{n-2}|q_{n-1}|^2-\sigma q_{n+2}|q_{n+1}|^2 \big],
\end{aligned}
\end{equation}
where $q_n$ is a complex-valued function,  $t\in \mathbb{R}$ represents the time variable, $n \in\mathbb{Z}$ is the space variable, the overbar means conjugate, and $\sigma=\pm1$ respectively correspond to the defocusing and focusing cases.
This equation was originally introduced as the discretization of the complex modified Korteweg-de Vries (cmKdV) equation~\cite{ck34}
\begin{align*}
q_t=q_{xxx}+6|q|^2 q_x,
\end{align*}
which is a prototypical integrable model for a lot of physical phenomena as diverse as the propagation of transverse waves in a molecular chain model~\cite{ck35}, the transmission of electromagnetic waves in liquid crystal (nematic) waveguides~\cite{ck36}, and  few-cycle optical pulses in cubic nonlinear media~\cite{ck37}. Meanwhile, the DcmKdV equation~(\ref{1.1}) are confirmed to be a integrable equations whose Lax pair is given as below~\cite{ck38}:
\begin{gather}
v_{n+1}=M_n v_n=(Z+Q_n)v_n,\label{1.2}\\
 \frac{\text{d}v_n}{\text{d}t}=L_nv_n,\label{1.3}
\end{gather}
\begin{equation}\notag
Z=\left( \begin{matrix}
	z&		0\\
	0&		z^{-1}\\
\end{matrix} \right),\ \ \ \
Q_n=\left( \begin{matrix}
	0&		q_n\\
	\sigma\bar{q}_n&		0\\
\end{matrix} \right),\ \ \ \
L_n=\left( \begin{matrix}
	l_n^{a}&		l_n^{b}\\
	l_n^{c}&		l_n^{d}\\
\end{matrix} \right) .
\end{equation}
where $v_n=(v_{n,1},v_{n,2})^\text{T}$ (the superscript $\text{T}$ on behalf of the transpose) is the eigenvector corresponds to the parameter $z$, and
\begin{align}
l_n^{a}=&-\frac{3}{4}z^{-4}+z^{-2} (\sigma q_{n-1}\bar{q}_n+2 ) +2\sigma q_n \bar{q}_{n-1}+ ( q_n \bar{q}_{n-1} ) ^2-\sigma q_n
\bar{q}_{n-2}-\sigma q_{n+1}\bar{q}_{n-1}\notag\\
&+q_n|q_{n-1}|^2 \bar{q}_{n-2}+q_{n+1}|q_n|^2 \bar{q}_{n-1}-\frac{3}{2}-\sigma z^2q_n \bar{q}_{n-1}+\frac{1}{4}z^4,\notag\\
l_n^{b}=&z^{-3}q_{n-1}+z^{-1}( q_{n-2}-\sigma q_{n-1}^{2}\bar{q}_n-\sigma q_{n-2}|q_{n-1}|^2-2q_{n-1} ) +z( q_{n+1}-\sigma
q_{n}^{2}\bar{q}_{n-1}-\sigma q_{n+1}|q_n|^2-2q_n ) +z^3q_n,\notag\\
l_n^{c}=&\sigma z^{3}\bar{q}_{n-1}+z( \sigma \bar{q}_{n-2}-\bar{q}_{n-1}^{2}q_n-\bar{q}_{n-2}|q_{n-1}|^2-2\sigma \bar{q}_{n-1} )
+z^{-1}( \sigma \bar{q}_{n+1}-\bar{q}_{n}^{2}q_{n-1}-\bar{q}_{n+1}|q_n|^2-2\sigma \bar{q}_n ) +\sigma z^{-3}\bar{q}_n,\notag\\
l_n^{d}=&-\frac{3}{4}z^{4}+z^{2} (\sigma \bar{q}_{n-1}q_n+2 ) +2\sigma \bar{q}_n q_{n-1}+ ( \bar{q}_n q_{n-1} ) ^2-\sigma \bar{q}_n
q_{n-2}-\sigma \bar{q}_{n+1} q_{n-1}\notag\\
&+\bar{q}_n|q_{n-1}|^2 q_{n-2}+\bar{q}_{n+1}|q_n|^2 q_{n-1}-\frac{3}{2}-\sigma z^{-2}\bar{q}_n q_{n-1}+\frac{1}{4}z^{-4}.\notag
\end{align}

So far, there have been meaningful progress for the investigation of the DcmKdV equation. In Ref.~\cite{ck38}, the Lax pair, conservation laws, the $N$-fold Darboux transformation (DT) have been presented, and their continuum limit for the defocusing case, and the discrete $W$-shape rational periodic solitary wave, first-order rogue waves and breather have been further derived by DT. For the focusing type, Ref.~\cite{ck39} not only has provided a series of nonlinear wave solutions including the periodic solutions, the anti-dark soliton solutions and the breather solutions through the constructed DT, but has exhibited the dynamic behavior of the two-soliton by numerical simulation. In Ref.~\cite{ck40}, the discrete rogue wave was shown to be expected as the solution of DcmKdV equation based on the analyzing of the modulation instability, and bell-shaped one-soliton, $W$-shaped soliton, three types of breathers, periodic solution and rogue wave are given through the established DT. Moreover, Ref.~\cite{ck41} has derived the
DcmKdV equation~(\ref{1.1}) from a discrete integrable hierarchy, and has proposed the corresponding nonlocal equation. It can be distinctly observed that the relevant researches about Eq.~(\ref{1.1}) are just monotonically focused on the classical DT method, which implies that further investigations for the equation are worthwhile and inevitable.

We adopt the RH approach to study the DcmKdV equation~(\ref{1.1}), and the work is organized as follows. In Section 2, we consider the conditions that the potential tends to 0 as the spatial variate $n$ goes to plus or minus infinity for the focusing case of the DcmKdV equation. In Section 2.1, we study the direct scattering problem, and in detail, we define eigenfunctions and scattering matrix starting from the spatial spectrum problem in Subsection 2.1.1 and 2.1.2, construct the summation equations of the modified eigenfunctions by using the method of Green's functions in Subsection 2.1.3, and prove the analyticity, asymptotics and symmetries of the modified eigenfunctions and scattering coefficients in Subsection 2.1.4-2.1.6. In Section 2.2, we investigate the inverse scattering problem in the case of simple poles, where, in Subsection 2.2.1, we establish the RH problems and the reconstruction formula, and present a algebraic-integral system,  and in Subsection 2.2.2, we provide the trace formulas. In Section 2.3, we take into account the temporal spectrum problem and complete the time evolution of the norming constants. In Section 2.4, we construct the $N$-soliton solutions from the obtained algebraic system under the reflectionless condition,  and exhibit the dynamic characteristics of the 1-soliton and 2-soliton through the figures. In Section 3, we consider the small norm nonzero BCs for the defocusing case, and adopt the same overall framework as in Section 2. The main specifical differences are that in Section 3.1, we define a uniformization variable to avoid the multivalued or the complexity of the Riemann surface, in subsection 3.1.5, we introduce a modified scattering problem, construct the summation equations of the corresponding modified eigenfunctions to study the analytic properties of the eigenfunctions, and discuss the behavior of the scattering coefficients at the branch points, and in section 3.2, we study the inverse problem on the plane of the uniformization variable instead of the spectrum parameter.

\section{ Zero boundary conditions}

\hspace{1.5em}In this section, we consider the case of $\sigma=-1$, and assume the potential $q_n$ satisfying $q_n \to 0$ as $n \to \pm \infty$ and $\sum_{n=-\infty}^{+\infty}{|q_n|}<\infty $. The dependence on time $t$ in the part of direct and inverse scattering problems is omitted for simplicity, and then the time evolution is completed in Section 2.3.

\subsection{Direct scattering problem}
\setcounter{equation}{0}
\subsubsection{Eigenfunctions}

\renewcommand{\theequation}{\arabic{section}.\arabic{subsection}.\arabic{equation}}\setcounter{equation}{0}

\hspace{1.5em}When $n\rightarrow \pm \infty$, the scattering problem can
reduce to
\begin{equation}\label{2.1.1}
v_{n+1}\sim Z v_n.
\end{equation}
Hence the eigenfunctions can be defined as
\begin{equation}\label{2.1.2}
\begin{aligned}
\phi  _n=\left(
	\phi _{n,1}\ \,		\phi _{n,2}\right)\sim Z^n,   \ \ \ \  n\rightarrow-\infty; \quad \quad \quad
\psi  _n=\left(
	\psi _{n,1}\ \,		\psi _{n,2}
 \right)\sim Z^n,   \ \ \ \  n\rightarrow+\infty.
\end{aligned}
\end{equation}

It is convenient to introduce the modified eigenfunctions by performing a transform:
\begin{equation}\label{2.1.3}
\Phi _n=\left(
	\Phi _{n,1}\ \,		\Phi _{n,2}\right)=\phi_n Z^{-n}, \quad \quad\quad
\Psi _n=\left(
	\Psi _{n,1}\ \,		\Psi _{n,2} \right)= \psi _n Z^{-n},
\end{equation}
which therefore have constant BCs:
\begin{equation}\label{2.1.4}
\begin{aligned}
\Phi _n\sim I,   \ \ \  \ n\rightarrow-\infty; \quad \quad \quad
\Psi _n\sim I,   \ \ \ \ n\rightarrow+\infty,
\end{aligned}
\end{equation}
here $I$ is the $2\times2$ identity matrix. Meanwhile, the modified scattering problems can be obtained:
\begin{align}
v_{n+1}-z^{-1}Zv_{n}&=z^{-1}Q_nv_{n},\label{2.1.5}\\
v_{n+1}-zZv_{n}&=zQ_nv_{n}.\label{2.1.6}
\end{align}
And the modified eigenfunctions $\Phi _{n,1}$, $\Psi _{n,1}$ satisfy the difference equation~(\ref{2.1.5}), while $\Phi _{n,2}$, $\Psi _{n,2}$
satisfy Eq.~(\ref{2.1.6}).

Due to the eigenfunctions $\phi_n$ and $\psi_n$ are solutions of Eq.~(\ref{1.2}), they satisfy recursive relations:
\begin{gather*}
  \text{det}\,\phi_{n+1}=(1+q_n \bar{q}_n)\,\text{det}\,\phi_n ,\\
  \text{det}\,\psi_{n+1}=(1+q_n \bar{q}_n)\,\text{det}\,\psi_n.
\end{gather*}
Therefore, combining with the BCs~(\ref{2.1.2}), we get the following by iterating:
\begin{align}
  \text{det}\,\phi_{n}&=\prod_{j=-\infty}^{n-1}(1+|q_j|^2)=\frac{\chi_{-\infty}}{\chi_{n}}>0,\label{2.1.7}\\
  \text{det}\,\psi_{n}&=\prod_{j=n}^{+\infty}(1+|q_j|^2)^{-1}=\chi_{n}^{-1}>0\label{2.1.8},
\end{align}
where
\begin{equation}\label{2.1.9}
\begin{aligned}
\chi_{n}=&\prod_{j=n}^{+\infty}(1+|q_j|^2),\\
\chi_{-\infty}=\lim_{n\rightarrow-\infty}&\chi_n=\prod_{j=-\infty}^{+\infty}(1+|q_j|^2).
\end{aligned}
\end{equation}

\subsubsection{Scattering matrix }

\hspace{1.5em} Eqs.~(\ref{2.1.7})-(\ref{2.1.8}) imply that both $\phi_n$ and $\psi_n$ are fundamental matrix solutions for Eq.~(\ref{1.2}), and then there is an
invertible second order matrix $S(z)$ which is independent of $n$ and satisfies
\begin{equation}\label{2.1.10}
\phi_n(z)=\psi_n(z) \,S(z),
\end{equation}
where $S(z)$ is called the scattering matrix and
\begin{equation}\notag
S(z)=\left( \begin{matrix}
	s_1(z)&		s_2(z)\\
	s_3(z)&		s_4(z)\\
\end{matrix} \right).
\end{equation}
Thus we get
\begin{gather}\label{2.1.11}
\Phi_n(z)Z^n=\Psi_n(z) Z^n\,S(z)
\end{gather}
and
\begin{gather}\label{2.1.12}
\text{det}\,S(z)=(\det\psi_n(z))^{-1}\det\phi_n(z)=\chi_{-\infty}.
\end{gather}
Then combining Eqs.~(\ref{2.1.10}) with~(\ref{2.1.8}) and  the transformation~(\ref{2.1.3}), there are
\begin{equation}\label{2.1.13}
\begin{aligned}
s_1(z)&=(\text{det}\,\psi_{n})^{-1}\text{det}\,(\phi_{n,1},\psi_{n,2})=\chi_n\,\text{det}\,(\phi_{n,1},\psi_{n,2})=\chi_n\,\text{det}\,(\Phi_{n,1},\Psi_{n,2}),\\
s_4(z)&=(\text{det}\,\psi_{n})^{-1}\text{det}\,(\psi_{n,1},\phi_{n,2})=\chi_n\,\text{det}\,(\psi_{n,1},\phi_{n,2})=\chi_n\,\text{det}\,(\Psi_{n,1},\Phi_{n,2}),\\
s_2(z)&=(\text{det}\,\psi_{n})^{-1}\text{det}\,(\phi_{n,2},\psi_{n,2})=\chi_n\,\text{det}\,(\phi_{n,2},\psi_{n,2})=\chi_nz^{-2n}\,\text{det}\,(\Phi_{n,2},\Psi_{n,2}),\\
s_3(z)&=(\text{det}\,\psi_{n})^{-1}\text{det}\,(\psi_{n,1},\phi_{n,1})=\chi_n\,\text{det}\,(\psi_{n,1},\phi_{n,1})=\chi_nz^{2n}\,\text{det}\,(\Psi_{n,1},\Phi_{n,1}).
\end{aligned}
\end{equation}

Define
\begin{equation}\notag
\alpha_n(z)=\frac{\Phi_{n,1}}{s_1(z)},\quad\quad \beta_n (z)=\frac{\Phi_{n,2}}{s_4(z)},
\end{equation}
and the reflection coefficients
\begin{equation}\label{2.1.14}
\gamma_1(z)=\frac{s_3(z)}{s_1(z)},\quad\quad \gamma_2(z)=\frac{s_2(z)}{s_4(z)},
\end{equation}
then Eqs.~(\ref{2.1.11}) can be written as
\begin{equation}\label{2.1.15}
\begin{aligned}
\alpha _n\left( z \right) &=\Psi _{n,1}\left( z \right) +z^{-2n}\gamma _1\left( z \right) \Psi _{n,2}\left( z \right),\\
\beta _n\left( z \right) &=\Psi _{n,2}\left( z \right) +z^{2n}\gamma _2\left( z \right) \Psi _{n,1}\left( z \right).
\end{aligned}
\end{equation}

\subsubsection{Summation equations}

\hspace{1.5em}We use the method of Green's functions to construct the summation equations for the solutions of Eqs.~(\ref{2.1.5}) and~(\ref{2.1.6}). For
Eq.~(\ref{2.1.5}), its solutions satisfy the summation equation
\begin{equation}\label{2.1.16}
v_n(z)=\upsilon +\sum_{j=-\infty}^{+\infty}{G_{n-j}(z)Q_jv_j(z)},
\end{equation}
where $G_n(z)$ are the Green's functions satisfying
\begin{equation}\label{2.1.17}
G_{n+1}(z)-z^{-1}ZG_n(z)=z^{-1}I\delta _{n,0},
\end{equation}
and the inhomogeneous term $\upsilon$ is the solution of
\begin{equation}\label{2.1.18}
\left( I-z^{-1}Z \right) \upsilon =0,
\end{equation}
here $\delta _{n,0}$ is the Kronecker delta.

To derive the Green's functions in Eq.~(\ref{2.1.16}), we first represent them and $\delta _{n,0}$ as Fourier integrals
 \begin{equation}\label{2.1.19}
G_n\left( z \right) =\frac{1}{2\pi i}\oint_{\left| s \right|=1}{\hat{G}\left( s \right)}s^{n-1}\text{d}s,\quad\quad
\delta _{n,0}=\frac{1}{2\pi i}\oint_{\left| s \right|=1}{s^{n-1}\text{d}s}.
\end{equation}
Substituting above integrals into Eq.~(\ref{2.1.17}) one obtain
 \begin{equation}\label{2.1.20}
\hat{G}\left( s \right) =\left( \begin{matrix}
	\frac{z^{-1}}{s-1}&		0\\
	0&		\frac{z^{-1}}{s-z^{-2}}\\
\end{matrix} \right),
\end{equation}
and consequently
 \begin{equation}\label{2.1.21}
G_n\left( z \right)=\frac{z^{-1}}{2\pi i}\oint_{\left| s \right|=1}{\left( \begin{matrix}
	\frac{1}{s-1}&		0\\
	0&		\frac{1}{s-z^{-2}}\\
\end{matrix} \right)s^{n-1}\text{d}s}.
\end{equation}
It is easy to see that there are poles 0, 1 and $z^{-2}$ in the integral above, and the poles 1 and $z^{-2}$ are inside or outside the contour
of integration generating two integral results. We consider contours that are perturbed away from $|s| = 1$ to avoid the poles located on
it. When $|z|\geq 1$, the contour enclosing $s=0, 1, z^{-2}$ can be considered, which leads to $G_n^{\langle1\rangle}\left( z \right)$; and when $|z|\leq
1$, the contour enclosing $s=0$ but neither 1 nor $z^{-2}$ can be considered, which leads to $G_n^{\langle2\rangle}\left(z \right)$. By residue theorem,
we derive that
 \begin{equation}\label{2.1.22}
G_n^{\langle1\rangle}\left( z \right)=\left\{ \begin{array}{l}
	z^{-1}\left( \begin{matrix}
	1&		0\\
	0&		z^{-2\left( n-1 \right)}\\
\end{matrix} \right) \ \ \ n> 0\\
	0\ \ \ \ \ \ \ \ \ \ \ \ \ \ \ \ \ \ \ \ \ \ \ \ \, \ \ \ n\le0\\
\end{array} \right.
;\quad \
G_n^{\langle2\rangle}\left( z \right)=\left\{ \begin{array}{l}
	-z^{-1}\left( \begin{matrix}
	1&		0\\
	0&		z^{-2\left( n-1 \right)}\\
\end{matrix} \right) \ \ \ n\le0\\
0\ \ \ \ \ \ \ \ \ \ \ \ \ \ \ \ \ \ \ \ \ \ \ \ \ \ \ \ \ \ n>0\\
\end{array} \right.,
\end{equation}
that is
 \begin{align}
 G_n^{\langle1\rangle}\left( z \right)&=\theta \left( n-1 \right) \left( \begin{matrix}
	z^{-1}&		0\\
	0&		z^{-2n+1}\\
\end{matrix} \right),\label{2.1.23}\\
G_n^{\langle2\rangle}\left( z \right)&=-\theta \left( -n \right) \left( \begin{matrix}
	z^{-1}&		0\\
	0&		z^{-2n+1}\\
\end{matrix} \right),\label{2.1.24}
\end{align}
 where
 \begin{equation}\notag
 \theta \left( n \right)=\left\{ \begin{array}{l}
	1\ \ \ \ \ \ \ \ n\ge 0\\
	0\ \ \ \ \ \ \ \ n<0 \\
\end{array} \right..
\end{equation}
Note that $Q_n \to \mathbf{0} $  as $n \to \pm \infty$ and that the BCs for $\Phi_{n,1}$ and $\Psi_{n,1}$ in Eq.~(\ref{2.1.4}) satisfy Eq.~(\ref{2.1.18}). Then the summation equations for
$\Phi_{n,1}$ and $\Psi_{n,1}$ can be given as
 \begin{align}
 \Phi_{n,1}(z)=\left( \begin{array}{c}
	1\\
	0\\
\end{array} \right) +\sum_{j=-\infty}^{+\infty}{G_{n-j}^{\langle1\rangle}(z)Q_j\Phi_{j,1}(z)},\label{2.1.25}\\
\Psi_{n,1}(z)=\left( \begin{array}{c}
	1\\
	0\\
\end{array} \right) +\sum_{j=-\infty}^{+\infty}{G_{n-j}^{\langle2\rangle}(z)Q_j\Psi_{j,1}(z)}.\label{2.1.26}
\end{align}

Similarly, solutions of Eq.~(\ref{2.1.6}) can be expressed as
\begin{equation}\label{2.1.27}
v_n(z)=\tilde{\upsilon} +\sum_{j=-\infty}^{+\infty}{\tilde{G}_{n-j}(z)Q_jv_j(z)},
\end{equation}
with Green's functions
\begin{align}
 \tilde{G}_n^{\langle1\rangle}\left( z \right)&=\theta \left( n-1 \right) \left( \begin{matrix}
	z^{2n-1}&		0\\
	0&		z\\
\end{matrix} \right),\label{2.1.28}\\
\tilde{G}_n^{\langle2\rangle}\left( z \right)&=-\theta \left( -n \right) \left( \begin{matrix}
	z^{2n-1}&		0\\
	0&		z\\
\end{matrix} \right),\label{2.1.29}
\end{align}
and the inhomogeneous term $\upsilon$ satisfying
\begin{equation}\label{2.1.30}
\left( I-zZ \right) \tilde{\upsilon} =0.
\end{equation}
Then the summation equations for $\Phi_{n,2}$ and $\Psi_{n,2}$ can be given as below according to the BCs for $\Phi_{n,2}$ and
$\Psi_{n,2}$ in Eq.~(\ref{2.1.4}):
\begin{align}
 \Phi_{n,2}(z)=\left( \begin{array}{c}
	0\\
	1\\
\end{array} \right) +\sum_{j=-\infty}^{+\infty}{\tilde{G}_{n-j}^{\langle1\rangle}(z)Q_j\Phi_{j,2}(z)},\label{2.1.31}\\
\Psi_{n,2}(z)=\left( \begin{array}{c}
	0\\
	1\\
\end{array} \right) +\sum_{j=-\infty}^{+\infty}{\tilde{G}_{n-j}^{\langle2\rangle}(z)Q_j\Psi_{j,2}(z)}.\label{2.1.32}
\end{align}

\subsubsection{Analyticity}

\hspace{1.5em}By studying the summation equations~(\ref{2.1.25})-(\ref{2.1.26}) and~(\ref{2.1.30})-(\ref{2.1.31}),   the existence and the analyticity of the modified eigenfunctions can be given in Proposition 2.1.

\paragraph{Proposition 2.1}  If $\lVert q_n \rVert _1=\sum_{n=-\infty}^{+\infty}{|q_n|}<\infty $, the modified eigenfunctions respectively
determined by Eqs.~(\ref{2.1.25}), (\ref{2.1.26}), (\ref{2.1.31}) and (\ref{2.1.32}) are unique in the space of bounded functions. Meanwhile, the
modified eigenfunctions $\Phi_{n,1}(z)$ and $\Psi_{n,2}(z)$ given in Eqs.~(\ref{2.1.25}) and~(\ref{2.1.32}) are analytic functions of $z$ for $|z| > 1$
and continuous for $|z| \ge 1$, while $\Psi_{n,1}(z)$ and $\Phi_{n,2}(z)$ given in Eqs.~(\ref{2.1.26}) and~(\ref{2.1.31}) are analytic functions of
$z$ for $|z| < 1$ and continuous for $|z| \le 1$.

\begin{proof}
 The modified eigenfunctions $\Phi_{n,1}(z)$ will be considered firstly. Define a recursive sequence
\begin{equation}\label{2.1.33}
\mu _{n}^{\langle 0 \rangle}\left( z \right) =\left( \begin{array}{c}
	1\\
	0\\
\end{array} \right),\quad\
\mu _{n}^{\langle k+1 \rangle}\left( z \right) =\sum_{j=-\infty}^{+\infty}{G_{_{n-j}}^{\langle1\rangle}Q_j}\mu _{j}^{\langle k \rangle}\left(
z \right),\ \, k=0, 1, 2\cdots
\end{equation}
Then the Neumann series $\sum_{k=0}^{+\infty}{\mu _{j}^{\langle k \rangle}\left( z \right)}$ can be shown to be a solution
of the summation equations~(\ref{2.1.25}):
\begin{align}\notag
\sum_{k=0}^{+\infty}{\mu _{j}^{\langle k \rangle}\left( z \right)} =\left( \begin{array}{c}
	1\\
	0\\
\end{array} \right) +\sum_{k=1}^{+\infty}{\mu _{j}^{\langle k \rangle}\left( z \right)}
=\left( \begin{array}{c}
	1\\
	0\\
\end{array} \right) +\sum_{j=-\infty}^{+\infty}{G_{_{n-j}}^{\langle1\rangle}Q_j}\sum_{k=0}^{+\infty}{\mu _{j}^{\langle k \rangle}\left( z
\right)}\notag.
\end{align}
We will show that the Neumann series $\sum_{k=0}^{+\infty}{\mu _{n}^{\langle k \rangle}\left( z \right)}$ converges to an analytic function
$\mu _{n}(z)$, and then prove that $\mu _{n}(z)$ is the unique solution of differential equation~(\ref{2.1.25}).
\\\indent The  expression in component form for  $\mu _{n}^{\left( k+1 \right)}\left( z \right)$ in Eq.~(\ref{2.1.33}) is
\begin{equation}\label{2.1.34}
\mu _{n}^{\langle k+1\rangle,\, \left(1 \right)}\left( z \right) =z^{-1}\sum_{j=-\infty}^{n-1}{q_j}\mu _{j}^{\langle k\rangle,\, \left(2
\right)}\left( z \right),\quad\quad
\mu _{n}^{\langle k+1\rangle,\, \left(2 \right)}\left( z \right) =z^{-2n+1}\sum_{j=-\infty}^{n-1}{z^{2j}\bar{q}_j}\mu _{j}^{\langle
k\rangle,\, \left(1 \right)}\left( z \right),
\end{equation}
where the superscript $(j)$ represents the $j$-th entry of the column vector. It is easy to obtain that $\mu _{n}^{\langle 2k+1\rangle,\,
\left(1 \right)}=0$ and $\mu _{n}^{\langle 2k\rangle,\, \left(2 \right)}=0$ since $\mu _{n}^{\langle 0\rangle,\, \left(2 \right)}=0$. And the
bounds for $\mu _{n}^{\langle 2k\rangle,\, \left(1 \right)}$ and $\mu _{n}^{\langle 2k+1\rangle,\, \left(2 \right)}$ can be found and
demonstrated by induction on $k$:
\\\indent When $|z|\ge 1$, we find $|\mu _{n}^{\langle 1\rangle,\, \left(2 \right)}|\le \sum_{j=-\infty}^{n-1}{|\bar{q}_j|}$ since $\mu
_{n}^{\langle 0\rangle,\, \left(1 \right)}=0 $. Suppose that
\begin{equation}\label{2.1.35}
|\mu _{n}^{\langle 2k\rangle,\, \left(1 \right)}|\le
\frac{\left(\sum_{j=-\infty}^{n-1}{|q_j|}\right)^{2k}}{k!\cdot k! },\quad\quad
|\mu _{n}^{\langle 2k+1\rangle,\, \left(2
\right)}|\le\frac{\left(\sum_{j=-\infty}^{n-1}{|q_j|}\right)^{2k+1}}{k!\cdot (k+1)!},
\end{equation}
Then we can prove that
\begin{align}\notag
|\mu _{n}^{\langle 2k+1\rangle,\, \left(2 \right)}|
&\le\sum_{j=-\infty}^{n-1}{|\bar{q}_j|}|\mu _{j}^{\langle 2k\rangle,\, \left(1 \right)}|
\le\sum_{j=-\infty}^{n-1}{|q_j|}\frac{\left(\sum_{l=-\infty}^{j-1}{|q_l|}\right)^k}{k!}\frac{\left(\sum_{l=-\infty}^{j-1}{|q_l|}\right)^k}{k!}\notag\\
&\le\frac{\left(\sum_{j=-\infty}^{n-1}{|q_j|}\right)^k}{k!}\sum_{j=-\infty}^{n-1}{|q_j|}\frac{\left(\sum_{l=-\infty}^{j-1}{|q_l|}\right)^k}{k!}
\le\frac{\left(\sum_{j=-\infty}^{n-1}{|q_j|}\right)^{2k+1}}{k!\cdot (k+1)!},
\notag\\
|\mu_{n}^{\langle 2k+2\rangle,\, \left(1 \right)}|&\le\sum_{j=-\infty}^{n-1}{|q_j|}|\mu _{j}^{\langle 2k+1\rangle,\, \left(2 \right)}|
\le\sum_{j=-\infty}^{n-1}{|q_j|}\frac{\left(\sum_{l=-\infty}^{j-1}{|q_l|}\right)^k}{k!}\frac{\left(\sum_{l=-\infty}^{j-1}{|q_l|}\right)^{k+1}}{(k+1)!}\notag\\
&\le\frac{\left(\sum_{j=-\infty}^{n-1}{|q_j|}\right)^{k+1}}{(k+1)!}\sum_{j=-\infty}^{n-1}{|q_j|}\frac{\left(\sum_{l=-\infty}^{j-1}{|q_l|}\right)^k}{k!}
\le\frac{\left(\sum_{j=-\infty}^{n-1}{|q_j|}\right)^{2(k+1)}}{(k+1)!\cdot (k+1)!},
\notag
\end{align}
where the inequality $\sum_{j=-\infty}^{n}{|q_j|}\left(\sum_{l=-\infty}^{j-1}{|q_l|}\right)^{k}
\le\frac{1}{k+1}\left(\sum_{j=-\infty}^{n}{|q_j|}\right)^{k+1}$ which has been proved in~\cite{ck27} can complete the last step of above two
derivation processes.
\\\indent Hence the assumption~(\ref{2.1.35}) is appropriate.
\\\indent Since the bounds in Eq.~(\ref{2.1.35}) are absolutely and uniformly (in $n$ and in $z$) summable with respect to $k$ when $\lVert q_n
\rVert _1<\infty $, the Neumann series $\sum_{k=0}^{+\infty}{\mu _{n}^{\langle k \rangle}\left( z \right)}$ can be proved to converge
absolutely and uniformly in $n$ and in $z$ when $|z|\ge 1$ . Due to the fact that a uniformly convergent series of analytic functions converages to an
analytic function in the interior of the domain, we can prove that the Neumann series $\sum_{k=0}^{+\infty}{\mu _{n}^{\langle k \rangle}\left(
z \right)}$ converges to a function $\mu _{n}(z)$ which is analytic in $|z|>1$ and continuous in $|z|\ge1$.
\\\indent To prove $\mu _{n}(z)$ is the unique solution of Eq.~(\ref{2.1.24}), we assume that there is another solution $\dot{\mu} _{n}(z)$.
Then
\begin{equation}\notag
\tau_n(z)=\mu_n(z)-\dot{\mu}_n(z)=\sum_{j=-\infty}^{+\infty}{G_{_{n-j}}^{\langle1\rangle}Q_j}\tau_j(z),
\end{equation}
that is
\begin{equation}\notag
\tau_n(z)=\left( \begin{array}{c}
	\tau_{n,1}(z)\\
	\tau_{n,2}(z)\\
\end{array} \right),
\end{equation}
where
\begin{align}
\tau_{n,1}(z)&=z^{-1}\sum_{j=-\infty}^{n-1}q_j\tau_{j,2}(z),\label{2.1.36}\\
\tau_{n,2}(z)&=-z^{-2n+1}\sum_{j=-\infty}^{n-1}z^{2j}\bar{q}_j\tau_{j,1}(z).\label{2.1.37}
\end{align}
Substituting Eq.~(\ref{2.1.37}) into~(\ref{2.1.36}) one can get
\begin{equation}\notag
\tau_{n,1}(z)=-\sum_{j=-\infty}^{n-1}z^{-2j}q_j\sum_{l=-\infty}^{j-1}z^{2l}\bar{q}_l\tau_{l,1}(z),
\end{equation}
hence for $|z|\geq1$ , $\tau_{n,1}(z)$ satisfies the inequality
\begin{equation}\notag
|\tau_{n,1}(z)|\leq \sum_{j=-\infty}^{n-1}|q_j|\sum_{l=-\infty}^{j-1}|q_l||\tau_{l,1}(z)|.
\end{equation}
Iterating once, there is
\begin{equation}\notag
|\tau_{n,1}(z)|\leq
\sum_{j=-\infty}^{n-1}|q_j|\sum_{l=-\infty}^{j-1}|\bar{q}_l|\sum_{k=-\infty}^{l-1}|q_k|\sum_{\xi=-\infty}^{k-1}|\bar{q}_\xi||\tau_{\xi,1}(z)|.
\end{equation}
Since $\mu_n(z)$ and $\dot{\mu}_n(z)$ are bounded functions, then $\exists\  C \geq 0$ s.t. $|\tau_{\xi,1}(z)|\leq C$ for all $\xi \in
\mathbb{Z}$.
Therefore
\begin{align}\notag
|\tau_{n,1}(z)|
\leq C \frac{\left(\sum_{j=-\infty}^{n-1}|q_j|\right)^2}{2!}\frac{\left(\sum_{j=-\infty}^{n-1}|q_j|\right)^2}{2!}.\notag
\end{align}
Then iterating $N$ times yields
\begin{equation}\label{2.1.38}
|\tau_{n,1}(z)|
\leq C
\frac{\left(\sum_{j=-\infty}^{n-1}|q_j|\right)^{N+1}}{\left(N+1\right)!}\frac{\left(\sum_{j=-\infty}^{n-1}|q_j|\right)^{N+1}}{\left(N+1\right)!}.
\end{equation}
When $N\rightarrow +\infty$, the right-hand side of Eq.~(\ref{2.1.38}) will tend to 0, hence $|\tau_{n,1}(z)|=0$. According to
Eq.~(\ref{2.1.37}), we obtain $|\tau_{n,2}(z)|=0$, therefore $\tau_{n}(z)=(0,0)^\text{T}$, i.e. $\mu _{n}(z)=\dot{\mu} _{n}(z)$.
\\\indent Thus we have proved $\mu _{n}(z)$ is the unique solution in the space of bounded functions. Since the modified eigenfunctions
$\Phi_{n,1}(z)$ also satisfy Eq.~(\ref{2.1.24}), we get $\Phi_{n,1}(z)=\mu _{n}(z)$.
\\\indent In summary, the Jost eigenfunction $\Phi_{n,1}(z)$ have been proved to be a analytic function of $z$ in $|z| > 1$ and continuous in
$|z| \ge 1$. And it is the unique solution of Eq.~(\ref{2.1.24}) in the space of bounded functions.
\\\indent Next we prove the properties of the Jost eigenfunction $\Psi_{n,1}(z)$. Since $\Psi_{n,1}(z)$ is the solution of Eq.~(\ref{2.1.5}),
then it satisfies the difference equation
\begin{equation}\notag
\prod_{k=n}^{+\infty}{\left( 1+|q_k|^2 \right)}\Psi_{n,1}(z) -zZ^{-1}\prod_{k=n+1}^{+\infty}{\left( 1+|q_k|^2
\right)}\Psi_{n+1,1}(z)=-zQ_n\prod_{k=n+1}^{+\infty}{\left( 1+|q_k|^2 \right)}\Psi_{n+1,1}(z).
\end{equation}
Therefore it is convenient to define the modified Jost eigenfunction $\hat{\Psi}_{n,1}(z)=\chi_n\Psi_{n,1}(z)$,
 and then $\hat{\Psi}_{n,1}(z)$ satisfies
\begin{equation}\label{2.1.39}
\hat{\Psi}_{n,1}(z) -zZ^{-1}\hat{\Psi}_{n+1,1}(z)=-zQ_n\hat{\Psi}_{n+1,1}(z).
\end{equation}
Note that $\chi_n$ converges absolutely when $\lVert q_n \rVert < \infty$, and $\chi_n\rightarrow1$ as $n\rightarrow+\infty$. Then
\begin{equation}\label{2.1.40}
\hat{\Psi}_{n,1}(z)\rightarrow\left(\begin{array}{c}
	1\\
	0\\
\end{array}\right) \quad \quad \text{as}\ \ n\rightarrow+\infty.
\end{equation}
\\\indent Similarly to the process of deriving the summation equation~(\ref{2.1.25}), using the method of Green's functions, one can obtain
\begin{equation}\label{2.1.41}
\hat{\Psi}_{n,1}(z)=\left( \begin{array}{c}
	1\\
	0\\
\end{array} \right) +\sum_{j=-\infty}^{+\infty}{\check{G}_{n-j}^{\langle 2\rangle}(z)Q_j\hat{\Psi}_{j+1,1}(z)},
\end{equation}
where
\begin{equation}\notag
\check{G}_n^{\langle 2\rangle}\left( z \right)=-\theta \left( -n \right) \left( \begin{matrix}
	z&		0\\
	0&		z^{-2n+1}\\
\end{matrix} \right).
\end{equation}
\\\indent Define a recursive sequence
\begin{equation}\notag
\hat{\mu} _{n}^{\left( 0 \right)}\left( z \right) =\left( \begin{array}{c}
	1\\
	0\\
\end{array} \right),\quad\
\hat{\mu} _{n}^{\left( k+1 \right)}\left( z \right) =\sum_{j=-\infty}^{+\infty}{\check{G}_n^{\langle 2\rangle}Q_j}\hat{\mu} _{j}^{\left( k
\right)}\left( z \right),\ \, k=0, 1, 2\cdots
\end{equation}
Then one can similarly prove that the Neumann series $\sum_{k=0}^{+\infty}\hat{\mu} _{n}^{\left( k \right)}\left( z \right)$ converges
absolutely and uniformly in $n$ and in $z$ when $|z|\leq1$, and it is the unique solution of Eq.~(\ref{2.1.41}) in the space of bounded
functions. Hence $\hat{\Psi}_{n,1}(z)$ have been shown to be an analytic function of $z$ in $|z|<1$ and continuous in $|z|\leq 1$. Since $\chi_n$
defined in Eq.~(\ref{2.1.9}) is not equal to 0, and $\lVert q_n \rVert _1<\infty $, then the Jost eigenfunction
$\Psi_{n,1}(z)=\chi_{n}^{-1}\hat{\Psi}_{n,1}(z)$ is an analytic function of $z$ in $|z|<1$ and continuous in $|z|\leq 1$.
\\\indent In the same way, we can demonstrate that the Jost eigenfunction $\Phi_{n,2}(z)$ is an analytic function of $z$ in $|z| < 1$ and
continuous in $|z| \leq 1$, and $\Psi_{n,2}(z)$ is an analytic function of $z$ in $|z| > 1$ and continuous in $|z| \ge 1$. And in the space of
bounded functions, the Jost
eigenfunctions $\Psi_{n,1}(z)$, $\Psi_{n,2}(z)$ and $\Phi_{n,2}(z)$ are the unique solutions of Eq.~(\ref{2.1.26}),~(\ref{2.1.31}) and
~(\ref{2.1.32}) respectively.
\\\indent To sum up, the Proposition 2.1 is proved.
\end{proof}

Using Eqs.~(\ref{2.1.13}) and the Proposition 2.1, the analytical properties of the scattering coefficients can be easily derived and be given in Proposition 2.2.

\paragraph{Proposition 2.2} If $\lVert q_n \rVert _1=\sum_{n=-\infty}^{+\infty}{|q_n|}<\infty $, then the scattering coefficient $s_1(z)$ is
an analytic function of $z$ for $|z| > 1$ and continuous for $|z| \ge 1$, and $s_4(z)$ is analytic for $|z| < 1$ and continuous for $|z| \le
1$. While the analytic domains of the scattering coefficients $s_2(z)$ and $s_3(z)$ cannot be determined by Eqs.~(\ref{2.1.13}), but they are
continuous to the contour $|z|=1$.

\subsubsection{Asymptotics}

\hspace{1.5em}Depending on the analytic domain of the eigenfunctions, the asymptotic behavior of the modified eigenfunctions will be analyzed by researching
their Laurent series expansions about 0 or infinity.

For the Jost eigenfunction $\Phi_{n,1}(z)$, which has been proved to be an analytic function of $z$ in $|z| > 1$, can be expanded in Laurent
series at the point $z=\infty$:
\begin{align}\notag
\Phi _{n,1}^{\left( 1 \right)}(z)&=\Phi _{n,1}^{\left( 1 \right) ,\, \left< 0 \right>}+\Phi _{n,1}^{\left( 1 \right) ,\, \left< -1
\right>}z^{-1}+O\left( z^{-2} \right),\notag\\
\Phi _{n,1}^{\left( 2 \right)}(z)&=\Phi _{n,1}^{\left( 2 \right) ,\, \left< 0 \right>}+\Phi _{n,1}^{\left( 2 \right) ,\, \left< -1
\right>}z^{-1}+O\left( z^{-2} \right).\notag
\end{align}
Since Eq.~(\ref{2.1.25}) can be written as
\begin{align}\notag
\left( \begin{array}{c}
	\Phi _{n,1}^{\left( 1 \right)}\left( z \right)\\
	\Phi _{n,1}^{\left( 2 \right)}\left( z \right)\\
\end{array} \right) &=\left( \begin{array}{c}
	1\\
	0\\
\end{array} \right) +\sum_{j=-\infty}^{n-1}{\left( \begin{array}{c}
	q_jz^{-1}\Phi _{j,1}^{\left( 2 \right)}\left( z \right)\\
	-z^{-2\left( n-j \right) +1}\bar{q}_j\Phi _{j,1}^{\left( 1 \right)}\left( z \right)\\
\end{array} \right) },
\end{align}
combining with the Laurent series expressions of $\Phi _{n,1}(z)$ and $\Phi _{n,2}(z)$, we can derive
\begin{equation}\label{2.1.42}
\begin{aligned}
\Phi _{n,1}(z) &=\left( \begin{array}{c}
	1\\
	0\\
\end{array} \right) +z^{-1}\left( \begin{array}{c}
	0\\
	-\bar{q}_{n-1}\\
\end{array} \right) +z^{-2}\left( \begin{array}{c}
	-\sum_{j=-\infty}^{n-1}{q_j\bar{q}_{j-1}}\\
	0\\
\end{array} \right) +O\left( z^{-3} \right),\\
& =\left( \begin{array}{c}
	1\\
	0\\
\end{array} \right) -\sum_{p=-1}^{-\infty}{z^{2p+1}\left( \begin{array}{c}
	0\\
	\sum_{l=n+p}^{n-1}{\bar{q}_l\Phi _{l,1}^{\left( 1 \right) ,\left< 2\left( n+p-l \right) \right>}}\\
\end{array} \right)}+\sum_{p=-1}^{-\infty}{z^{2p}\left( \begin{array}{c}
	\sum_{j=-\infty}^{n-1}{q_j\Phi _{j,1}^{\left( 2 \right) ,\left< 2p+1 \right>}}\\
	0\\
\end{array} \right)}.
\end{aligned}
\end{equation}

For the asymptotic behavior of the Jost eigenfunction $\Psi_{n,1}(z)$, we first consider the Laurent series expansions of the modified Jost
eigenfunction $\hat{\Psi}_{n,1}(z)$ about $z=0$:
\begin{align}\notag
\hat{\Psi} _{n,1}^{\left( 1 \right)}(z)&=\hat{\Psi} _{n,1}^{\left( 1 \right) ,\, \left< 0 \right>}+\hat{\Psi} _{n,1}^{\left( 1 \right) ,\, \left<
1 \right>}z+O\left( z^{2} \right),\notag\\
\hat{\Psi} _{n,1}^{\left( 2 \right)}(z)&=\hat{\Psi} _{n,1}^{\left( 2 \right) ,\, \left< 0 \right>}+\hat{\Psi} _{n,1}^{\left( 2 \right) ,\, \left<
1 \right>}z+O\left( z^{2} \right).\notag
\end{align}
It is easy to write Eq.~(\ref{2.1.41}) as
\begin{align}\notag
\left( \begin{array}{c}
	\hat{\Psi} _{n,1}^{\left( 1 \right)}\left( z \right)\\
	\hat{\Psi} _{n,1}^{\left( 2 \right)}\left( z \right)\\
\end{array} \right) &=\left( \begin{array}{c}
	1\\
	0\\
\end{array} \right) +\sum_{j=n}^{+\infty}{\left( \begin{array}{c}
	-q_jz\hat{\Psi} _{j+1,1}^{\left( 2 \right)}\left( z \right)\\
	z^{-2\left( n-j \right) +1}\bar{q}_j\hat{\Psi} _{j+1,1}^{\left( 1 \right)}\left( z \right)\\
\end{array} \right) },
\end{align}
hence the asymptotic expression for $\hat{\Psi}_{n,1}(z)$ about $z=0$ can be obtained as below:
\begin{equation}\label{2.1.43}
\begin{aligned}
\hat{\Psi} _{n,1}(z) =&\left( \begin{array}{c}
	1\\
	0\\
\end{array} \right) +z\left( \begin{array}{c}
	0\\
	\bar{q}_{n}\\
\end{array} \right) +z^{2}\left( \begin{array}{c}
	-\sum_{j=n}^{+\infty}{q_j\bar{q}_{j+1}}\\
	0\\
\end{array} \right) +O\left( z^{3} \right),\\
 =&\left( \begin{array}{c}
	1\\
	0\\
\end{array} \right) +\sum_{p=1}^{+\infty}{z^{2p-1}\left( \begin{array}{c}
	0\\
	\sum_{l=n}^{n+p-1}{\bar{q}_l\Hat{\Psi} _{l+1,1}^{\left( 1 \right) ,\left< 2\left( n+p-l-1 \right) \right>}}\\
\end{array} \right)}-\sum_{p=1}^{+\infty}{z^{2p}\left( \begin{array}{c}
	\sum_{j=n}^{+\infty}{q_j\Hat{\Psi} _{j+1,1}^{\left( 2 \right) ,\left< 2p-1 \right>}}\\
	0\\
\end{array} \right)}.
\end{aligned}
\end{equation}
Then accordingly we get
\begin{equation}\label{2.1.44}
\begin{aligned}
\Psi _{n,1}(z)  &=\left( \begin{array}{c}
	\chi_n^{-1}\\
	0\\
\end{array} \right) +z\left( \begin{array}{c}
	0\\
	\chi_n^{-1}\bar{q}_{n}\\
\end{array} \right) +z^{2}\left( \begin{array}{c}
	-\chi_n^{-1}\sum_{j=n}^{+\infty}{q_j\bar{q}_{j+1}}\\
	0\\
\end{array} \right) +O\left( z^{3} \right),\\
&=\left( \begin{array}{c}
	\chi_n^{-1}\\
	0\\
\end{array} \right) +\sum_{p=0}^{+\infty}{z^{2p+1}\left( \begin{array}{c}
	0\\
	\chi_n^{-1}\sum_{l=n}^{n+p}{\bar{q}_l\Psi _{l+1,1}^{\left( 1 \right) ,\left< 2\left( n+p-l \right) \right>}}\\
\end{array} \right)}-\sum_{p=1}^{+\infty}{z^{2p}\left( \begin{array}{c}
	\chi_n^{-1}\sum_{j=n}^{+\infty}{q_j\Psi _{j+1,1}^{\left( 2 \right) ,\left< 2p-1 \right>}}\\
	0\\
\end{array} \right)}.
\end{aligned}
\end{equation}

By similar calculations, the asymptotic behavior of the Jost eigenfunction $\Phi_{n,2}(z)$ about the point $z=0$ can be derived as
\begin{equation}\label{2.1.45}
\begin{aligned}
\Phi _{n,2}(z)& =\left( \begin{array}{c}
	0\\
	1\\
\end{array} \right) +z\left( \begin{array}{c}
	q_{n-1}\\
	0\\
\end{array} \right) +z^{2}\left( \begin{array}{c}
	0\\
	-\sum_{j=-\infty}^{n-1}{q_{j-1}\bar{q}_{j}}\\
\end{array} \right) +O\left( z^{3} \right),\\
&=\left( \begin{array}{c}
	0\\
	1\\
\end{array} \right) +\sum_{p=1}^{+\infty}{z^{2p-1}\left( \begin{array}{c}
	\sum_{l=n-p}^{n-1}{q_l\Phi _{l,2}^{\left( 2 \right),\left< 2\left( p-n+l \right) \right>}} \\
	0\\
\end{array} \right)}-\sum_{p=1}^{+\infty}{z^{2p}\left( \begin{array}{c}
	0\\
	\sum_{j=-\infty}^{n-1}{\bar{q}_j\Phi _{j,2}^{\left( 1 \right) ,\left< 2p-1 \right>}}\\
\end{array} \right)},
\end{aligned}
\end{equation}
and the asymptotic behavior of $\Psi_{n,2}(z)$ about the point $z=\infty$ is
\begin{equation}\label{2.1.46}
\begin{aligned}
\Psi _{n,2}(z)  &=\left( \begin{array}{c}
	0\\
	\chi_n^{-1}\\
\end{array} \right) +z^{-1}\left( \begin{array}{c}
	-\chi_n^{-1}q_{n}\\
	0\\
\end{array} \right) +z^{-2}\left( \begin{array}{c}
	0\\
	-\chi_n^{-1}\sum_{j=n}^{+\infty}{q_{j+1}\bar{q}_{j}}\\
\end{array} \right) +O\left( z^{-3} \right),\\
&=\left( \begin{array}{c}
	0\\
	\chi_n^{-1}\\
\end{array} \right) -\sum_{p=0}^{-\infty}{z^{2p-1}\left( \begin{array}{c}
	\chi_n^{-1}\sum_{l=n}^{n-p}{q_l\Psi _{l+1,2}^{\left( 2 \right),\left< 2\left( p-n+l \right) \right>}} \\
	0\\
\end{array} \right)}+\sum_{p=-1}^{-\infty}{z^{2p}\left( \begin{array}{c}
	0\\
	\chi_n^{-1}\sum_{j=n}^{+\infty}{\bar{q}_j\Psi _{j+1,2}^{\left( 1 \right) ,\left< 2p+1 \right>}}\\
\end{array} \right)}.
\end{aligned}
\end{equation}

Then combining the Laurent series expansions~(\ref{2.1.42}),~(\ref{2.1.44})-(\ref{2.1.46}) with Eqs.~(\ref{2.1.13}), the asymptotic behavior of the scattering coefficients $s_1(z)$ and $s_4(z)$ can be respectively
derived:
\begin{align}
s_1(z)&=1+O(z^{-2p}),\quad\quad\ \text{as}\ |z|\rightarrow\infty,\label{2.1.47}\\
s_4(z)&=1+O(z^{2p}),\quad\quad\quad  \text{as}\ |z|\rightarrow 0,\label{2.1.48}
\end{align}
where $p\in\mathbb{Z}_+$.

According to Eqs.~(\ref{2.1.13}),~(\ref{2.1.42}) and (\ref{2.1.44})-(\ref{2.1.48}), the following Lemma can be obtained.

\paragraph{Proposition 2.3} The functions $\Psi_{n,1}(z)$, $\Psi_{n,2}(z)$, $\alpha_{n}(z)$ and $\beta_{n}(z)$  have the following asymptotic
behaviors:
\begin{align*}
\Psi_{n,1}(z)&\rightarrow\left( \begin{array}{c}
	a_{n}^{-1}\\
	0\\
\end{array} \right),\quad \beta_{n}(z)\rightarrow\left( \begin{array}{c}
	0\\
	1\\
\end{array} \right),\quad \quad \quad \text{as}\ |z|\rightarrow 0,\\
\Psi_{n,2}(z)&\rightarrow\left( \begin{array}{c}
	0\\
	a_{n}^{-1}\\
\end{array} \right),\quad \alpha_{n}(z)\rightarrow\left( \begin{array}{c}
	1\\
	0\\
\end{array} \right),\quad \quad \quad \text{as}\ |z|\rightarrow \infty.
\end{align*}
and
\begin{align*}
\beta_{n}(z) =\frac{\Phi_{n,2}(z)}{s_4(z)}=\left( \begin{array}{c}
	0\\
	1\\
\end{array} \right) +z\left( \begin{array}{c}
	q_{n-1}\\
	0\\
\end{array} \right) +\text{O}(z^{2}).
\end{align*}

Hence we can derive that
\begin{align}\label{2.1.49}
q_n(z)=\lim_{z\to 0}z^{-1}\beta^{(1)}_{n+1}(z).
\end{align}

The scattering problem can have the discrete spectrum that consists of the eigenvalues satisfying the decay of the eigenfunctions at infinity,
which corresponds to the bound states. As a matter of fact, it can be made up of eigenvalues that let $s_1(z_l)=0$ or $s_4(\tilde{z}_j)=0$.
Here we suppose that $s_1(z), \ s_4(z)\neq0$ on $|z|=1$, which means that the number of eigenvalues is finite.

At the discrete eigenvalues, we can get from Eq.~(\ref{2.1.11}) that there are
\begin{align}
\Phi_{n,1}(z_l)&=z_l^{-2n} s_3(z_l)\Psi_{n,2}(z_l),\label{2.1.50}\\
\Phi_{n,2}(\tilde{z}_j)&=\tilde{z}_j^{2n} s_2(\tilde{z}_j)\Psi_{n,1}(\tilde{z}_j).\label{2.1.51}
\end{align}

\subsubsection{Symmetries }

\hspace{1.5em}The symmetry of the eigenfunctions will be given in Proposition 2.4 and 2.5.

\paragraph{Proposition 2.4} The eigenfunctions $\phi_{n}(z)$ and $\psi_{n}(z)$ have the following symmetry:
\begin{gather*}
\phi_{n}(z)=D^{-1}\bar{\phi}_{n}(\bar{z}^{-1})D,\quad \psi_{n}(z)=D^{-1}\bar{\psi}_{n}(\bar{z}^{-1})D,
\end{gather*}
where
\begin{gather*}
D=\left( \begin{matrix}
	0&		-1\\
	1&		0\\
\end{matrix} \right) .
\end{gather*}
\begin{proof} First, it is easily to derive that $M_{n}(z)=D^{-1}\bar{M}_{n}(\bar{z}^{-1})D$. And we can get
$\bar{\phi}_{n+1}(\bar{z}^{-1})=\bar{M}_n(\bar{z}^{-1}) \bar{\phi}_{n}(\bar{z}^{-1})$
from Eq.~(\ref{1.2}). Then there is
\begin{align*}
D^{-1}\bar{\phi}_{n+1}(\bar{z}^{-1})D=D^{-1}\bar{M}_n(\bar{z}^{-1})D
D^{-1}\bar{\phi}_{n}(\bar{z}^{-1})D={M}_n(z)D^{-1}\bar{\phi}_{n}(\bar{z}^{-1})D,
\end{align*}
that is to say, $D^{-1}\bar{\phi}_{n}(\bar{z}^{-1})D$ and ${\phi}_{n}(z)$ satisfy the same differential equation.
\\\indent Second, according to the asymptotic property~(\ref{2.1.4}), we can derive that $D^{-1}\bar{\phi}_{n}(\bar{z}^{-1})D\sim I$, that is
${\phi}_{n}(z)$ and $D^{-1}\bar{\phi}_{n}(\bar{z}^{-1})D$  satisfy the same asymptotic property.
\\\indent In conclusion, there is $\phi_{n}(z)=D^{-1}\bar{\phi}_{n}(\bar{z}^{-1})D$. In the same way,
$\psi_{n}(z)=D^{-1}\bar{\psi}_{n}(\bar{z}^{-1})D$ can be proved.
\end{proof}

According to Eqs.~(\ref{2.1.42}), and~(\ref{2.1.44})-(\ref{2.1.46}), the Proposition 2.5 can be obtained.

\paragraph{Proposition 2.5} The modified eigenfunctions have the following symmetry:
\begin{alignat*}{4}
\Phi_{n,1}^{(1)}(-z)&=\Phi_{n,1}^{(1)}(z),&\quad \ \  & \Phi_{n,2}^{(2)}(-z)=\Phi_{n,2}^{(2)}(z),&\quad\ \  & \Psi_{n,1}^{(1)}(-z)=\Psi_{n,1}^{(1)}(z),&\quad\ \ &
\Psi_{n,2}^{(2)}(-z)=\Psi_{n,2}^{(2)}(z),\\
\Phi_{n,1}^{(2)}(-z)&=-\Phi_{n,1}^{(2)}(z),&&\Phi_{n,2}^{(1)}(-z)=-\Phi_{n,2}^{(1)}(z),&&
 \Psi_{n,1}^{(2)}(-z)=-\Psi_{n,1}^{(2)}(z),&&\Psi_{n,2}^{(1)}(-z)=-\Psi_{n,2}^{(1)}(z).
\end{alignat*}

Combining Eqs.~(\ref{2.1.10})-(\ref{2.1.11}) with the Proposition 2.4 and 2.5, and according to Eqs.~(\ref{2.1.47}) and~(\ref{2.1.48}), we derive the symmetry properties of
the scattering coefficients, which will be given in Proposition 2.6.

\paragraph{Proposition 2.6} The scattering matrix $S(z)$ has the following symmetry properties:

{\it Symmetry 1}:
 \begin{align*}
S(z)=D^{-1}\bar{S}(\bar{z}^{-1})D,
\end{align*}
i.e., the scattering coefficients and the reflection coefficients satisfy
\begin{equation*}
 s_1(z)=\bar{s}_4(\bar{z}^{-1}),\quad\quad s_2(z)=-
\bar{s}_3(\bar{z}^{-1}),\quad\quad \gamma_2(z)=- \bar{\gamma}_1(\bar{z}^{-1}).
\end{equation*}

{\it Symmetry 2}:
 \begin{alignat*}{3}
 &s_1(-z)=s_1(z),&\quad\quad\quad& s_4(-z)=s_4(z),\\
&s_1'(-z)=-s_1(z),&\quad\quad\quad& s_4'(-z)=-s_4(z),\\
&s_3(-z_l)=-s_1(z_l),&\quad\quad\quad& s_2(-\tilde{z}_j)=-s_2(\tilde{z}_j).
\end{alignat*}

\begin{proof} For {\it Symmetry 1}, since we can get $D^{-1}\bar{S}(\bar{z}^{-1})D=D^{-1}\bar{\psi}^{-1}_n(\bar{z}^{-1})D\,D^{-1}\bar{\phi}_n(\bar{z}^{-1})D $
from Eq.~(\ref{2.1.10}), combined with the Proposition 2.4, there is $D^{-1}\bar{S}(\bar{z}^{-1})D=\psi^{-1}_n(z) \phi_n(z) =S(z)$.
\\\indent For {\it Symmetry 2}, It follows from Eqs.~(\ref{2.1.47}) and~(\ref{2.1.48}) that $s_1(z)$ and $s_4(z)$ are even functions of $z$, and consequently,
their first derivatives are odd functions. By Eqs.~(\ref{2.1.50}) and~(\ref{2.1.51}), we get $\sigma_3 \Phi_{n,1}(z_l)= z_l^{-2n}
s_3(z_l)\sigma_3\Psi_{n,2}(z_l)$, and $\sigma_3\Phi_{n,2}(\tilde{z}_j)=\tilde{z}_j^{2n} s_2(\tilde{z}_j)\sigma_3\Psi_{n,1}(\tilde{z}_j)$,
where $\sigma_3=\text{diag}(1,\,-1) $. Then using the Proposition 2.5, there are $\Phi_{n,1}(-z_l)= (-z_l)^{-2n}\big(- s_3(z_l)\big)\Psi_{n,2}(-z_l)$ and
$\Phi_{n,2}(-\tilde{z}_j)=(-\tilde{z}_j)^{2n}\big(- s_2(-\tilde{z}_j)\big)\Psi_{n,1}(\tilde{z}_j)$. i.e., $s_3(-z_l)=-s_3(z_l)$ and
$s_2(-\tilde{z}_j)=-s_2(\tilde{z}_j)$.
\end{proof}

It follows that $s_4'( z ) =-z^{-2}\overline{ s}_1'( \bar{z}^{-1} )$ by the Proposition 2.6.

Then there are $4N$ discrete eigenvalues according to the Proposition 2.6, and the set of them are $\Omega=\{\pm z_l, \ \pm \tilde{z}_l , \
l=1, 2, \cdots,  N:|z_l|>1 \ \text{and} \ \tilde{z}_l=\bar{z}^{-1}_l \}$.

\subsection{  Inverse scattering problem}
\renewcommand{\theequation}{\arabic{section}.\arabic{subsection}.\arabic{equation}}\setcounter{equation}{0}

\subsubsection{RH problem and reconstruction formula }

\hspace{1.5em}According to the Proposition 2.3, we first introduce
\begin{gather}\label{2.2.1}
 \tilde{\Psi}_{n}(z)= \left( \begin{matrix}
	1&		0\\
	0&		\chi_n\\
\end{matrix} \right)\Psi_{n}(z),\quad\quad \tilde{\alpha}_{n}(z)= \left( \begin{matrix}
	1&		0\\
	0&		\chi_n\\
\end{matrix} \right)\alpha_{n}(z),\quad\quad\tilde{\beta}_{n}(z)= \left( \begin{matrix}
	1&		0\\
	0&		\chi_n\\
\end{matrix} \right)\beta_{n}(z).
\end{gather}
Thus
\begin{align*}
\tilde{\Psi}_{n,1}(z)&\rightarrow\left( \begin{array}{c}
	\chi_{n}^{-1}\\
	0\\
\end{array} \right),\quad \tilde{\beta}_{n}(z)\rightarrow\left( \begin{array}{c}
	0\\
	\chi_{n}\\
\end{array} \right),\quad \quad \quad \text{as}\ |z|\rightarrow 0,\\
\tilde{\Psi}_{n,2}(z)&\rightarrow\left( \begin{array}{c}
	0\\
	1\\
\end{array} \right),\quad\quad \tilde{\alpha}_{n}(z)\rightarrow\left( \begin{array}{c}
	1\\
	0\\
\end{array} \right),\quad \quad \quad\ \, \text{as}\ |z|\rightarrow \infty,
\end{align*}
and Eqs.~(\ref{2.1.15}) can be rewritten as
\begin{align*}
 \tilde{\alpha }_n\left( z \right) &=\tilde{\Psi} _{n,1}\left( z \right) +z^{-2n}\gamma _1\left( z \right) \tilde{\Psi} _{n,2}\left( z
 \right),\\
\tilde{\beta} _n\left( z \right) &=\tilde{\Psi} _{n,2}\left( z \right) +z^{2n}\gamma _2\left( z \right) \tilde{\Psi }_{n,1}\left( z \right).
\end{align*}

In addition, from Proposition 2.1 and 2.2 we get that  $\tilde{\Psi}_{n,2}(z)$ is analytic in $|z|>1$ and $\tilde{\Psi}_{n,1}(z)$  is analytic in
$|z|<1$. Meanwhile, $\tilde{\alpha}_{n}(z)$  is analytic in $|z|>1$ except not at the zeros of $s
_1(z)$, and $\tilde{\beta}_{n}(z)$ is analytic in $|z|<1$ except not at the zeros of $s
_4(z)$.

Define the counterclockwise directed curve $\Sigma=\{z: |z |=1\}$ and the piecewise function
\begin{gather}\label{2.2.2}
R_n\left(z\right)=\left\{ \begin{array}{l}R^{+}_n\left(z\right)=
	\big( \begin{matrix}
	\tilde{\Psi}_{n,1}(z)&		\tilde{\beta}_{n}(z)\\
\end{matrix} \big) ,\quad \ |z|<1,\\
R^{-}_n\left(z\right)=\big( \begin{matrix}
	\tilde{\alpha}_{n}(z)&		\tilde{\Psi}_{n,2}(z)\\
\end{matrix} \big) ,\quad\  |z|>1.\\
\end{array} \right.
\end{gather}
 Then we obtain the following generalized RH problem:

\vspace{5mm}{\bf RHP 2.1}\quad Find a $2\times2$ matrix $R_n\left(z\right)$ such that\\
(1)\ $R_n\left(z\right)$ is analytic on $\mathbb{C}\setminus\Sigma\cup\Omega$,\\
(2)\ $R^{-}_n\left(z\right)=R^{+}_n\left(z\right)(I+H_n\left(z\right))$ on $\Sigma$, where $H_n(z)=\left( \begin{matrix}
	-\gamma _1\left( z \right) \gamma _2\left( z \right)&		-z^{2n}\gamma _2\left( z \right)\\
	z^{-2n}\gamma _1\left( z \right)&		0\\
\end{matrix} \right) $,\\
(3)\ $R_n\left(z\right)\rightarrow I$ as $\left|z  \right|\rightarrow \infty$.

Combining Eqs.~(\ref{2.1.49}) with~(\ref{2.2.1}) and~(\ref{2.2.2}) yields
\begin{gather}\label{2.2.3}
q_n(z)=\lim_{z\to 0}z^{-1}\Big(R^{+}_{n+1}\left(z\right)\Big)^{(1,\,2)}.
\end{gather}

Assume that the scattering coefficients $s_1(z)$ and $s_4(z)$ have only simple zeros, i.e., the poles of $R_n\left(z\right)$ are all simple. By
removing the singularity of $R_n\left(z\right)$ at discrete eigenvalues, a RH problem about $\tilde{R}_n\left(z\right)$ can be obtained, in
which $\tilde{R}_n\left(z\right)=R_n\left(z\right)-\sum_{l=1}^N{\frac{\text{Res}\left(R^{-}_n\left(z\right),z_l
\right)}{z-z_l}}-\sum_{l=1}^N{\frac{\text{Res}\left(R^{-}_n\left(z\right),-z_l
\right)}{z+z_l}}-\sum_{l=1}^N{\frac{\text{Res}\left(R^{+}_n\left(z\right),\tilde{z}_l
\right)}{z-\tilde{z}_l}}-\sum_{l=1}^N{\frac{\text{Res}\left(R^{+}_n\left(z\right),-\tilde{z}_l  \right)}{z+\tilde{z}_l}}$.

\vspace{5mm}{\bf RHP 2.2}\quad Find a $2\times2$ matrix $\tilde{R}_n\left(z\right)$ such that\\
(1)\ $\tilde{R}_n\left(z\right)$ is analytic on $\mathbb{C}\setminus\Sigma$,
\begin{flalign}\label{2.2.4}
&\text{(2)}\ \tilde{R}^{-}_n\left(z\right)-\tilde{R}^{+}_n\left(z\right)=R^{+}_n\left(z\right)H_n\left(z\right)\ \text{on}\ \Sigma,&
\end{flalign}
(3)\ $\tilde{R}_n\left(z\right)\rightarrow I$ as $\left|z  \right|\rightarrow \infty$.

\paragraph{Proposition 2.7} The residues of $R_n\left(z\right)$ at discrete eigenvalues can be expressed as
\begin{align*}
  \text{Res}\left(R^{+}_n\left( z\right),\pm \tilde{z}_l  \right)&=\left( 0\, ,\, \tilde{C}_l\tilde{z}_l^{2n} \tilde{\Psi} _{n,1}\left( \pm
  \tilde{z}_l \right)\right) ,\\
  \text{Res}\left(R^{-}_n\left(z\right),\pm z_l  \right)&=\left(C_l  z_l^{-2n}  \tilde{\Psi} _{n,2}\left( \pm z_l \right),\, 0\right),
\end{align*}
where the norming constants can be written as
\begin{gather*}
C_l=\frac{s_3(z_l)}{s_1'(z_l)},\quad\quad  \tilde{C}_l=\frac{s_2(\tilde{z}_l)}{s_4'(\tilde{z}_l)}= \bar{z}_l^{-2}\bar{C}_l.
\end{gather*}

\begin{proof} According to Eqs.~(\ref{2.1.15}) , we first derive
 \begin{align*}
  &\text{Res}\left(\alpha_n\left(z\right),z_l  \right)=\frac{\Phi_{n,1}(z_l)}{s_1'(z_l)}=\frac{s_3(z_l)}{s_1'(z_l)}z_l^{-2n}  \Psi
  _{n,2}\left( z_l \right)=C_l z_l^{-2n}  \Psi _{n,2}\left( z_l \right) ,\\
  &\text{Res}\left(\alpha_n\left(z\right),-z_l  \right)=\frac{\Phi_{n,1}(-z_l)}{s_1'(-z_l)}=\frac{s_3(-z_l)}{s_1'(-z_l)}(-z_l)^{-2n}  \Psi
  _{n,2}\left( -z_l \right)=C_l^{-} z_l^{-2n}  \Psi _{n,2}\left( -z_l \right) ,\\
  &\text{Res}\left(\beta_n\left(z\right),\tilde{z}_l
  \right)=\frac{\Phi_{n,2}(\tilde{z}_l)}{s_4'(\tilde{z}_l)}=\frac{s_2(\tilde{z}_l)}{s_4'(\tilde{z}_l)}\tilde{z}_l^{2n}  \Psi _{n,1}\left(
  \tilde{z}_l \right)=\tilde{C}_l\tilde{z}_l^{2n}  \Psi _{n,1}\left( \tilde{z}_l \right),\\
  &\text{Res}\left(\beta_n\left(z\right),-\tilde{z}_l
  \right)=\frac{\Phi_{n,2}(-\tilde{z}_l)}{s_4'(-\tilde{z}_l)}=\frac{s_2(-\tilde{z}_l)}{s_4'(-\tilde{z}_l)}-\tilde{z}_l^{2n}  \Psi _{n,1}\left(
  -\tilde{z}_l \right)=\tilde{C}_l^{-}(-\tilde{z}_l)^{2n}  \Psi _{n,1}\left( -\tilde{z}_l \right).
\end{align*}
By the Proposition 2.6, it is easy to derive that $\dfrac{s_3(-z_l)}{s_1'(-z_l)}=\dfrac{s_3(-z_l)}{s_1'(-z_l)}$,
$\dfrac{s_2(-\tilde{z}_l)}{s_4'(-\tilde{z}_l)}=\dfrac{s_2(\tilde{z}_l)}{s_4'(\tilde{z}_l)}$ and
$\dfrac{s_2(\tilde{z}_l)}{s_4'(\tilde{z}_l)}= \bar{z}_l^{-2}\dfrac{\bar{s}_3(z_l)}{\bar{s}_1'(z_l)}$, which  means that $C_1^{-}=C_1$,
$\tilde{C}_1^{-}=\tilde{C}_1$ and $\tilde{C}_l= \bar{z}_l^{-2}\bar{C}_l$.
\\\indent Therefore,
\begin{align*}
 \text{Res}\left(R^{+}_n\left(z\right)\, ,\,\pm z_l  \right)&=\Big(0\, ,\,\text{Res}\big(\tilde{\beta}_n(z)\, ,\,\pm z_l\big)\Big)=\left(
   \begin{matrix}
	1&		0\\
	0&		\chi_n\\
\end{matrix} \right)\Big(0,\,\text{Res}\big(\beta_n(z)\, ,\,\pm z_l\big)\Big)\\
&=\left( \begin{matrix}
	1&		0\\
	0&		\chi_n\\
\end{matrix} \right)\Big(0\, ,\,\tilde{C}_l\tilde{z}_l^{2n}  \Psi _{n,1}\left(\pm \tilde{z}_l \right) \Big)=\left( 0\, ,\,
\tilde{C}_l\tilde{z}_l^{2n}  \tilde{\Psi} _{n,1}\left(\pm \tilde{z}_l \right)\right),\\
  \text{Res}\left(R^{-}_n\left(z\right)\, ,\,\pm z_l  \right)&=\Big(\text{Res}\big(\tilde{\alpha}_n(z)\,,\,\pm z_l\big)\,,\, 0\Big)=\left(
  \begin{matrix}
	1&		0\\
	0&		\chi_n\\
\end{matrix} \right)\Big(\text{Res}\big(\alpha_n(z)\,,\,\pm z_l\big)\, ,\, 0\Big)\\
&=\left( \begin{matrix}
	1&		0\\
	0&		\chi_n\\
\end{matrix} \right)\Big(C_l z_l^{-2n}  \Psi _{n,2}\left(\pm z_l \right)\, ,\, 0\Big)=\left(C_l z_l^{-2n}  \tilde{\Psi} _{n,2}\left(\pm z_l
\right)\, , \, 0\right)   .
\end{align*}
\end{proof}

The jump condition Eq.~(\ref{2.2.4}) can be rewritten as
\begin{equation}\label{2.2.5}
(\tilde{R}^{+}_n\left(z\right)-I)-(\tilde{R}^{-}_n\left(z\right)-I)=-R^{+}_n\left(z\right)H_n\left(z\right).
\end{equation}
Define integral operators $\tilde{P}$ and $P$ as follows:
\begin{gather*}
\tilde{P}[f](z)=\frac{1}{2\pi i}\lim_{\substack{z'\to z\\ |z'|>1}}\oint_{|\rho |=1}{\frac{f\left( \rho \right)}{\rho -z'}}\text{d}\rho,  \quad\quad
|z|\leq1,\\
P[f](z)=\frac{1}{2\pi i}\lim_{\substack{z'\to z\\ |z'|<1}}\oint_{|\rho |=1}{\frac{f\left( \rho \right)}{\rho
-z'}}\text{d}\rho,\quad\quad|z|\geq1,
\end{gather*}
and respectively applying them to both sides of Eq.~(\ref{2.2.5}) yields
\begin{gather*}
\tilde{R}^{+}_n\left(z\right)=I-\frac{1}{2\pi i}\lim_{\substack{z'\to z\\ |z'|>1}}\oint_{|\rho
|=1}{\frac{R^{+}_n\left(\rho\right)H_n\left(\rho\right)}{\rho -z'}}\text{d}\rho,\\
\tilde{R}^{-}_n\left(z\right)=I-\frac{1}{2\pi i}\lim_{\substack{z'\to z\\ |z'|<1}}\oint_{|\rho
|=1}{\frac{R^{+}_n\left(\rho\right)H_n\left(\rho\right)}{\rho -z'}}\text{d}\rho,
\end{gather*}
which is based on the Cauchy's integral formula and the Cauchy's theorem.
Accordingly, there are
\begin{align}
R^{+}_n\left(z\right)=&I+\sum_{l=1}^N\Big({\frac{\text{Res}\left(R^{-}_n\left(z\right),z_l
\right)}{z-z_l}}+{\frac{\text{Res}\left(R^{-}_n\left(z\right),-z_l \right)}{z+z_l}}+{\frac{\text{Res}\left(R^{+}_n\left(z\right),\tilde{z}_l
\right)}{z-\tilde{z}_l}}+{\frac{\text{Res}\left(R^{+}_n\left(z\right),-\tilde{z}_l  \right)}{z+\tilde{z}_l}}\Big)\notag\\
&-\frac{1}{2\pi i}\lim_{\substack{z'\to z\\ |z'|>1}}\oint_{|\rho |=1}{\frac{R^{+}_n\left(\rho\right)H_n\left(\rho\right)}{\rho
-z'}}\text{d}\rho,\label{2.2.6}\\
R^{-}_n\left(z\right)=&I+\sum_{l=1}^N\Big({\frac{\text{Res}\left(R^{-}_n\left(z\right),z_l
\right)}{z-z_l}}+{\frac{\text{Res}\left(R^{-}_n\left(z\right),-z_l \right)}{z+z_l}}+{\frac{\text{Res}\left(R^{+}_n\left(z\right),\tilde{z}_l
\right)}{z-\tilde{z}_l}}+{\frac{\text{Res}\left(R^{+}_n\left(z\right),-\tilde{z}_l  \right)}{z+\tilde{z}_l}}\Big)\notag\\
&-\frac{1}{2\pi i}\lim_{\substack{z'\to z\\ |z'|<1}}\oint_{|\rho |=1}{\frac{R^{+}_n\left(\rho\right)H_n\left(\rho\right)}{\rho
-z'}}\text{d}\rho.\label{2.2.7}
\end{align}

According to Eqs.~(\ref{2.2.3}) and~(\ref{2.2.7}), and combining with the Proposition 2.5 and the Proposition 2.7, the reconstruction formula of
the potential function $q_n(z)$ can be derived as below:
\begin{gather}\label{2.2.8}
q_n=-2\sum_{l=1}^N{\tilde{C}_l\tilde{z}_l^{2n} \tilde{\Psi}^{(1)} _{n+1,1}\left( \tilde{z}_l \right)}+\frac{1}{2\pi i}\oint_{|\rho
|=1}{{\rho}^{2n}\gamma_2(\rho)\tilde{\Psi}^{(1)} _{n+1,1}(\rho)}\text{d}\rho.
\end{gather}

\subsubsection{ Trace formulas }

\hspace{1.5em}Recall that $s_1(z)$ and $s_4(z)$ are analytic for $|z| > 1$ and $|z| < 1$, with zeros at $\{\pm z_l, | \ l=1, 2, \cdots,  N\}$ and $\{\pm
\tilde{z}_l |\ l=1, 2, \cdots,  N\}$, respectively.
If we define
\begin{gather*}
\tilde{s}_1(z)=\prod_{l=1}^N{\frac{z^2-\tilde{z}_{l}^{2}}{z^2-z_{l}^{2}}}s_1(z),\quad\quad
\tilde{s}_4(z)=\prod_{l=1}^N{\frac{z^2-z_{l}^{2}}{z^2-\tilde{z}_{l}^{2}}}s_4(z),
\end{gather*}
then accordingly, the function $\tilde{s}_1(z)$ is analytic for $|z| > 1$ while $\tilde{s}_4(z)$ is analytic for $|z| < 1$ , and they have no
zeros. In addition, due to Eqs.~(\ref{2.1.47}) and~(\ref{2.1.48}), both $\tilde{s}_1(z)$ and $\tilde{s}_4(z)$ are even functions of $z$, and
$\tilde{s}_1(z)\rightarrow1$ as $|z|\rightarrow\infty$.

According to Eqs.~(\ref{2.1.10}), (\ref{2.1.14}) and the Proposition 2.6, we can derive
\begin{gather*}
\tilde{s}_1(z)\tilde{s}_4(z)=\chi_{-\infty}(1+|\gamma_1(z)|^2)^{-1}
\end{gather*}
for $|z|=1$ with $\sigma=-1$,
i.e., $\log\big(\tilde{s}_1(z)\big)+\log\big(\tilde{s}_4(z)\big)=\log\big(\chi_{-\infty}\big)-\log\big(1+|\gamma_1(z)|^2\big).$

Define
\begin{gather*}
\varsigma(z)=\left\{ \begin{array}{l}
	\varsigma^+(z)=\tilde{s}_4(z), \quad\quad|z|<1,\\
	\varsigma^-(z)=\tilde{s}_1(z), \quad\quad|z|>1.\\
\end{array} \right.
\end{gather*}
Then a RH problem about $\varsigma(z)$ can be given as below:

\vspace{5mm}{\bf RHP 2.3}\quad Find a scalar function $\varsigma(z)$ such that\\
(1)\ $\varsigma(z)$ is analytic in $\mathbb{C}\setminus\Sigma$,
\begin{flalign*}
&\text{(2)}\ \log\big(\varsigma^+(z)\big)+\log\big(\varsigma^-(z)\big)=\log(\chi_{-\infty})-\log\big(1+|\gamma_1(z)|^2\big)\ \text{on}\ \Sigma,&
\end{flalign*}
(3)\ $\varsigma(z)\rightarrow 1$ as $\left|z  \right|\rightarrow \infty$.

 Applying the integral operators $P$ and $\tilde{P}$ to the jump condition can respectively yield
 \begin{align*}
\log\tilde{s}_1(z)&= \log\big(\varsigma^-(z)\big)=\lim_{\substack{z'\to z\\ |z'|>1}}\frac{1}{2\pi i}\oint_{|\rho
|=1}{\frac{\log\big(1+|\gamma_1(z)|^2\big)}{\rho-z}}\text{d}\rho,\\
\log\tilde{s}_4(z)&= \log\big(\varsigma^+(z)\big)=-\lim_{\substack{z'\to z\\ |z'|>1}}\frac{1}{2\pi i}\oint_{|\rho
|=1}{\frac{\log\big(1+|\gamma_1(z)|^2\big)}{\rho-z}}\text{d}\rho,
\end{align*}
and then
\begin{align*}
 \log\tilde{s}_1(-z)&=\lim_{\substack{z'\to z\\ |z'|>1}}\frac{1}{2\pi i}\oint_{|\rho
 |=1}{\frac{\log\big(1+|\gamma_1(z)|^2\big)}{\rho+z}}\text{d}\rho,\\
  \log\tilde{s}_4(-z)&=-\lim_{\substack{z'\to z\\ |z'|>1}}\frac{1}{2\pi i}\oint_{|\rho
 |=1}{\frac{\log\big(1+|\gamma_1(z)|^2\big)}{\rho+z}}\text{d}\rho.
\end{align*}
Since  $\tilde{s}_1(z)$ is a even function of $z$, consequently
  \begin{align*}
\log\tilde{s}_1(z)&=\frac{1}{2}\big(\log\tilde{s}_1(z)+\log\tilde{s}_1(-z)\big)=\lim_{\substack{z'\to z\\ |z'|>1}}\frac{1}{2\pi i}\oint_{|\rho
|=1}{\frac{\rho \log\big(1+|\gamma_1(z)|^2\big)}{\rho^2-z^2}}\text{d}\rho,\\
\log\tilde{s}_4(z)&=\frac{1}{2}\big(\log\tilde{s}_4(z)+\log\tilde{s}_4(-z)\big)=-\lim_{\substack{z'\to z\\ |z'|>1}}\frac{1}{2\pi i}\oint_{|\rho
|=1}{\frac{\rho \log\big(1+|\gamma_1(z)|^2\big)}{\rho^2-z^2}}\text{d}\rho.
\end{align*}

Using the relation between $\tilde{s}_1(z)$ and $s_1(z)$, we then obtain the trace formulas:
 \begin{alignat*}{2}
s_1(z)&= \prod_{l=1}^N{\frac{z^2-z_{l}^{2}}{z^2-\tilde{z}_{l}^{2}}}\exp\Big[\frac{1}{2\pi i}\oint_{|\rho |=1}{\frac{\rho
\log\big(1+|\gamma_1(z)|^2\big)}{\rho^2-z^2}}\text{d}\rho\Big],&\quad \ \quad&|z|>1,\\
s_4(z)&= \prod_{l=1}^N{\frac{z^2-\tilde{z}_{l}^{2}}{z^2-z_{l}^{2}}}\exp\Big[\frac{-1}{2\pi i}\oint_{|\rho |=1}{\frac{\rho
\log\big(1+|\gamma_1(z)|^2\big)}{\rho^2-z^2}}\text{d}\rho\Big], &&|z|<1.
\end{alignat*}

\subsection{ Time evolution}
\renewcommand{\theequation}{\arabic{section}.\arabic{subsection}.\arabic{equation}}\setcounter{equation}{0}

\hspace{1.5em}The time evolution of the eigenfunctions can be determined by the time-dependence equation~(\ref{1.3}), whose asymptotic form can be expressed
as
 \begin{gather}\label{2.3.1}
\frac{\text{d}v_n}{\text{d}t}=\left( \begin{matrix}
	\omega \left( z \right)&		0\\
	0&		\omega \left( z^{-1} \right)\\
\end{matrix} \right) v_n,\quad \quad\ n\rightarrow\pm\infty,
\end{gather}
in which
\begin{gather*}
\omega \left( z \right)=-\frac{3}{4}z^{-4}+2z^{-2}+\frac{1}{4}z^{4}-\frac{3}{2}.
\end{gather*}

It is not difficult to derive that the fundamental matrix solution of Eq.~(\ref{2.3.1}) can be written as
\begin{gather*}
\tilde{v}_n=\left( \begin{matrix}
	e^{\omega( z )t}&	0	\\
	0&	e^{\omega ( z^{-1})t}	\\
\end{matrix} \right) .
\end{gather*}
We define the time-dependence eigenfunctions as
\begin{gather}\label{2.3.2}
W_n(z,t)=\phi _n(z,t)\tilde{v}_n(z,t),\quad\quad N_n(z,t)=\psi _n(z,t)\tilde{v}_n(z,t),
\end{gather}
which are simultaneous solutions of the Lax pair~(\ref{1.2})-(\ref{1.3}) and have the BCs as
 \begin{gather*}
W _n(z,t)\sim Z^n \tilde{v}_n,   \ \ \ \  n\rightarrow-\infty; \quad \quad \quad
N _n(z,t)\sim Z^n\tilde{v}_n,   \ \ \ \  n\rightarrow+\infty.
\end{gather*}
From Eq.~(\ref{2.3.2}) one has
\begin{align*}
  \text{det}\,W_{n}(z,t)&=\text{det}\,\phi _n(z,t)\,\text{det}\,\tilde{v}_n(z,t)=\frac{\chi_{-\infty}}{\chi_n}>0,\\
  \text{det}\,N_{n}(z,t)&=\text{det}\,\psi _n(z,t)\,\text{det}\,\tilde{v}_n(z,t)=\frac{1}{\chi_n}>0.
\end{align*}
That is to say, both $W_{n}(z,t)$ and $N_{n}(z,t)$ are fundamental matrix solutions of Eqs.~(\ref{1.2})-(\ref{1.3}).

Therefore, there is the scattering matrix $\tilde{S}(z)$ which is independent of $n$ and $t$ and satisfies
 \begin{gather}\label{2.3.3}
W_n(z,t)=N_n(z,t)\tilde{S}(z) .
\end{gather}
 And
 \begin{gather*}
 \tilde{S}(z)=\tilde{v}^{-1}_nS(z)\tilde{v}_n=\left( \begin{matrix}
	s_1(z)&	s_2(z)e^{-\tilde{\omega}(z)t}	\\
	s_3(z)e^{\tilde{\omega}(z)t}&	s_4(z)	\\
\end{matrix} \right)
\end{gather*}
can be derived by combining Eqs.~(\ref{2.3.3}), (\ref{2.3.2}) with~(\ref{2.1.10}), where $\tilde{\omega}(z)=\omega(z)-\omega(z^{-1}).$

Since $\tilde{S}(z)$ is independent of $t$, we then obtain
\begin{gather*}
 \frac{\partial}{\partial t}s_1(z)=0,\quad \frac{\partial}{\partial t}s_4(z)=0,\quad   \frac{\partial}{\partial
 t}\big(s_2(z)e^{-\tilde{\omega}(z)t})=0,\quad 	
	\frac{\partial}{\partial t}\big(s_3(z)e^{\tilde{\omega}(z)t}\big)	=0.
\end{gather*}
Therefore
\begin{gather*}
 s_1(z,t)=s_1(z,0),\quad   s_2(z,t)= s_2(z,0)e^{\tilde{\omega}(z)t},\quad 	
	s_3(z,t)= s_3(z,0)s_3(z)e^{-\tilde{\omega}(z)t}, \quad s_4(z,t)=s_4(z,0).
\end{gather*}
That is, $s_1(z)$ and $s_4(z)$ are $t$-independent, and accordingly the eigenvalues are constant with the time evolution of solutions.

The time evolution of the reflection coefficients and the norming constants can be thus given by
\begin{align*}
&\gamma_1(z,t)=\gamma_1(z,0)e^{-\tilde{\omega}(z)t},\quad   \gamma_2(z,t)=\gamma_2(z,0)e^{\tilde{\omega}(z)t},\\
&C_l(t)=C_l(0)e^{-\tilde{\omega}(z_l)t},\quad \quad \quad
\tilde{C}_l(t)=\tilde{C}_l(0)e^{\tilde{\omega}(\tilde{z}_l)t}= \bar{z}_l^{-2}\bar{C}_l(0)e^{\tilde{\omega}(\tilde{z}_l)t}.
\end{align*}

\subsection{  $N$-soliton solution}
\renewcommand{\theequation}{\arabic{section}.\arabic{subsection}.\arabic{equation}}\setcounter{equation}{0}

\hspace{1.5em} We consider the reflectionless case, i.e., $\gamma_1(z)=\gamma_2(z)=0$ on $|z|=1$. In this situation the integrals vanish in both
 Eqs.~(\ref{2.2.6}) and~(\ref{2.2.7}), and then an algebraic system as below can be derived from the algebraic-integral
 system~(\ref{2.2.6})-(\ref{2.2.7}) by using the Proposition 2.5 and Proposition 2.7:
\begin{align}
\tilde{\Psi}_{n,1}^{\left( 1 \right)}\left( \tilde{z}_j \right)
&=1+2\sum_{l=1}^N{\frac{z_{l}^{-2n+1}C_l}{\tilde{z}_{j}^{2}-z_{l}^{2}}}\tilde{\Psi}_{n,2}^{\left( 1 \right)}\left( z_l \right),\label{2.4.1}\\
\tilde{\Psi}_{n,2}^{\left( 1 \right)}\left( z_j \right)
&=2\sum_{l=1}^N{\frac{z_j\tilde{z}_{l}^{2n}\tilde{C}_l}{z_{j}^{2}-\tilde{z}_{l}^{2}}}\tilde{\Psi}_{n,1}^{\left( 1 \right)}\left( \tilde{z}_l
\right).\label{2.4.2}
\end{align}
Accordingly, the reconstruction formula~(\ref{2.2.8}) becomes
\begin{gather}
q_n=-2\sum_{l=1}^N{\tilde{C}_l\tilde{z}_l^{2n} \tilde{\Psi}^{(1)} _{n+1,1}\left( \tilde{z}_l \right)}.
\end{gather}

If we set $G=I+\big(G^{(j,k)}\big)$, in which $G^{(j,k)}=-4\tilde{z}_{k}^{2(n+1)}\tilde{C}_k \sum_{l=1}^N{\dfrac{C_l
z_l^{-2n}}{(\tilde{z}_{j}^{2}-z_{l}^{2})(z_{l}^{2}-\tilde{z}_{k}^{2})}}$, and $X=\big(X_1,\ X_2,\cdots,$ $
X_N\big)=\big(\tilde{\Psi}_{n+1,1}^{\left( 1 \right)}\left( \tilde{z}_1 \right),\ \tilde{\Psi}_{n+1,1}^{\left( 1 \right)}\left( \tilde{z}_2
\right),\ \cdots,\ \tilde{\Psi}_{n+1,1}^{\left( 1 \right)}\left( \tilde{z}_N \right)\big)$, then there is
$GX=1$  according to Eqs.~(\ref{2.4.1}) and~(\ref{2.4.2}).

 By using the Cramer's rule, we consequently derive the $N$-soliton solution:
 \begin{equation}\label{2.4.4}
 q_n=2 \frac{\det\tilde{G}}{\det G},
 \end{equation}
 in which
 \begin{equation*}
 \tilde{G}=\left( \begin{matrix}
	0&	F	\\
	Y&	G	\\
\end{matrix} \right),
  \end{equation*}
and $F=(\tilde{C}_1\tilde{z}_1^{2n},\ \tilde{C}_2\tilde{z}_2^{2n},\ \cdots,\ \tilde{C}_N\tilde{z}_N^{2n})$, while $Y$ is a column vector with $N$ entries that are all 1.

\subsubsection{ 1-soliton }

\hspace{1.5em}When $N=1$, i.e., the set of discrete eigenvalues is $\Omega=\{\pm z_1,\pm\tilde{z}_1\}$ , the solution~(\ref{2.4.4}) reduces to
\begin{equation}\label{2.4.5}
 q_n=\frac{-2 \tilde{C}_1 \tilde{z}_1^{2n}}{1+4C_1\tilde{C}_1 z_1^{-2n}\tilde{z}_1^{2(n+1)}(z_1^2-\tilde{z}_1^2)^{-2}}.
 \end{equation}

We set $z_1=\text{e}^{\nu_1+i \eta_1}$, and $\tilde{z}_1=\text{e}^{-\nu_1+i \eta_1}$. Therefore, considering the time evolution of the norming
constants, 1-soliton solution of the following form can be derived:
\begin{equation}\label{2.4.6}
 q_n=-\frac{\bar{C}_1(0)}{|C_1(0)|}\sinh(2 \nu_1) \sech\big(2 \nu_1(n+1)-\varrho-2 \zeta_1 t\big)\,\text{e}^{2 i\left(\eta_1(n+1)+\zeta_2
 t\right)},
 \end{equation}
in which
\begin{gather*}
\varrho=\log|C_1(0)|-\log\sinh2\nu_1,\\ \zeta_1=2\sinh2\nu_1\cos2\eta_1-\sinh4\nu_1\cos4\eta_1,\\ \zeta_2=\cosh4\nu_1\sin4\eta_1 -2
\cosh2\nu_1\sin2\eta_1.
\end{gather*}
And the shape of the 1-soliton can be plainly presented in Figs. 1.

\begin{figure}[H]
\setcounter{subfigure}{0}
\centering
{\includegraphics[width=9cm]{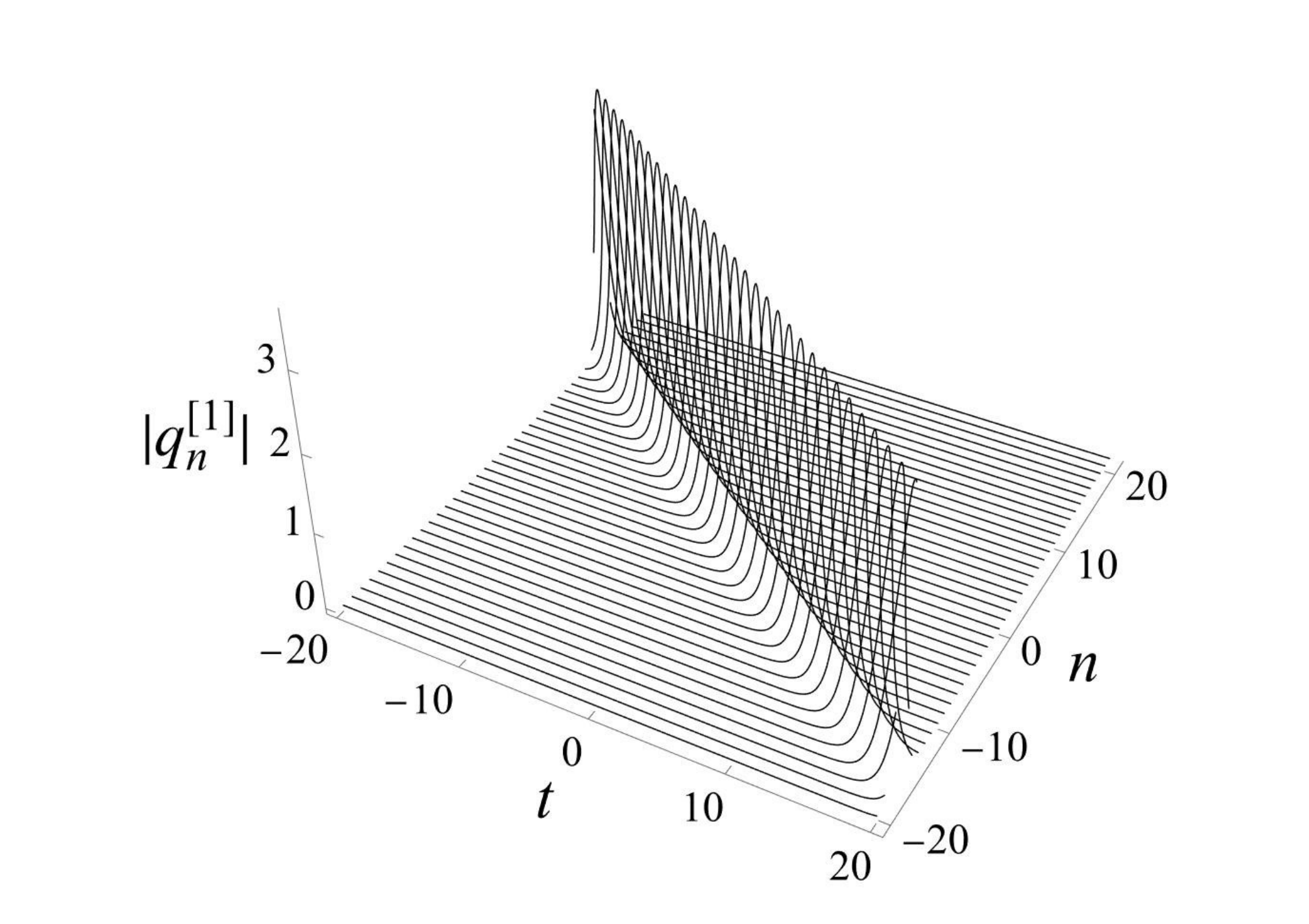}}
\flushleft{\footnotesize
\textbf{Figs.~$2.1$.} The bright 1-soliton of Eq.~(\ref{1.1}) with $C_1(0)=1, \nu_1=1$, and $\eta_1=2 $. }
\end{figure}

\subsubsection{ 2-soliton}

\hspace{1.5em}When $N=2$, i.e., the set of discrete eigenvalues is $\Omega=\{\pm z_1,\pm\tilde{z}_1,\,\pm z_2,\pm\tilde{z}_2\}$, we set
$z_1=\text{e}^{\nu_1+i \eta_1}$, $\tilde{z}_1=\text{e}^{-\nu_1+i \eta_1}, z_2=\text{e}^{\nu_2+i \eta_2}$ and $\tilde{z}_2=\text{e}^{-\nu_2+i
\eta_2}$, and the corresponding 2-soliton solution can be derived from Eq.~(\ref{2.4.4}) by taking the appropriate parameter values and shown in
Fig. 2.

\begin{figure}[H]
\setcounter{subfigure}{0}
\centering
{\includegraphics[width=9cm]{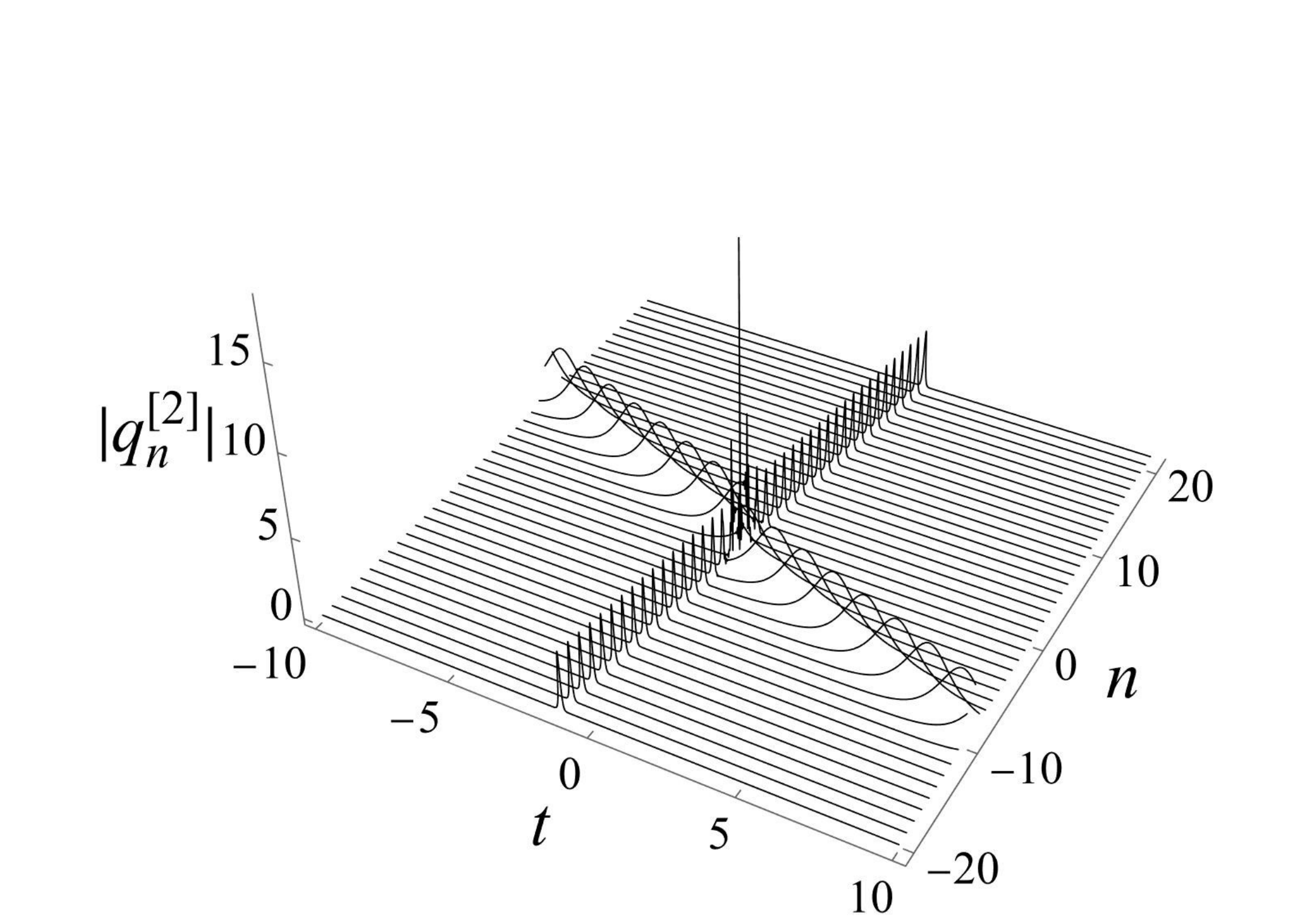}}
\flushleft{\footnotesize
\textbf{Figs.~$2.2$.} The bright 2-soliton of Eq.~(\ref{1.1}) with $C_1(0)=1+i, C_2(0)=1-i, \nu_1=1, \nu_2=1$, and $\eta_1=1, \eta_2=2 $. }
\end{figure}

\section{Non-zero boundary conditions}
\renewcommand{\theequation}{\arabic{section}.\arabic{subsection}.\arabic{equation}}\setcounter{equation}{0}

\hspace{1.5em}In this section, we consider the case that $\sigma=1$ and the potential $q_n\in\mathcal{H}_k$, where
\begin{equation}
\mathcal{H}_k=\{f_n:\sum_n^{\infty}{(|n|+1)^k\left( |f_n-q_r|+|f_{-n}-q_l| \right)}<\infty, \forall n\in\mathbb{Z}\}, \quad\quad k=0,1.
\end{equation}
Here $q_l$ and $q_r$ are two constants that can be unequal, and we set $q_{l/r}=Q e^{i \theta_{l/r}}$, where $0<Q<1$ and $Q\in\mathbb{R}$.
We also assume that the potential $q_n$ satisfies the small norm condition, i.e. $\underset{n\in \mathbb{Z}}{\sup}\,|q_n|<1$.

We omit the dependence on time $t$ in the part of direct and inverse scattering problems for simplicity. In addition, the same notations that
are not redefined in this section will follow the definitions in Section 2.

\subsection{Direct scattering problem}
\setcounter{equation}{0}
\subsubsection{Eigenfunctions}

\hspace{1.5em}When $n \rightarrow \infty$, the scattering problem~(\ref{1.2}) will reduce to
\begin{equation}\label{3.1.1}
v_{n+1}\sim M_{l/r}v_n=(Z+Q_{l/ r}) v_n,\quad\quad n \to \mp \infty,
\end{equation}
where
\begin{equation*}
Q_{l/r}=\left( \begin{matrix}
	0&		q_{l/r}\\
	\bar{q}_{l/r}&		0\\
\end{matrix} \right),
\end{equation*}
and the slash in the subscript denotes ``or''.

Thus there are two matrix solutions of Eq.~(\ref{3.1.1}) satisfying
\begin{gather}
\phi_n\sim M_l^n \quad \quad n\rightarrow-\infty, \label{3.1.2} \\
\psi_n\sim M_r^n \quad \quad n\rightarrow+\infty. \label{3.1.3}
\end{gather}

We set $\tau=(1-Q^2)^{\frac{1}{2}}$, hence $\tau \in (0,1)$, and it is easy to derive that $M_{l/r}$ has two eigenvalues $\tau \gamma$ and $\tau \gamma^{-1}$, which
can satisfy
\begin{gather}\label{3.1.4}
\gamma+\gamma^{-1}=\frac{1}{\tau}(z+z^{-1}).
\end{gather}
That is
\begin{gather}\label{3.1.5}
\gamma=\eta\pm\sqrt{\eta^2-1},\quad \quad \gamma^{-1}=\eta\mp\sqrt{\eta^2-1},
\end{gather}
where $\eta=\dfrac{z+z^{-1}}{2\tau}$. And the matrix composed of corresponding eigenvectors is
\begin{gather}\label{3.1.6}
V_{l/r}=(\gamma \tau-z)\sigma_1+\sigma_3 Q_{l/r}\sigma_1,
\end{gather}
where
\begin{gather*}
\sigma_1=\left( \begin{matrix}
	0&	1	\\
	1&	0	\\
\end{matrix} \right) .
\end{gather*}
Therefore, there is
\begin{gather}\label{3.1.7}
M_{l/r}=\tau  V_{l/r}\,\Gamma\, V_{l/r}^{-1},
\end{gather}
where $\Gamma=\text{diag}(\gamma,\,\gamma^{-1})$.

We define the eigenfunctions as
\begin{gather}
\Phi_n=(\Phi_{n,1},\,\Phi_{n,2})=\phi_n V_{l}, \label{3.1.8}\\
\Psi_n=(\Psi_{n,1},\,\Psi_{n,2})=\psi_n V_{r}.\label{3.1.9}
\end{gather}
Then combining Eqs.~(\ref{3.1.2}),~(\ref{3.1.3}) with~(\ref{3.1.7}),~(\ref{3.1.8}) and~(\ref{3.1.9}) yields
\begin{gather}
\Phi_n\sim \tau^n \, V_{l} \, \Gamma^n \quad \quad n\rightarrow-\infty, \label{3.1.10}\\
\Psi_n\sim \tau^n \, V_{r} \, \Gamma^n \quad \quad n\rightarrow+\infty.\label{3.1.11}
\end{gather}

\subsubsection{Conformal mappings and uniformization}

\hspace{1.5em}We consider the conformal mappings $\gamma=\eta\pm\sqrt{\eta^2-1}$ and $\eta=\dfrac{z+z^{-1}}{2\tau}$. It is easy to see that the
correspondence between the variables $z$ and $\gamma$ is 1-\,to\,-2, hence the two-sheeted Riemann surface can be introduced as
$(\gamma-\eta)^2-\eta^2+1=0$, which consists of two complex $\eta$-planes glued together, with $\eta=\pm1$ as the branch points and the
segment $(-1,1)$ as the branch cut. Corresponding to the complex $z$-plane, the branch points are the solutions of $z+z^{-1}=\pm 2\tau $, i.e.
$\pm z_0$ and $\pm \bar{z}_0$, where $z_0=\tau+i Q$.

For fear of the complexity of multi-value or the Rieman surface, the conformal mapping
\begin{gather}\label{3.1.12}
\lambda(z)=\frac{\gamma(z)} { z}
\end{gather}
can be created to introduce the uniformization variable $\lambda$, and the inverse scattering problem will be discussed later on the complex
$\lambda$-plane.

Due to $\gamma(\pm z_0)=\gamma(\pm \bar{z}_0)=\pm 1$, we derive that $\lambda(\pm z_0)=\bar{\lambda}_0$ and $\lambda(\pm \bar{z}_0)=\lambda_0$
from the mapping~(\ref{3.1.12}), where $\lambda_0=\tau+i Q$. That is to say, this mapping transforms four branch points $\pm z_0$ and $\pm
\bar{z}_0$ on the $z$-plane into two points $\bar{\lambda}_0$ and $\lambda_0$ on the $\lambda$-plane.

Combining the mapping~(\ref{3.1.12}) with the relation~(\ref{3.1.4}) yields
\begin{gather}\label{3.1.13}
\frac{\lambda-\tau}{\lambda(\tau \lambda-1)}=z^2,\quad \ \frac{\lambda(\lambda-\tau)}{\tau \lambda-1}=\gamma^2,\quad \
\frac{\lambda-\tau}{\tau \lambda-1}=z \gamma.
\end{gather}

\paragraph{Proposition 3.1} Based on the three mappings, there are equivalence relations between the variables $\gamma$, $\eta$, $z$ and
$\lambda$:
\begin{align}
&|\lambda|=1\Leftrightarrow|\gamma|=1\Leftrightarrow \eta\in [-1,1]\Leftrightarrow  z\in\{z: |z|=1 \ \text{and} \ |\text{Re} z|\leq \tau\},
\label{3.1.14}\\
& z\in\{z:|z|=1 \ \text{and} \ |\text{Re} z|> \tau \} \Leftrightarrow \lambda \in \{ \lambda:|\lambda-1/\tau|=Q/\tau \ \text{and}
 \ \lambda \neq \lambda_0, \bar{\lambda}_0\},\label{3.1.15} \\
 &|\lambda|<1\Leftrightarrow|\gamma|<1,\label{3.1.16} \\
&|z|<1\Leftrightarrow (|\lambda|^2-1)(|\lambda-1/\tau|^2-Q^2/\tau^2)<0.\label{3.1.17}
\end{align}
\begin{proof} \textbf{(a)}\ The equivalence relation~(\ref{3.1.14}) is proved in the following five parts:
\textbf{(a.1)}\ `` $|\lambda|=1 \Leftrightarrow |\gamma|=1 $ '':
From the relations~(\ref{3.1.13}), we get $|\gamma|=1\Leftrightarrow|\gamma|^2=1\Leftrightarrow |\lambda| |\lambda-\tau|=|\tau \lambda-1
|\Leftrightarrow (|\lambda|^2-1)(|\lambda-\tau|^2+Q^2)=0\Leftrightarrow |\lambda|=1$ or $(|\lambda-\tau|^2+Q^2)=0$. Owing to $0<Q<1$, then
$|\lambda-\tau|^2+Q^2>0$. Therefore, we obtain that $|\gamma|=1\Leftrightarrow |\lambda|=1$.
\textbf{(a.2)}\ `` $|\gamma|=1\Rightarrow \eta\in [-1,1]$ '':
Since $|\gamma|=1$, let  $\gamma=a+i b$, where $a^2+b^2=1$,then $-1\leq a\leq1$. Thus according to $
\eta=\dfrac{z+z^{-1}}{2\tau}=\dfrac{\gamma+\gamma^{-1}}{2}=a$, we obtain $-1\leq  \eta \leq1$.
\textbf{(a.3)}\  ``  $\eta\in [-1,1]\Rightarrow|\gamma|=1$ '':
 Due to $\eta\in [-1,1]$, then $\gamma=\eta\pm i \sqrt{1-\eta^2}$. Therefore, $|\gamma|=\sqrt{\eta^2+(1-\eta^2)}=1$.
\textbf{(a.4)}\  `` $\eta\in [-1,1]\Rightarrow |z|=1 \ \text{and} \ |\text{Re}\, z|\leq \tau$ '':
Let $z=x+i y$. Then $\eta=\dfrac{z+z^{-1}}{2\tau}=\dfrac{x}{2\tau}(1+\dfrac{1}{x^2+y^2})+i \dfrac{y}{2\tau}(1-\dfrac{1}{x^2+y^2})$. Since
$\eta\in [-1,1]$ , i.e. $\eta\in \mathbb{R}$, can yield $y=0$ or $x^2+y^2=1$. If $y=0$, then $\eta\in [-1,1]\Rightarrow -2<-2\tau\leq
x+\dfrac{1}{x}\leq 2\tau<2$, which is in contradiction with the fact that $x+\dfrac{1}{x}\geq 2$ or $\leq -2$; while if $x^2+y^2=1$, i.e.
$|z|=1$, then $\eta\in [-1,1]\Rightarrow -1\leq \dfrac{x}{2\tau}\cdot 2 \leq 1 \Rightarrow -\tau\leq x \leq \tau$, i.e. $|\text{Re} z|\leq
\tau$. To sum up, there is $|z|=1 \ \text{and} \ |\text{Re}\, z|\leq \tau$ under the condition of $\eta\in [-1,1]$.
\textbf{(a.5)}\   ``$z\in\{z: |z|=1 \ \text{and} \ |\text{Re} z|\leq \tau\}\Rightarrow\eta\in [-1,1]$ '':
Since that $|z|=1 \Rightarrow x^2+y^2=1$, and $|\text{Re} z|\leq \tau \Rightarrow -\tau\leq x \leq \tau$, then
$|\eta|=\left|\dfrac{x}{\tau}\right|\leq 1$, i.e. $\eta\in [-1,1]$.
\textbf{(b)} The proof of the relation~(\ref{3.1.15}):
From the relations~(\ref{3.1.13}), we get $|z|=1\Leftrightarrow|z|^2=1\Leftrightarrow|\lambda-\tau|=|\lambda| |\tau \lambda-1 |\Leftrightarrow
(|\lambda|^2-1)(|\lambda-1/\tau|^2-Q^2/\tau^2)=0\Leftrightarrow |\lambda|=1$ or $\lambda \in \{ \lambda\in \mathbb{C}:|\lambda-1/\tau|=Q/\tau
\ \text{and} \ \lambda \neq \lambda_0, \bar{\lambda}_0\}$. Due to $ z\in\{z:|z|=1 \ \text{and} \ |\text{Re} z|\leq
\tau\}\Leftrightarrow|\lambda|=1$, hence $z\in\{z:|z|=1 \ \text{and} \ |\text{Re} z|> \tau \} \Leftrightarrow \lambda \in \{
\lambda:|\lambda-1/\tau|=Q/\tau \ \text{and} \ \lambda \neq \lambda_0, \bar{\lambda}_0\}$.
\textbf{(c)} The proof of the relation~(\ref{3.1.16}):
Similar to (a.1), there is $|\gamma|<1\Leftrightarrow(|\lambda|^2-1)(|\lambda-\tau|^2+Q^2)<0$. Since $|\lambda-\tau|^2+Q^2>0$, then
$|\gamma|<1\Leftrightarrow|\lambda|<1$.
\textbf{(d)} The proof of the relation~(\ref{3.1.17}):
Similar to (b), there is $|z|<1\Leftrightarrow (|\lambda|^2-1)(|\lambda-1/\tau|^2-Q^2/\tau^2)<0$.
\end{proof}

From Eqs.~(\ref{3.1.10})-(\ref{3.1.11}), we can see that the asymptotic behaviors of $\Phi_n$ and $\Psi_n$ depend on $\tau^n$.  Hence the
continuous spectrum of the scattering problem~(\ref{1.2}) can be defined as any $z\in \mathbb{C}$ such that $\tau^{-n}\Phi_n(z)$ and
$\tau^{-n}\Psi_n(z)$ are bounded for $n \in \mathbb{Z}$, which corresponds to the set of $z$ satisfying $|\gamma(z)|=1$. Excluding the branch
points, the continuous spectrum can be obtained from the Proposition 3.1 and expressed as
\begin{gather}\label{3.1.18}
\mathcal{Z}=\{ z\in  \mathbb{C}: |z|=1 \ \text{and} \ |\text{Re}\, z|<\tau \}.
\end{gather}
Recall that $\lambda(\pm z_0)=\bar{\lambda}_0$ and $\lambda(\pm \bar{z}_0)=\lambda_0$. Meanwhile $z({\lambda}_0)=\pm \bar{z}_0$ and $z({\bar{\lambda}}_0)=\pm z_0$ can be obtained from the relation $\dfrac{\lambda-\tau}{\lambda(\tau \lambda-1)}=z^2$ in Eq.~(\ref{3.1.13}).
Then it follows from the Proposition 3.1 that the continuous spectrum can be mapped to the unit circle of the $\lambda$-plane except for the branch points $\lambda_0$ and $\bar{\lambda}_0$.

As for the discrete spectrum, it consists of special values of $z_j \in \mathbb{C}$ for which $\Phi_n(z_j)$ and $\Psi_n(z_j)$ vanish as
$n\rightarrow \pm \infty$~\cite{ck42}. Consequently, the fact that the discrete eigenvalues lie on the circle $|z|=1$ can be proved as follows:
\begin{proof}
 For any discrete eigenvalue $z_j$, from the scattering problem~(\ref{1.2}), we can derive that
 \begin{equation*}
 \begin{split}
   |\Phi_{n+1,1}^{(1)}(z_j)|^2-|\Phi_{n+1,1}^{(2)}(z_j)|^2-(1-|q_n|^2)(|\Phi_{n,1}^{(1)}(z_j)|^2-|\Phi_{n,1}^{(2)}(z_j)|^2)\\
 =(|z_j|^2-1)(|\Phi_{n+1,1}^{(2)}(z_j)|^2+ (1-|q_n|^2)|\Phi_{n,1}^{(1)}(z_j)|^2).
  \end{split}
 \end{equation*}
  Then multiplying both sides of above equation by $c_{n+1}=\prod_{k=n+1}^\infty(1-|q_{k}|^2)$, which is bounded for any $n \in \mathbb{Z}$,
  yields
  \begin{equation*}
   \begin{split}
  (c_{n+1}|\Phi_{n+1,1}^{(1)}(z_j)|^2-c_{n}|\Phi_{n,1}^{(1)}(z_j)|^2)-(c_{n+1}|\Phi_{n+1,1}^{(2)}(z_j)|^2-c_{n}|\Phi_{n,1}^{(2)}(z_j)|^2) \\
  =(|z_j|^2-1)(c_{n+1}|\Phi_{n+1,1}^{(2)}(z_j)|^2+ c_{n}|\Phi_{n,1}^{(1)}(z_j)|^2).
  \end{split}
  \end{equation*}
Summing over all $n \in \mathbb{Z}$, since $\Phi_n(z_j)$ and $\Psi_n(z_j)$ decay
sufficiently fast as $n\rightarrow \pm \infty$, we derive
\begin{equation*}
  (|z_j|^2-1)\sum_{n=- \infty}^{+\infty}c_{n}(|\Phi_{n,1}^{(2)}(z_j)|^2 +|\Phi_{n,1}^{(1)}(z_j)|^2)=0.
  \end{equation*}
 Under the small norm condition $\underset{n\in \mathbb{Z}}{\sup}\,|q_n|<1$, one has $c_{n}>0$ for all $n \in \mathbb{Z}$. Then
  $\sum\limits_{n=- \infty}^{+\infty}c_{n}(|\Phi_{n,1}^{(2)}(z_j)|^2 +|\Phi_{n,1}^{(1)}(z_j)|^2)>0$.  Therefore, we obtain that $|z_j|=1$.
\end{proof}
For simplicity, we assume that there are no discrete eigenvalues embedding in the discrete spectrum. Therefore, the set of discrete
eigenvalues is
\begin{gather*}
\mathcal{L}=\{ z_j\in  \mathbb{C},\ j=1,2,\dots J: |z_j|=1 \ \text{and} \ |\text{Re}\, z_j|>\tau \}.
\end{gather*}

From the Proposition 3.1 it follows that the discrete spectrum can be mapped to the circle $|\lambda-1/\tau|=Q/\tau$ excluding the branch
points $\lambda_0$ and $\bar{\lambda}_0$ on the $\lambda$-plane, thus it can be expressed as
\begin{gather}\label{3.1.19}
\Omega=\{ \lambda_j\in \mathbb{C},\  j=1,2,\dots J:|\lambda_j-1/\tau|=Q/\tau \ \text{and} \ \lambda_j \neq \lambda_0, \bar{\lambda}_0\}.
\end{gather}

In order to facilitate the discussion of the properties of the eigenfunctions on the $\lambda$-plane, we consider the asymptotic behavior
~(\ref{3.1.10})\,-\,(\ref{3.1.11}) and introduce the modified eigenfunctions as
\begin{align}
(U_n\left( \lambda \right)\ \tilde{U}_n\left(\lambda \right))& =\tau^{-n}\Lambda \left( \gamma \right)\Phi_n(z)\Gamma^{-n}\Lambda ^{-1}\left(
\gamma \right),\label{3.1.20}\\
(W_n\left( \lambda \right)\ \tilde{W}_n\left(\lambda \right)) &=\tau^{-n}\Lambda \left( \gamma \right)\Psi_n(z)\Gamma^{-n}\Lambda ^{-1}\left(
\gamma \right),\label{3.1.21}
\end{align}
where
\begin{gather}\label{3.1.22}
\Lambda \left( \gamma \right)=\left(\begin{matrix}
                                1 & 0\\
                                0 & \gamma
                              \end{matrix}\right).
\end{gather}
Then according to the mapping~(\ref{3.1.12}) and the relations~(\ref{3.1.13}), the asymptotic behavior of the modified eigenfunctions can be
written as
\begin{gather}
(U_n\left( \lambda \right)\ \tilde{U}_n\left(\lambda \right)) \sim \left(\begin{matrix}
                                q_l & \tau-z/\gamma\\
                                \tau \gamma^2-z\gamma & -\bar{q_l}
                              \end{matrix}\right)=\left(\begin{matrix}
                                q_l & \tau-1/\lambda\\
                                \lambda-\tau & -\bar{q_l}
                              \end{matrix}\right)\quad \quad n\rightarrow-\infty, \label{3.1.23}\\
(\tilde{W}_n\left( \lambda \right)\ W_n\left(\lambda \right)) \sim \left(\begin{matrix}
                                q_r & \tau-z/\gamma\\
                                \tau \gamma^2-z\gamma & -\bar{q_r}
                              \end{matrix}\right)=\left(\begin{matrix}
                                q_r & \tau-1/\lambda\\
                               \lambda-\tau & -\bar{q_r}
                              \end{matrix}\right)\quad \quad n\rightarrow+\infty.\label{3.1.24}
\end{gather}
Combining the transformations~(\ref{3.1.20})-(\ref{3.1.21}) with the scattering problem~(\ref{1.2}) yields the modified scattering problems:
\begin{gather}
  v_{n+1}(\lambda)=\frac{\gamma}{\tau} K_n v_{n}(\lambda),\label{3.1.25}\\
 v_{n+1}(\lambda)=\frac{1}{\tau \gamma}K_n v_{n}(\lambda),\label{3.1.26}
\end{gather}
where
\begin{gather*}
K_n=\left(\begin{matrix}
z & q_n/\gamma\\\gamma \bar{q}_n & 1/z\end{matrix}\right).
\end{gather*}
Thereinto, $\tilde{U}_n\left(\lambda \right)$ and $W_n\left(\lambda \right)$ satisfy Eq.~(\ref{3.1.25}), while $U_n\left( \lambda \right)$ and
$\tilde{W}_n\left( \lambda \right)$ satisfy~(\ref{3.1.26}).

\subsubsection{Scattering matrix }

\hspace{1.5em}From the asymptotic behaviors~(\ref{3.1.10})-(\ref{3.1.11}), we get
\begin{gather}
\det(\tau^{-n}\Phi_n)\sim  \det V_{l} =(\gamma \tau-z)^2+Q^2 \quad \quad n\rightarrow-\infty, \label{3.1.27}\\
\det(\tau^{-n}\Psi_n)\sim \det V_{r} = (\gamma \tau-z)^2+Q^2 \quad \quad n\rightarrow+\infty.\label{3.1.28}
\end{gather}
Taking the determinant of scattering problem~(\ref{1.2}) and iterating continuously, one can derive the determinant of the two eigenfunction
matrices as below by considering their respective BCs~(\ref{3.1.27})-(\ref{3.1.28}):
\begin{gather}
\det\Phi_n=-\tau^{2n}\big((\gamma \tau-z)^2+Q^2\big)\prod_{j=-\infty}^{n-1}{\frac{1-|q_j|^2}{1-Q^2}} , \label{3.1.29}\\
\det\Psi_n=-\tau^{2n}\big((\gamma \tau-z)^2+Q^2\big)\prod_{j=n}^{+\infty}{\frac{1-Q^2}{1-|q_j|^2}} .\label{3.1.30}
\end{gather}

\paragraph{Proposition 3.2} Both $\Phi_n(z)$ and $\Psi_n(z)$ are fundamental matrix solutions of the scattering problem~(\ref{1.2}) for all
$z$ except at the branch points $z=\pm z_0$ and $z=\pm \bar{z}_0$.
\begin{proof}
 Under the conditions $\underset{n\in \mathbb{Z}}{\sup}\,|q_n|<1$ and $0<Q<1$, we get that $\det\Phi_n=0$ or $\det\Psi_n=0$ only at the
 zero of $(\gamma \tau-z)^2+Q^2$. By the relation~(\ref{3.1.4}) and the mapping~(\ref{3.1.12}), we get
 \begin{equation}\label{3.1.31}
  (\gamma \tau-z)^2+Q^2=(\gamma\tau-z)(\tau/ \gamma-1/z)+Q^2=\tau(\gamma /z+z/\gamma-2\tau)=\tau(\lambda+1/\lambda-2\tau).
  \end{equation}
  If we let $\lambda=a+i b$, then $\lambda+1/\lambda=a\big(1+(a^2+b^2)^{-1}\big)+i b\big(1-(a^2+b^2)^{-1}\big)$. Hence it follows from Eq.~(\ref{3.1.31}) that $(\gamma \tau-z)^2+Q^2=0\Leftrightarrow \lambda+1/\lambda=2\tau
  \Leftrightarrow b\big(1-(a^2+b^2)^{-1}\big)=0$ and $a\big(1+(a^2+b^2)^{-1}\big)=2\tau$.
   \\\indent Assuming that $b=0$, then $a+1/a=2\tau$ , which contradicts $0<\tau<1$. Thus we get $1-(a^2+b^2)^{-1}=0$, i.e. $a^2+b^2=1$, and
   then $a=\tau$ and $b=\pm Q$, i.e. $\lambda=\lambda_0$ or $\bar{\lambda}_0$. Therefore, according to the correspondence of the branch
   points, $(\gamma \tau-z)^2+Q^2=0\Leftrightarrow z=\pm z_0$ or $\pm \bar{z}_0$. That is to say, $\det\Phi_n (z)\ne  0$ and $\det\Psi_n(z)
   \ne 0$ except at the branch points.
\end{proof}

It follows from the Proposition 3.2 that there exists an $n$-independent invertible matrix $S(z)$ called the scattering matrix and satisfying
\begin{gather}\label{3.1.32}
\Phi_n(z)=\Psi_n(z)S(z),\quad\quad\  z\neq \pm z_0,\, \pm\bar{z}_0,
\end{gather}
where
\begin{gather}\label{3.1.33}
S(z)=\left( \begin{matrix}
s_1(z)	&	s_2(z)	\\
s_3(z)	&	s_4(z)	\\
\end{matrix} \right) .
\end{gather}
 Eq.~(\ref{3.1.32}) implies that the scattering coefficients can be expressed as
\begin{equation}\label{3.1.34}
\begin{aligned}
s_1(z)=\big(\text{det}\Psi_n(z)\big)^{-1}\text{det}\big(\Phi_{n,1}(z),\Psi_{n,2}(z)\big),\quad \
s_2(z)=\big(\text{det}\Psi_n(z)\big)^{-1}\text{det}\big(\Phi_{n,2}(z),\Psi_{n,2}(z)\big),\\
s_3(z)=\big(\text{det}\Psi_n(z)\big)^{-1}\text{det}\big(\Psi_{n,1}(z),\Phi_{n,1}(z)\big),\quad \
s_4(z)=\big(\text{det}\Psi_n(z)\big)^{-1}\text{det}\big(\Psi_{n,1}(z),\Phi_{n,2}(z)\big).\\
\end{aligned}
\end{equation}
And combining Eqs.~(\ref{3.1.32}) with~(\ref{3.1.29}) and~(\ref{3.1.30}) can yield
\begin{gather}\label{3.1.35}
\text{det} S(z)=\big(\text{det}\Psi_n(z)\big)^{-1}\text{det} \Phi_n(z)=\chi_{- \infty}=\underset{n\rightarrow -\infty}{\lim}\chi_n,
\end{gather}
in which
\begin{gather}\label{3.1.36}
\chi_n=\prod_{j=n}^{+\infty}{\frac{1-|q_j|^2}{1-Q^2}}   .
\end{gather}
 By the transformation (\ref{3.1.20})\,-\,(\ref{3.1.21}), Eq.~(\ref{3.1.32}) becomes
 \begin{gather}\label{3.1.37}
(U_n\left( \lambda \right)\ \tilde{U}_n\left(\lambda \right))=(\tilde{W}_n\left( \lambda \right)\ W_n\left(\lambda \right))\, \Gamma^n
T(\lambda) \Gamma^{-n},
\end{gather}
in which
\begin{gather}\label{3.1.38}
T(\lambda)=\Lambda S(z) \Lambda^{-1}=\left(\begin{matrix} s_1(\lambda) &  \tilde{s}_2(\lambda)\\ \tilde{s}_3(\lambda) &
s_4(\lambda)\end{matrix}\right),
\end{gather}
where
\begin{equation}\label{3.1.39}
\begin{aligned}
 s_1(\lambda)&=s_1(z),\quad \quad \quad \ \  s_4(\lambda)= s_4(z),\\
 \tilde{s}_2(\lambda)&=\gamma^{-1}s_2(z),\quad \quad  \tilde{s}_3(\lambda)=\gamma s_3(z).
 \end{aligned}
\end{equation}
Since according to Eqs.~(\ref{3.1.20})-(\ref{3.1.21}),~(\ref{3.1.29})-(\ref{3.1.31}) and~(\ref{3.1.36}), we get
\begin{align}
\text{det}(U_n\left( \lambda \right)\ \tilde{U}_n\left(\lambda \right)) &=\tau^{-2n}\det\Phi_n =-\tau(\lambda+1/\lambda-2\tau)\chi_{-\infty}/\chi_n ,
\label{3.1.40}\\
\text{det}(\tilde{W}_n\left( \lambda \right)\ W_n\left(\lambda \right))& =\tau^{-2n}\det\Psi_n =-\tau(\lambda+1/\lambda-2\tau)/\chi_n ,\label{3.1.41}
\end{align}

it follows from Eqs.~(\ref{3.1.37}) and~(\ref{3.1.13}) that
\begin{equation}\label{3.1.42}
\begin{aligned}
s_1(\lambda)&=\frac{\text{det}\big(U_{n}(\lambda),W_{n}(\lambda)\big)}{\text{det}(\tilde{W}_n(\lambda)\ W_n(\lambda))}
=\chi_n\frac{ \text{det}\big(W_{n}(\lambda),U_{n}(\lambda)\big)}{\tau(\lambda+1/\lambda-2\tau)},\\
s_4(\lambda)&=\frac{\text{det}\big(\tilde{W}_n(\lambda),\tilde{U}_n(\lambda)\big)}{\text{det}(\tilde{W}_n(\lambda)\ W_n(\lambda))}
=\chi_n\frac{ \text{det}\big(\tilde{U}_n(\lambda),\tilde{W}_n(\lambda)\big)}{\tau(\lambda+1/\lambda-2\tau)},\\
\tilde{s}_2(\lambda)&=\gamma^{-2n}\frac{\text{det}\big(\tilde{U}_n(\lambda),W_n(\lambda) \big)}{\text{det}(\tilde{W}_n(\lambda)\
W_n(\lambda))}=\frac{\chi_n}{\big(\gamma^2(\lambda)\big)^{n}}\,\frac{\text{det}\big(W_n(\lambda), \tilde{U}_n(\lambda)
\big)}{\tau(\lambda+1/\lambda-2\tau)} ,\\
\tilde{s}_3(\lambda)&=\gamma^{2n}\frac{\text{det}\big(U_n(\lambda),\tilde{W}_n(\lambda) \big)}{\text{det}(\tilde{W}_n(\lambda)\
W_n(\lambda))}=\big(\gamma^2(\lambda)\big)^n\chi_n\,\frac{\text{det}\big(U_n(\lambda),\tilde{W}_n(\lambda)
\big)}{\tau(\lambda+1/\lambda-2\tau)}.
\end{aligned}
\end{equation}

\subsubsection{Summation equations}

\hspace{1.5em}In this subsection, based on the modified scattering problems~(\ref{3.1.25}) and~(\ref{3.1.26}), we will consider the summation equations
satisfied by the modified eigenfunctions by Green's function method.

We first consider the modified eigenfunctions $\tilde{U}_n\left(\lambda \right)$ and $W_n\left(\lambda \right)$ which satisfy
Eq.~(\ref{3.1.25}):
\begin{gather}\label{3.1.43}
v_{n+1}(\lambda)=\frac{\gamma}{\tau}K_n v_{n}(\lambda)=\frac{\gamma}{\tau}\left(\begin{matrix}
z & \dfrac{q_{l/r}}{\gamma}\\ \gamma \bar{q}_{l/r} & z^{-1}\end{matrix}\right)v_{n}+\frac{\gamma}{\tau}\Gamma^{-1} (Q_n-Q_{l/r}) v_{n},
\end{gather}
that is,
\begin{gather*}
\text{L}v_{n}=(Q_n-Q_{l/r}) v_{n},
\end{gather*}
where $\text{L}$ is a operator such that
\begin{gather*}
 \text{L}v_{n}=\tau \gamma^{-1}\Gamma v_{n+1}(\lambda)-\Gamma \left(\begin{matrix}
z & \dfrac{q_{l/r}}{\gamma}\\ \gamma \bar{q}_{l/r} & z^{-1}\end{matrix}\right)v_{n}.
\end{gather*}

If there are Green's functions $G_n^{l/r}$ and inhomogeneous terms $\upsilon_{l/r}$ satisfying
\begin{gather*}
\text{L}G_n^{l/r}=\delta_{n,0} I,\quad \quad \text{L}\upsilon_{l/r}=0,
\end{gather*}
then the solutions of Eq.~(\ref{3.1.43}) can be expressed as the summation equations
\begin{gather}\label{3.1.44}
v_n(\lambda)=\upsilon_{l/r} +\sum_{j=-\infty}^{+\infty}{G_{n-j}^{l/r}(\lambda)(Q_n-Q_{l/r}) v_j(\lambda)}.
\end{gather}

For the inhomogeneous terms $\upsilon_{l/r}$, the BCs~(\ref{3.1.23})-(\ref{3.1.24}) can be considered, and it can be found
that the vectors $\upsilon_{l/r}=(\tau-1/\lambda \ \ -\bar{q}_{l/r})^T$ satisfy $\text{L}\upsilon_{l/r}=0$.

For the Green's functions $G_n^{l/r}$, they can be derived by the Fourier transforms and the residue theorem, and be expressed as
\begin{equation}\label{3.1.45}
\begin{aligned}
G_{n}^{l}\left( \lambda \right) =\theta \left( n-1 \right) \frac{\gamma ^2}{\tau^2\left( 1-\gamma ^2 \right)}\left[ \left( \begin{matrix}
	\frac{1-\tau\lambda}{\lambda}&		q_l\\
	\bar{q}_l&		Q^2\frac{\lambda}{1-\tau\lambda}\\
\end{matrix} \right) -\gamma ^{2\left( n-1 \right)}\left( \begin{matrix}
	\frac{Q^2}{\lambda-\tau}&		q_l\\
	\bar{q}_l&		\lambda -\tau\\
\end{matrix} \right) \right], \\
G_{n}^{r}\left( \lambda \right) =-\theta \left( -n \right) \frac{\gamma ^2}{\tau^2\left( 1-\gamma ^2 \right)}\left[ \left( \begin{matrix}
	\frac{1-\tau\lambda}{\lambda}&		q_r\\
	\bar{q}_r&		Q^2\frac{\lambda}{1-\tau\lambda}\\
\end{matrix} \right) -\gamma ^{2\left( n-1 \right)}\left( \begin{matrix}
	\frac{Q^2}{\lambda -\tau}&		q_r\\
	\bar{q}_r&		\lambda -\tau\\
\end{matrix} \right) \right] ,
\end{aligned}
\end{equation}
which can be proved as follows:
\begin{proof}
Similar to the process in the section 2,  we first represent the Green's functions $G_n^{l/r}$ and $\delta_{n,0}$ as Fourier integrals
\begin{equation*}
  G_n^{l/r}\left( z \right) =\frac{1}{2\pi i}\oint_{\left| s \right|=1}{\hat{G}^{l/r}\left( s \right)}s^{n-1}\text{d}s,\quad\quad
\delta _{n,0}=\frac{1}{2\pi i}\oint_{\left| s \right|=1}{s^{n-1}\text{d}s}.
\end{equation*}
Therefore, $\text{L}G_n^{l/r}=\delta_{n,0}$ can be written as
\begin{equation*}
\frac{\Gamma}{2\pi i}\oint_{\left| s \right|=1}{\left[ s\tau\gamma ^{-1}I-\left( \begin{matrix}
	z&		\dfrac{q_{l/r}}{\gamma}\\
	\gamma \bar{q}_{l/r}&		z^{-1}\\
\end{matrix} \right) \right]\hat{G}^{l/r}\left( s \right)}s^{n-1} \text{d}s=\frac{1}{2\pi i}\oint_{\left| s \right|=1}{s^{n-1}\text{d}s}.
\end{equation*}
Consequently, we can derive that
\begin{equation*}
\hat{G}^{l/r}\left( s \right)=\frac{1}{\tau(s-1)(s-\gamma^2)}\left( \begin{matrix}
	s-\frac{\gamma}{\tau z}&	\frac{q_{l/r}}{\tau}\gamma^2\\
	\frac{\bar{q}_{l/r}}{\tau}\gamma ^2&		( s-\frac{\gamma z}{\tau} ) \gamma ^2\\
\end{matrix} \right),
\end{equation*}
and accordingly,
\begin{equation*}
 G_n^{l/r}\left( \lambda \right) =\frac{1}{2\pi i}\oint_{\left| s \right|=1}\frac{ s^{n-1}}{\tau(s-1)(s-\gamma^2)}\zeta_{l/r}(s)\text{d}s,
\end{equation*}
where
\begin{equation*}
\zeta_{l/r}(s)=\left( \begin{matrix}
	s-\frac{\gamma}{\tau z}&	\frac{q_{l/r}}{\tau}\gamma^2\\
	\frac{\bar{q}_{l/r}}{\tau}\gamma ^2&		( s-\frac{\gamma z}{\tau} ) \gamma ^2\\
\end{matrix} \right).
\end{equation*}
Next we pay attention to the integral above. To avoid the singularities on the circle $|s|=1$, there are two contours $C^{\text{out}}$ and
$C^{\text{in}}$ that perturbed away from $|s|=1$ can be considered, where the contour $C^{\text{out}}$ encloses the singularities on $|s|=1$
while $C^{\text{in}}$ does not. When $|\lambda|\leq 1$, i.e. $|\gamma|\leq 1$, the contour $C^{\text{out}}$ can enclose $s=0, 1, \gamma^2$,
and by the residue theorem, we derive that
\begin{equation*}
G_n^{\text{out}}\left( \lambda \right) =\frac{1}{2\pi i}\oint_{\left| s \right|=1}\frac{
s^{n-1}}{\tau(s-1)(s-\gamma^2)}\zeta_{l/r}(s)\text{d}s=\left\{ \begin{array}{r}
	\frac{\zeta_{l/r}\left( 1 \right) -\gamma ^{2\left( n-1 \right)}\zeta_{l/r}( \gamma ^2 )}{\tau \left( 1-\gamma ^2 \right)},\ \ \ \ n\ge 1\\
	0,\ \ \ \ \ \ \ \ \ \ \quad \quad  \ n<1\\
\end{array} \right. .
\end{equation*}
When $|\lambda|\geq 1$, i.e. $|\gamma|\geq 1$, the contour $C^{\text{in}}$ can enclose $s=0$ but neither 1 nor $\gamma^2$, and similar to
above, we obtain that
\begin{equation*}
G_n^{\text{in}}\left( \lambda \right) =\frac{1}{2\pi i}\oint_{\left| s \right|=1}\frac{ s^{n-1}}{\tau(s-1)(s-\gamma^2)}\zeta_{l/r}(s)\text{d}s=
\left\{ \begin{array}{r}
	0,\ \ \ \ \ \ \ \ \ \ \ \quad \quad  \ n\ge 1\\
	-\frac{\zeta_{l/r}\left( 1 \right) -\gamma ^{2\left( n-1 \right)}\zeta_{l/r}( \gamma ^2 )}{\tau\left( 1-\gamma ^2 \right)},\,\,\,\,\,\,\,\,n<1\\
\end{array} \right. .
\end{equation*}
Taking into account Eq.~(\ref{3.1.44}) and the asymptotic behaviors~(\ref{3.1.23})-(\ref{3.1.24}) can yield that $G_n^{\text{out}}$ corresponds to
$G_n^{l}$ associated with the modified eigenfunctions $\tilde{U}_n$ and that $G_n^{\text{in}}$ corresponds to $G_n^{r}$ associated with $W_n$.
That is,
\begin{equation*}
G_n^{l}=\theta(n-1)\frac{\zeta_{l}\left( 1 \right) -\gamma ^{2\left( n-1 \right)}\zeta_{l}( \gamma ^2 )}{\tau\left( 1-\gamma ^2 \right)},\quad \quad
G_n^{r}=-\theta(-n)=\frac{\zeta_{r}\left( 1 \right) -\gamma ^{2\left( n-1 \right)}\zeta_{r}( \gamma ^2 )}{\tau\left( 1-\gamma ^2 \right)}.
\end{equation*}
Finally, Eqs.~(\ref{3.1.45}) can be obtained by using the relations~(\ref{3.1.13}).
\end{proof}

Similarly, the modified eigenfunctions $U_n\left(\lambda \right)$ and $\tilde{W}_n\left(\lambda \right)$, which are the solutions of
Eq.~(\ref{3.1.26}), can satisfy the summation equations
\begin{gather}\label{3.1.46}
v_n(\lambda)=\tilde{\upsilon}_{l/r} +\sum_{j=-\infty}^{+\infty}{\tilde{G}_{n-j}^{l/r}(\lambda)(Q_n-Q_{l/r}) v_j(\lambda)},
\end{gather}
in which the inhomogeneous terms $\tilde{\upsilon}_{l/r}=(q_{l/r} \ \ \lambda - \tau)^T$ and the Green’s functions can be expressed as
\begin{equation}\label{3.1.47}
\begin{aligned}
\tilde{G}_{n}^{l}\left( \lambda \right) =\theta \left( n-1 \right) \frac{\gamma ^{-2}}{\tau^2\left( \gamma ^2-1 \right)}\left[ \left( \begin{matrix}
	\frac{Q^2}{\lambda -\tau}&		q_l\\
	\bar{q}_l&		\lambda -\tau\\
\end{matrix} \right) -\gamma ^{-2\left( n-1 \right)}\left( \begin{matrix}
	\frac{1-\tau\lambda}{\lambda}&		q_l\\
	\bar{q}_l&		Q^2\frac{\lambda}{1-\tau\lambda}\\
\end{matrix} \right) \right] ,\\
\tilde{G}_{n}^{r}\left( \lambda \right) =-\theta \left( -n \right) \frac{\gamma ^{-2}}{\tau^2\left( \gamma ^2-1 \right)}\left[ \left( \begin{matrix}
	\frac{Q^2}{\lambda -\tau}&		q_r\\
	\bar{q}_r&		\lambda -\tau\\
\end{matrix} \right) -\gamma ^{-2\left( n-1 \right)}\left( \begin{matrix}
	\frac{1-\tau \lambda}{\lambda}&		q_r\\
	\bar{q}_r&		Q^2\frac{\lambda}{1-\tau \lambda}\\
\end{matrix} \right) \right].
\end{aligned}
\end{equation}

Note that according to the BCs~(\ref{3.1.23})-(\ref{3.1.24}), $U_n\left(\lambda \right)$ and $\tilde{U}_n\left(\lambda
\right)$ satisfy the summation equations with the ``$l$'' signs, while $\tilde{W}_n\left(\lambda \right)$ and $W_n\left(\lambda \right)$
  satisfy the equations with the ``$r$'' signs.

From Eqs.~(\ref{3.1.45}) and~(\ref{3.1.47}), we can see that the term $1-\gamma ^2$ gives singularities for $\gamma=\pm 1$, which corresponds
to the branch points $\lambda_0$ and $\bar{\lambda}_0$ on the $\lambda$-plane. However, as $\lambda\rightarrow \tau \pm iQ$, one can derive
that
\begin{align*}
 G_{n}^{l}\thicksim \theta \left( n-1 \right) \frac{n-1}{\tau^2}\left( \begin{matrix}
	\mp iQ&		q_l\\
	\bar{q}_l&		\pm iQ\\
\end{matrix} \right),\quad \quad
 G_{n}^{r}\thicksim - \theta \left( -n \right) \frac{n-1}{\tau^2}\left( \begin{matrix}
	\mp iQ&		q_r\\
	\bar{q}_r&		\pm iQ\\
\end{matrix} \right),\\
\tilde{G}_{n}^{l}\thicksim \theta \left( n-1 \right) \frac{n-1}{\tau^2}\left( \begin{matrix}
	\mp iQ&		q_l\\
	\bar{q}_l&		\pm iQ\\
\end{matrix} \right),\quad \quad
\tilde{ G}_{n}^{r}\thicksim - \theta \left( -n \right) \frac{n-1}{\tau^2}\left( \begin{matrix}
	\mp iQ&		q_r\\
	\bar{q}_r&		\pm iQ\\
\end{matrix} \right),
 \end{align*}
which means that the Green’s functions we derive are well defined at the branch points and linearly growing in $n$. Hence, if we further
require that the potential function $q_n$ belongs to $\mathcal{H}_1$, then the associated modified eigenfunctions are well defined at the
branch points as well.

\subsubsection{Analyticity}

 \hspace{1.5em}In this subsection, such as the method in Ref.~\cite{ck42}, we introduce a modified scattering problem to study the analytic properties of the
 eigenfunctions.

 First we let $\hat{M}_n=\Lambda M_n \Lambda^{-1}$, and from the scattering problem~(\ref{1.2}),  $\hat{\Phi}_n=\Lambda \Phi_n \Lambda^{-1}$
 and $\hat{\Psi}_n=\Lambda \Psi_n \Lambda^{-1}$  can be obtained to be the solutions of $\hat{v}_{n+1}=\hat{M}_n \hat{v}_n$ and expressed as
 follows according to Eqs.~(\ref{3.1.10})-(\ref{3.1.11}) and~(\ref{3.1.6}):
 \begin{gather}
\hat{\Phi}_n\sim \tau^n \,\Lambda V_{l} \Lambda^{-1}\, \Gamma^n=\tau^n\left(\begin{matrix}
                                q_l & \tau-z/\gamma\\
                                \tau \gamma^2-z\gamma & -\bar{q_l}
                              \end{matrix}\right) \Gamma^n \quad \quad n\rightarrow-\infty, \label{3.1.48} \\
\hat{\Psi}_n\sim \tau^n \,\Lambda V_{r} \Lambda^{-1}\, \Gamma^n=\tau^n\left(\begin{matrix}
                                q_r & \tau-z/\gamma\\
                                \tau\gamma^2-z\gamma & -\bar{q_r}
                              \end{matrix}\right) \Gamma^n \quad \quad n\rightarrow+\infty.\label{3.1.49}
\end{gather}
If we introduce
  \begin{gather}\label{3.1.50}
 \hat{N}_n=\left(\begin{matrix}
   {r}_n & \tau-z/\gamma\\
     \tau\gamma^2-z\gamma & -\bar{r}_n
      \end{matrix}\right)=\left(\begin{matrix}
   {r}_n & \tau-\lambda^{-1}\\
     \lambda-\tau & -\bar{r}_n
      \end{matrix}\right),
 \end{gather}
 in which the modified potentials ${r}_n$ satisfies
 \begin{gather}\label{3.1.51}
{r}_n \in \mathcal{H}_0, \ \  \, \text{and} \ \, \ |r_n|=Q \ \ \, \forall n\in \mathbb{Z},
  \end{gather}
  and define
  \begin{equation}\label{3.1.52}
  \begin{aligned}
  \check{\Phi}_n=\hat{N}_n^{-1} \hat{\Phi}_n , \\
  \check{\Psi}_n=\hat{N}_n^{-1} \hat{\Psi}_n ,
  \end{aligned}
  \end{equation}
  then combining $\hat{v}_{n+1}=\hat{M}_n \hat{v}_n$ one can show that $\check{\Phi}_n$ and $\check{\Psi}_n$ satisfy the modified scattering
  problem
  \begin{gather}\label{3.1.53}
\check{v}_{n+1}=\check{M}_n \check{v}_n,
  \end{gather}
  where
 \begin{gather}\label{3.1.54}
\check{M}_n =(\hat{N}_{n+1}^{-1}-\hat{N}_n^{-1})\hat{M}_n \hat{N}_n+ \hat{N}_n^{-1}\hat{M}_n \hat{N}_n.
  \end{gather}

 Due to
\begin{gather*}
 \hat{N}_n^{-1}=\frac{1}{(\gamma \tau-z)^2+Q^2}\left(\begin{matrix}
  \bar{ r}_n & \tau-z/\gamma\\
     \tau\gamma^2-z\gamma & -r_n
      \end{matrix}\right),
\end{gather*}
it can be found that $\hat{N}_{n+1}^{-1} $ and $\hat{N}_n^{-1} $ have the same limit, which contributes to the decay of the first term on the
right-hand side of Eq.~(\ref{3.1.54}) as $n\rightarrow \pm \infty $. In fact,
\begin{gather*}
  (\hat{N}_{n+1}^{-1}-\hat{N}_n^{-1})\hat{M}_n \hat{N}_n
  =\frac{1}{(\gamma \tau-z)^2+Q^2}B_n^{\langle1\rangle} ,
\end{gather*}
where
\begin{align*}
B_n^{\langle1\rangle} =\left( \begin{matrix}
	\left( z\left( r_n-q_n \right) +\gamma \tau q_n \right) \left( \bar{r}_{n+1}-\bar{r}_n \right)&		\gamma ^{-1}\left( z\left( \gamma \tau
-z \right) -q_n\bar{r}_n \right) \left( \bar{r}_{n+1}-\bar{r}_n \right)\\
	-\gamma \left( z^{-1}\left( \gamma \tau -z \right) +r_n\bar{q}_n \right) \left( r_{n+1}-r_n \right)&		\left( z^{-1}\left(
\bar{r}_n-\bar{q}_n \right) +\gamma ^{-1}\tau \bar{q}_n \right) \left( r_{n+1}-r_n \right)\\
\end{matrix} \right).
\end{align*}

Since one has $\hat{N}_n \sim \Lambda V_{l/r} \Lambda^{-1} $ from Eq.~(\ref{3.1.50}), for the second term on the right-hand side of Eq.~(\ref{3.1.54}), we can derive that $\hat{N}_n^{-1}\hat{M}_n \hat{N}_n \sim \tau \Gamma$ as $n\rightarrow \pm \infty$ according to Eq.~(\ref{3.1.7}).
Therefore, this term can be decomposed into two pieces, one of which is $\tau \Gamma$ and the other decays as $n\rightarrow \pm \infty$:
\begin{align*}
\hat{N}_n^{-1}\hat{M}_n \hat{N}_n=\tau \Gamma
+ \frac{1}{(\gamma \tau-z)^2+Q^2}B_n^{\langle2\rangle},
\end{align*}
where
\begin{align*}
 B_n^{\langle2\rangle}=\left( \begin{matrix}
	( \gamma \tau -z ) ( q_n\bar{r}_n+r_n\bar{q}_n-2Q^2 )&		\gamma ^{-1}\big( ( \gamma \tau -z ) ^2( \bar{q}_n-\bar{r}_n ) +\bar{r}_n(
Q^2-q_n\bar{r}_n ) \big)\\
	\gamma \big( ( \gamma \tau -z ) ^2( q_n-r_n ) +r_n( Q^2-r_n\bar{q}_n )\big )&		-\big( \gamma \tau -z ) (
q_n\bar{r}_n+r_n\bar{q}_n-2Q^2 \big)\\
\end{matrix} \right).
\end{align*}

Then by Eq.~(\ref{3.1.54}), it can be concluded that
\begin{gather}\label{3.1.55}
\check{M}_n=\tau \Gamma+\frac{1}{\tau(\lambda+1/\lambda-2\tau)}B_n,
  \end{gather}
where $B_n=B_n^{\langle1\rangle}+B_n^{\langle2\rangle}$ vanishes as $n \rightarrow \pm \infty$.

Let us introduce the modified eigenfunctions as
\begin{equation}\label{3.1.56}
\begin{aligned}
(\hat{U}_n\left( \lambda \right)\ \hat{\tilde{U}}_n\left(\lambda \right))& =\tau^{-n} \check{\Phi}_n(z)\Gamma^{-n},\\
(\hat{\tilde{W}}_n\left( \lambda \right)\ \hat{W}_n\left(\lambda \right))& =\tau^{-n}\check{\Psi}_n(z)\Gamma^{-n}.
 \end{aligned}
 \end{equation}
 Then from Eqs.~(\ref{3.1.53}) and~(\ref{3.1.55}) one can derive that
 \begin{equation}\label{3.1.57}
\begin{aligned}
 \hat{U}_{n+1}&=\gamma^{-1}\Gamma \hat{U}_{n}+\hat{B}_n \hat{U}_{n}, & \hat{\tilde{U}}_{n+1}&=\gamma\Gamma \hat{\tilde{U}}_{n}+\gamma^2 \hat{B}_n
 \hat{\tilde{U}}_{n},\\
 \hat{\tilde{W}}_{n+1}&=\gamma^{-1}\Gamma \hat{W}_{n}+\hat{B}_n \hat{W}_{n}, & \hat{W}_{n+1}&=\gamma\Gamma \hat{\tilde{W}}_{n}+\gamma^2 \hat{B}_n
 \hat{\tilde{W}}_{n},
\end{aligned}
 \end{equation}
 where $\hat{B}_n=\dfrac{\tau^{-1}\gamma^{-1}}{\tau(\lambda+1/\lambda-2\tau)}B_n=\dfrac{1}{\tau(\lambda+1/\lambda-2\tau)}\tilde{B}_{n}$ and
 $\tilde{B}_{n}$ has entries
 \begin{equation}\label{3.1.58}
\begin{aligned}
 (\tilde{B}_n)_{11}&= \tau^{-1}\lambda^{-1}\big((\tau\lambda-1)(p_n+ \bar{p}_n)+\bar{h}_n (\tau\lambda q_n -g_n) \big) ,\\
 (\tilde{B}_n)_{12}&=\tau \gamma^{-2} \big((\tau\gamma-z)^2 \bar{g}_n -\bar{r}_n p_n +\bar{h}_n (\tau z \gamma-z^2-q_n\bar{r}_n)\big),\\
 (\tilde{B}_n)_{21}&= \tau^{-1}\big( (\tau\gamma-z)^2 g_n -r_n \bar{p}_n  -h_n(\tau\lambda-1+ r_n\bar{q}_n )\big)
 ,\\
 (\tilde{B}_n)_{22}&=-\tau^{-1}\lambda^{-1}(\tau\lambda-1)(p_n+ \bar{p}_n)+h_n\big(\gamma^{-2}\bar{q}_n-\tau^{-1}(z \gamma)^{-1}
 \bar{g}_n\big),
\end{aligned}
 \end{equation}
 in which the notations are introduced as
 \begin{align}\label{3.1.59}
   p_n=q_n \bar{r}_n -Q^2, \quad \quad g_n=q_n-r_n, \quad \quad h_n=r_{n+1}-r_{n}.
    \end{align}
It can be easily shown that except for the points 0 and $\infty$, all entries of $\tilde{B}_n$ are bounded functions of $\lambda$ for $|\lambda| \geq1$ and all entries of
$\gamma^2\tilde{B}_n$ for $|\lambda|\leq1$ by using the relations~(\ref{3.1.13}). Therefore, if we exclude $\lambda=0, \ \infty$, and the
points where $\dfrac{1}{\tau(\lambda+1/\lambda-2\tau)}$ diverges, i.e. the branch points,  $\hat{B}_n(\lambda)$ is a bounded function of $\lambda$
for $|\lambda|\geq1$ and $\gamma^2 \hat{B}_n(\lambda)$ for $|\lambda|\leq1$.  That is, there is a matrix $\dot{B}_n$ independent of $\lambda$ such
that $ \|\hat{B}_n(\lambda)\| \leq \|\dot{B}_n\|$ for $\lambda\in \{\lambda\in
\mathbb{C}:|\lambda|\geq1\}\setminus(\mathcal{B}_\varepsilon(\lambda_0)\bigcup \mathcal{B}_\varepsilon(\bar{\lambda}_0))\bigcup
\mathcal{B}_\varepsilon(\infty))$ and $ \|\gamma^2 \hat{B}_n(\lambda)\| \leq \|\dot{B}_n\|$ for $\lambda\in \{\lambda\in
\mathbb{C}:|\lambda|\leq1\}\setminus(\mathcal{B}_\varepsilon(\lambda_0)\bigcup \mathcal{B}_\varepsilon(\bar{\lambda}_0))\bigcup
\mathcal{B}_\varepsilon(0))$, where $\mathcal{B}_\varepsilon(\lambda_1)=\{\lambda\in \mathbb{C}:|\lambda-\lambda_1|<\varepsilon\}$. Meanwhile,
since $q_n, r_n \in \mathcal{H}_0$ implies all potentials $p_n$, $g_n$ and $h_n$ vanish as $n\rightarrow\pm \infty$ and are absolutely
summable for $n$, we obtain $\hat{B}_n$ and $\gamma^2 \hat{B}_n$ approach 0 as $n\rightarrow \pm \infty$ and
$\sum_{n=-\infty}^{+\infty}{\|\dot{B}_n\|}<\infty$.

Combining Eqs.~(\ref{3.1.20})-(\ref{3.1.21}) with~(\ref{3.1.56}) and~(\ref{3.1.52}) yields
\begin{equation}\label{3.1.60}
\begin{aligned}
(\hat{U}_n\left( \lambda \right)\ \hat{\tilde{U}}_n\left(\lambda \right)) &=\hat{N}_n^{-1}(U_n\left( \lambda \right)\ \tilde{U}_n\left(\lambda
\right)),\\
(\hat{\tilde{W}}_n\left( \lambda \right)\ \hat{W}_n\left(\lambda \right)) &=\hat{N}_n^{-1}(\tilde{W}_n\left( \lambda \right)\ W_n\left(\lambda
\right)),
  \end{aligned}
 \end{equation}
which, by Eq.~(\ref{3.1.50}) and the asymptotic behaviors~(\ref{3.1.23})-(\ref{3.1.24}), imply
\begin{equation}\label{3.1.61}
(\hat{U}_n\left( \lambda \right)\ \hat{\tilde{U}}_n\left(\lambda \right)) \sim I, \quad n\rightarrow - \infty;\quad\quad
(\hat{\tilde{W}}_n\left( \lambda \right)\ \hat{W}_n\left(\lambda \right))\sim I, \quad n\rightarrow +\infty.
 \end{equation}

As in the previous section, we apply the Green's function method to find the summation equations for the modified eigenfunctions $\hat{U}_n$,
$\hat{\tilde{U}}_n$, $\hat{\tilde{W}}_n$, and $\hat{W}_n$. From the difference equations~(\ref{3.1.57}), we can write
 \begin{equation}\label{3.1.62}
\begin{aligned}
\hat{U}_n(\lambda)&=\left( \begin{array}{c}
	1\\
	0\\
\end{array} \right)  +\sum_{j=-\infty}^{+\infty}{\check{G}_{n-j}^{l}(\lambda)\hat{B}_j \hat{U}_j(\lambda)},\quad \ &
\hat{\tilde{U}}_n(\lambda)&=\left( \begin{array}{c}
	0\\
	1\\
\end{array} \right)  +\sum_{j=-\infty}^{+\infty}{\check{\tilde{G}}_{n-j}^{l}(\lambda)\gamma^2 \hat{B}_j \hat{\tilde{U}}_n(\lambda)},\\
\hat{\tilde{W}}_n(\lambda)&=\left( \begin{array}{c}
	1\\
	0\\
\end{array} \right)  +\sum_{j=-\infty}^{+\infty}{\check{G}_{n-j}^{r}(\lambda)\hat{B}_j \hat{W}_j(\lambda)},\quad \ &
\hat{W}_n(\lambda)&=\left( \begin{array}{c}
	0\\
	1\\
\end{array} \right)  +\sum_{j=-\infty}^{+\infty}{\check{\tilde{G}}_{n-j}^{r}(\lambda)\gamma^2 \hat{B}_j \hat{\tilde{W}}_n(\lambda)},
 \end{aligned}
 \end{equation}
in which the Green's functions satisfy
\begin{equation}\label{3.1.63}
\check{G}_{n+1}^{l/r}-\gamma^{-1}\Gamma \check{G}_{n}^{l/r}=\delta_{n,0}I, \quad \  \check{\tilde{G}}_{n+1}^{l/r}-\gamma\Gamma
\check{\tilde{G}}_{n}^{l/r}=\delta_{n,0}I.
\end{equation}
By the Fourier transform, the Green's functions can be expressed as the integrals
\begin{equation}\label{3.1.64}
\begin{aligned}
\check{G}_n^{l/r}\left( \lambda \right) &=\frac{1}{2\pi i}\oint_{\left| s \right|=1}\left(\begin{matrix}
                                                                                                     (s-1)^{-1} & 0 \\
                                                                                                     0 & (s-\gamma^{-2})^{-1}
                                                                                                   \end{matrix}\right) s^{n-1}\text{d}s,\\
   \check{\tilde{G}}_n^{l/r}\left( \lambda \right) &=\frac{1}{2\pi i}\oint_{\left| s \right|=1}\left(\begin{matrix}
                                                                                                     (s-\gamma^{2})^{-1} & 0 \\
                                                                                                     0 & (s-1)^{-1}
                                                                                                   \end{matrix}\right) s^{n-1}\text{d}s .
\end{aligned}
 \end{equation}

 Similar to the last subsection, to avoid the singularities on the circle $|s|=1$, there are two contours perturbed away from $|s|=1$ can be
 considered, one of which encloses the singularities on $|s|=1$ while the other does not. In view of the asymptotic behavior~(\ref{3.1.61})
 and the summation equations~(\ref{3.1.62}), when $|\lambda|\geq1$, i.e. $|\gamma|\geq1$, we consider the contour enclosing the singularities
 $0$, $1$ and $\gamma^{-2}$ for $\check{G}_n^{l}$, while the contour enclosing $s=0$ but neither 1 nor $\gamma^2$ for $\check{\tilde{G}}_n^{r}$;
 when $|\lambda|\leq1$, i.e. $|\gamma|\leq1$, the contour enclosing $0$, $1$ and $\gamma^{2}$ can be considered for $\check{\tilde{G}}_n^{l}$,
 while the contour enclosing $s=0$ but neither 1 nor $\gamma^{-2}$ for $\check{G}_n^{r}$. Therefore, we derive that
 \begin{equation}\label{3.1.65}
\begin{aligned}
\check{G}_n^{l}\left( \lambda \right)&=\theta(n-1)\left(\begin{matrix} 1 & 0 \\
                                                       0 & \gamma^{2(1-n)}
                                                     \end{matrix}\right),\quad \quad &
 \check{\tilde{G}}_n^{l}\left( \lambda \right)&=\theta(n-1)\left(\begin{matrix}
                                                       \gamma^{2(n-1)} & 0 \\
                                                       0 & 1
                                                     \end{matrix}\right),   \\
 \check{G}_n^{r}\left( \lambda \right)&=-\theta(-n)\left(\begin{matrix}
                                                       1 & 0 \\
                                                       0 & \gamma^{2(1-n)}
                                                     \end{matrix}\right),\quad \quad &
 \check{\tilde{G}}_n^{r}\left( \lambda \right)&=-\theta(-n)\left(\begin{matrix}
                                                       \gamma^{2(n-1)} & 0 \\
                                                       0 & 1
                                                     \end{matrix}\right).
\end{aligned}
 \end{equation}

 Then according to the summation equations~(\ref{3.1.62}) and the relations~(\ref{3.1.60}), we can derive the following Proposition.
 \paragraph{Proposition 3.2}
The modified eigenfunctions determined by Eq.~(\ref{3.1.62}) are unique in the space of bounded functions. Meanwhile, the modified eigenfunctions
$U_n\left( \lambda \right)$, $W_n\left( \lambda \right)$, and the scattering coefficient $s_1 ( \lambda )$ are analytic functions of $\lambda
$ outside the circle $|\lambda|=1$, except possibly not at $\lambda=\infty$; $\tilde{U}_n\left(\lambda \right)$, $\tilde{W}_n\left( \lambda
\right)$, and $s_4 ( \lambda )$ are analytic inside the circle $|\lambda|=1$, except possibly not at $\lambda=0$.
\begin{proof}
we consider the modified eigenfunctions $U_n\left( \lambda \right)$ for $|\lambda|\geq1$.
Define a recursive sequence
\begin{equation*}
\mu _{n}^{\langle 0 \rangle}\left( \lambda \right) =\left( \begin{array}{c}
	1\\
	0\\
\end{array} \right),\quad\
\mu _{n}^{\langle k+1 \rangle}\left( \lambda \right) =\sum_{j=-\infty}^{+\infty}{\check{G}_{n-j}^{l}(\lambda)\hat{B}_j}\mu _{j}^{\langle k
\rangle}\left( \lambda \right),\ \, k=0, 1, 2\cdots
\end{equation*}
Then the Neumann series $\sum_{k=0}^{+\infty}{\mu _{n}^{\langle k \rangle}\left( \lambda \right)}$ can be shown to be a solution for the
first equation of Eq.~(\ref{3.1.62}):
\begin{align*}
\sum_{k=0}^{+\infty}{\mu _{n}^{\langle k \rangle}\left( \lambda \right)} =\left( \begin{array}{c}
	1\\
	0\\
\end{array} \right) +\sum_{k=1}^{+\infty}{\mu _{j}^{\langle k \rangle}\left( \lambda \right)}
=\left( \begin{array}{c}
	1\\
	0\\
\end{array} \right) +\sum_{j=-\infty}^{+\infty}{{\check{G}_{n-j}^{l}(\lambda)\hat{B}_j}}\sum_{k=0}^{+\infty}{\mu _{j}^{\langle k \rangle}\left(
\lambda \right)}.
\end{align*}
We will show that the Neumann series $\sum_{k=0}^{+\infty}{\mu _{n}^{\langle k \rangle}\left( \lambda \right)}$ converges to an analytic
function $\mu _{n}(\lambda)$, and then prove that $\mu _{n}(z)$ is the unique solution for the first equation of Eq.~(\ref{3.1.62}).
We first use induction to prove that for$\lambda\in \{\lambda\in
\mathbb{C}:|\lambda|\geq1\}\setminus(\mathcal{B}_\varepsilon(\lambda_0)\bigcup \mathcal{B}_\varepsilon(\bar{\lambda}_0))\bigcup
\mathcal{B}_\varepsilon(\infty))$ :
\begin{equation*}
|\mu _{n}^{\langle k\rangle}(\lambda)|\le \frac{(\sum_{j=-\infty}^{n-1}{\|\dot{B}_j\|})^k}{k!}.
\end{equation*}
Introducing the $L^1$ vector norm and the corresponding subordinate matrix norm, the conclusion is obviously true when $k=0$. Suppose that the
conclusion holds when $k=l$, then there is
\begin{equation*}
\|\mu _{n}^{\langle j+1 \rangle}\left( \lambda \right)\| \leq \sum_{j=-\infty}^{n-1}{\|H_j\| \|\hat{B}_j\|}\|\mu _{j}^{\langle j \rangle}\left(
\lambda \right)\| \le \sum_{j=-\infty}^{n-1}{\|H_j\| \|\dot{B}_j\|}\frac{(\sum_{s=-\infty}^{j-1}{\|\dot{B}_j\|})^j}{j!},
\end{equation*}
where $H_j=\text{diag}(1,\gamma^{2(1-n+j)})$. For $j\le n-1$, since $|\lambda|\geq1$ implies $|\gamma|^{2(1-n+j)} \le 1 $ , we then have
$\|H_j\|=1$. Therefore,
\begin{equation*}
\|\mu _{n}^{\langle j+1 \rangle}\left( \lambda \right)\|  \le \sum_{j=-\infty}^{n-1}{
\|\dot{B}_j\|}\frac{(\sum_{s=-\infty}^{j-1}{\|\dot{B}_j\|})^j}{j!}\le
\sum_{j=-\infty}^{n-1}{\frac{(\sum_{s=-\infty}^{j-1}{\|\dot{B}_j\|})^{j+1}}{(j+1)!}}.
\end{equation*}
We thus complete the proof of the induction part.
Recalling $\sum_{j=-\infty}^{+\infty}{\|\dot{B}_j\|}<\infty$ , we can derive that $\frac{(\sum_{j=-\infty}^{n-1}{\|\dot{B}_j\|})^k}{k!}$ is
absolutely and uniformly (in $n$) summable in $k$, and then the Neumann series $\sum_{k=0}^{+\infty}{\mu _{n}^{\langle k \rangle}\left(
\lambda\right)}$ converges absolutely and uniformly with respect to $n$ and to $\lambda\in \{\lambda\in
\mathbb{C}:|\lambda|\geq1\}\setminus(\mathcal{B}_\varepsilon(\lambda_0)\bigcup \mathcal{B}_\varepsilon(\bar{\lambda}_0))\bigcup
\mathcal{B}_\varepsilon(\infty))$  for all $\varepsilon>0$. Consequently,  $\sum_{k=0}^{+\infty}{\mu _{n}^{\langle k \rangle}\left( \lambda
\right)}$ converges to a function $\mu _{n}(\lambda)$ which is analytic for $\{|\lambda|>1\}\setminus\{\infty\}$ according to the fact that a
uniformly convergent series of analytic functions converages to an analytic function in the interior of the domain.
We next prove the $\mu _{n}(z)$ is the unique solution. Suppose that there is another solution $ \dot{\mu} _{n}(z)$, then
\begin{align*}
\|\mu _n-\dot{\mu} _{n}\|
\le \sum_{j=-\infty}^{+\infty}{\|{\check{G}_{n-j}^{l}\| \|\hat{B}_j}\|}\|\mu _j-\dot{\mu} _{j}\|\le\sum_{j=-\infty}^{n-1}{ \|\dot{B}_j\|}\|\mu
_j-\dot{\mu} _{j}\|.
\end{align*}
Since $\mu_n(\lambda)$ and $\dot{\mu}_n(\lambda)$ are bounded functions, then $\exists\  C \geq 0$ s.t. $\|\mu _n-\dot{\mu} _{n}\|\leq C$ for
all $n\in \mathbb{Z}$.
Therefore, iterating once yields
\begin{align*}
\|\mu _n-\dot{\mu} _{n}\|
\le \sum_{j=-\infty}^{n-1}{ \|\dot{B}_j\|}\sum_{s=-\infty}^{j-1}{ \|\dot{B}_s\|}\|\mu _s-\dot{\mu} _{s}\|\le
\frac{C}{2}(\sum_{j=-\infty}^{n-1}{ \|\dot{B}_j\|})^2.
\end{align*}
And N iterations leads to
\begin{align*}
\|\mu _n-\dot{\mu} _{n}\|
\le \frac{C}{N!}\sum_{j=-\infty}^{n-1}{ \|\dot{B}_j\|}(\sum_{s=-\infty}^{j-1}{ \|\dot{B}_s\|})^N\le
\frac{C}{(N+1)!}(\sum_{j=-\infty}^{n-1}{ \|\dot{B}_j\|})^{N+1}.
\end{align*}
When $N\rightarrow +\infty$, the right-hand side of above equation will tend to 0, hence $\|\mu _n-\dot{\mu} _{n}\|=0$. That is, $\mu
_{n}(\lambda)=\dot{\mu} _{n}(\lambda)$. Thus the uniqueness of the solution for the first equation of Eq.~(\ref{3.1.62}) is proved. It follows
that $\hat{U}_n\left( \lambda \right)=\mu _{n}(\lambda)$, and they have the same analyticity. In the same way, we can prove that $\hat{W}_n$
is analytic in $\{|\lambda|>1\}\setminus\{\infty\}$ , while $\hat{\tilde{U}}_n$ and $\hat{\tilde{W}}_n$ are analytic in $\{|\lambda|<1\}\setminus\{0\}$  by Eqs.~(\ref{3.1.62}), and they
are unique solutions to the  summation equations they satisfy.
From Eq.~(\ref{3.1.50}) and the relation~(\ref{3.1.60}), we obtain that the modified eigenfunctions $U_n$ and $W_n$ are analytic in
$|\lambda|>1$ with at most a pole at $\lambda=\infty$, and that $\tilde{U}_n$ and $\tilde{W}_n$ are analytic in $|\lambda|<1$ with at most a
pole at $\lambda=0$. Then according to Eqs.~(\ref{3.1.42}), the scattering coefficient $s_1(\lambda)$  has the same analyticity with the
modified eigenfunctions $U_n$ and $W_n$, and $s_4(\lambda)$  has the same analyticity with $\tilde{U}_n$ and $\tilde{W}_n$.
\end{proof}
In fact, the asymptotic behavior~(\ref{3.1.76}) and~(\ref{3.1.81}) in subsection 3.1.5 implys that  $\infty$ and $0$ are poles of $U_n(\lambda)
$ and $\tilde{U}_n (\lambda)$ respectively, while that $W_n$, $s_1(\lambda)$ are well defined at infinity and $\tilde{W}_n$, $s_4(\lambda)$
are well defined at 0.

 Although we excluded the branch points in the above discussion, it was shown in subsection 3.1.4 that the eigenfunctions $U_n$, $W_n$,
 $\tilde{U}_n$ and $\tilde{W}_n$ are well defined at the branch points. By Eqs.~(\ref{3.1.42}), the scattering coefficients can be written as
\begin{align*}
s_1(\lambda)&=\frac{a_1}{\lambda-\lambda_0}+O(1),\quad\quad \tilde{s}_3(\lambda)=\frac{a_3}{\lambda-\lambda_0}+O(1) ,\quad\quad \text{as}\
\lambda\rightarrow \lambda_0,\\
s_1(\lambda)&=\frac{\tilde{a}_1}{\lambda-\bar{\lambda}_0}+O(1),\quad\quad
\tilde{s}_3(\lambda)=\frac{\tilde{a}_3}{\lambda-\bar{\lambda}_0}+O(1) ,\quad\quad \text{as}\ \lambda\rightarrow \lambda_0,
\end{align*}
where
\begin{alignat*}{2}
a_1&=\chi_n\frac{i \lambda_0 \text{det}\big(U_{n}(\lambda_0),W_{n}(\lambda_0)\big)}{2  \tau Q},\quad && a_3=\chi_n\frac{i \lambda_0 \text{det}
\big(\tilde{W}_{n}(\lambda_0), U_{n}(\lambda_0) \big)}{2 \tau Q},\\
\tilde{a}_1&=\chi_n\frac{i\bar{\lambda}_0 \text{det}\big(W_{n}(\bar{\lambda}_0),U_{n}(\bar{\lambda}_0)\big)}{2 \tau Q},&\quad&
\tilde{a}_3=\chi_n\frac{i\bar{\lambda}_0 \text{det} \big(U_{n}(\bar{\lambda}_0),\tilde{W}_{n}(\bar{\lambda}_0) \big)}{2  \tau Q}.
\end{alignat*}

Consequently,  if $U_n(\lambda)$ and $W_n(\lambda)$ are linearly independent at the branch points $\lambda_0$ and $\bar{\lambda}_0$, then
$a_1$, $\tilde{a}_1\neq0$, which  implies that $\lambda_0$ and $\bar{\lambda}_0$ are two poles of $s_1(\lambda)$. Conversely, the case when
either $a_1$ or $\tilde{a}_1$ vanish can lead to what is called a {\it virtual level} in scattering theory.

In addition, according to Eq.~(\ref{3.1.42}) one has
 \begin{align*}
(\tilde{W}_n\left( \lambda_0 \right)\ W_n\left(\lambda_0 \right))=(\tilde{W}_n\left( \bar{\lambda}_0 \right)\ W_n\left(\bar{\lambda}_0
\right))=0.
\end{align*}
That is, $\tilde{W}_n$ and $W_n$ are proportional to each other at $\lambda=\lambda_0$ or $\bar{\lambda}_0$.
Comparing the behavior~(\ref{3.1.23})-(\ref{3.1.24}) yields
\begin{align*}
\tilde{W}_n\left( \lambda_0 \right)=- i e^{i\theta_r}\ W_n\left(\lambda_0 \right),\quad \quad\tilde{W}_n\left( \bar{\lambda}_0 \right)=i
e^{i\theta_r}W_n\left(\bar{\lambda}_0 \right).
\end{align*}
It follows that $a_3=i e^{i\theta_r}a_1$ and $\tilde{a}_3=-i e^{i\theta_r}\tilde{a}_1$.

\subsubsection{Asymptotics}

\hspace{1.5em}From the relations~(\ref{3.1.12}) and~(\ref{3.1.13}) one has
\begin{equation}\label{3.1.66}
\begin{alignedat}{4}
  z^2\sim& \frac{\tau}{\lambda},&\quad \  \, \quad& \gamma^2\sim\tau\lambda,&\quad \  \, \quad& z\gamma\sim\tau,&\quad\  \, \quad&  \text{as}
  \ \lambda\rightarrow0,\\
  z^2\sim& \frac{1}{\tau\lambda},&& \gamma^2\sim\frac{\lambda}{\tau},&&  z\gamma\sim\frac{1}{\tau},&  &\text{as} \ \lambda\rightarrow\infty,\\
  z^2\sim& -\frac{\lambda-\tau}{\tau Q^2},&&\gamma^2\sim -\frac{\tau(\lambda-\tau)}{ Q^2},&& z\gamma\sim-\frac{\lambda-\tau}{ Q^2},&&\text{as}
  \ \lambda\rightarrow\tau,\\
  z^2\sim& \frac{Q^2}{\tau\lambda-1},&& \gamma^2\sim \frac{Q^2}{\tau^2(\tau\lambda-1)},&&
  z\gamma\sim\frac{Q^2}{\tau(\tau\lambda-1)},&&\text{as} \ \lambda\rightarrow\tau^{-1}.
\end{alignedat}
\end{equation}
Thus, the expansions for the elements in Eqs.~(\ref{3.1.25})-(\ref{3.1.26}) at $\lambda=0$ and $\infty$, respectively, can be written as follows:
\begin{equation}\label{3.1.67}
\begin{alignedat}{9}
\gamma^2&=\tau\lambda+O(\lambda^2),&\quad &  &&z\gamma= \tau +O(\lambda),&\quad &  &&\gamma^{-2}=\frac{1}{\tau\lambda} +O(1),&\quad &
&&(z\gamma)^{-1}= \frac{1}{\tau} +O(\lambda),&\quad &  &&\lambda\rightarrow 0\\
\gamma^2&=\frac{\lambda}{\tau}+O(1),&\quad &  &&z\gamma= \frac{1}{\tau} +O(\frac{1}{\lambda}),&\quad &  &&\gamma^{-2}=\frac{\tau}{\lambda}
+O(\frac{1}{\lambda^2}),&\quad &  &&(z\gamma)^{-1}= \tau +O(\frac{1}{\lambda}),&\quad &  &&\lambda\rightarrow \infty,
\end{alignedat}
\end{equation}

We write the WKB expansion of the entries of the modified eigenfunctions $\tilde{U}_n(\lambda)$ and $\tilde{W}_n(\lambda)$ at $\lambda=0$, and
$U_n(\lambda)$ and $W_n(\lambda)$ at $\lambda=\infty$ as
\begin{alignat}{2}
\tilde{U}_n^{(1)}(\lambda)&=\sum_{k=-1}^{\infty}\tilde{U}_n^{(1),\langle k\rangle}\lambda^k,&\quad \quad&
\tilde{U}_n^{(2)}(\lambda)=\sum_{k=0}^{\infty}\tilde{U}_n^{(2),\langle k\rangle}\lambda^k,\label{3.1.68}\\
\tilde{W}_n^{(1)}(\lambda)&=\sum_{k=0}^{\infty}\tilde{W}_n^{(1),\langle k\rangle}\lambda^k,&&
\tilde{W}_n^{(2)}(\lambda)=\sum_{k=0}^{\infty}\tilde{W}_n^{(2),\langle k\rangle}\lambda^k,\label{3.1.69}\\
U_n^{(1)}(\lambda)&=\sum_{k=0}^{\infty}U_n^{(1),\langle k\rangle}\lambda^{-k},&&
 U_n^{(2)}(\lambda)=\sum_{k=-1}^{\infty}U_n^{(2),\langle k\rangle}\lambda^{-k},\label{3.1.70}\\
W_n^{(1)}(\lambda)&=\sum_{k=0}^{\infty}W_n^{(1),\langle k\rangle}\lambda^{-k},&&
 W_n^{(2)}(\lambda)=\sum_{k=0}^{\infty}W_n^{(2),\langle k\rangle}\lambda^{-k}.\label{3.1.71}
\end{alignat}

Substituting Eqs.~(\ref{3.1.68}) and (\ref{3.1.71}) into the scattering problem~(\ref{3.1.25}), and combining the expansions~(\ref{3.1.67})
yields
\begin{alignat}{2}
&\tau \tilde{U}_{n+1}^{(1),\langle -1\rangle}\lambda^{-1}+O(1)=\tau \tilde{U}_{n}^{(1),\langle -1\rangle}\lambda^{-1}+O(1),&\quad&
\lambda\rightarrow 0,\label{3.1.72}\\
&\tau \tilde{U}_{n+1}^{(2),\langle -1\rangle}\lambda^{-1}+\tau \tilde{U}_{n+1}^{(2),\langle 0\rangle}+O(\lambda)=\tau \bar{q}_n
\tilde{U}_{n}^{(1),\langle -1\rangle}+\tilde{U}_{n}^{(2),\langle -1\rangle}+O(\lambda),&&\lambda\rightarrow 0,\label{3.1.73}\\
&\tau W_{n+1}^{(1),\langle 0\rangle}+O(\lambda^{-1})=\tau^{-1}W_n^{(1),\langle 0\rangle}+q_n W_n^{(2),\langle
0\rangle}+O(\lambda^{-1}),&&\lambda\rightarrow\infty, \label{3.1.74}\\
&O(1)=\tau^{-1}\bar{q}_n W_n^{(1),\langle 0\rangle}\lambda+ W_n^{(2),\langle 0\rangle}\lambda+O(1),&&\lambda\rightarrow \infty.\label{3.1.75}
\end{alignat}
Comparing the coefficients of the different powers of $\lambda$ on both sides of Eqs.~(\ref{3.1.72})-(\ref{3.1.73}) and taking account the
asymptotic behavior~(\ref{3.1.23}) gives that $\tilde{U}_{n}^{(1),\langle -1\rangle}=-1$,  $\tilde{U}_{n}^{(2),\langle -1\rangle}=0$,  and
$\tilde{U}_{n}^{(2),\langle 0\rangle}=-\bar{q}_{n-1}$.  Equally, from Eqs .~(\ref{3.1.74})-(\ref{3.1.75}) one has $W_n^{(2),\langle
0\rangle}=-\dfrac{\bar{q}_n}{\tau}$, and $ W_{n+1}^{(1),\langle 0\rangle}=\dfrac{1-|q_n|^2}{\tau^2}W_{n}^{(1),\langle 0\rangle}$, which
implies that $ W_{n}^{(1),\langle 0\rangle}=\dfrac{\tau}{\chi_n}$ and that $W_n^{(2),\langle 0\rangle}=-\dfrac{\bar{q}_n}{\chi_n} $ by considering the
asymptotic behavior~(\ref{3.1.24}).
Therefore,
\begin{align}\label{3.1.76}
\tilde{U}_n(\lambda)\sim \left(\begin{array}{c}
                                  -\lambda^{-1} \\
                                  -\bar{q}_{n-1}
                                \end{array}\right),  \quad  \quad \text{as} \ \lambda\rightarrow0;\quad\quad\quad
 W_n(\lambda)\sim \dfrac{1}{\chi_n}\left(\begin{array}{c}
                                  \tau \\
                                  -\bar{q}_n
                                \end{array}\right),    \quad  \quad \text{as} \ \lambda\rightarrow\infty.
\end{align}

In the same way, substituting Eqs.~(\ref{3.1.69}) and~(\ref{3.1.70}) into (\ref{3.1.26}) and combining (\ref{3.1.67}) yields
\begin{alignat}{2}
&O(1)=\tilde{W}_{n}^{(1),\langle 0\rangle}\lambda^{-1}+\frac{q_n}{\tau}\tilde{W}_{n}^{(2),\langle 0\rangle}\lambda^{-1}+O(1),&\quad \ \quad&
\lambda\rightarrow 0, \label{3.1.77}\\
&\tau \tilde{W}_{n+1}^{(2),\langle 0\rangle}+O(\lambda)=\bar{q}_n \tilde{W}_{n}^{(1),\langle
0\rangle}+\frac{1}{\tau}\tilde{W}_{n}^{(2),\langle 0\rangle}+O(\lambda),&&\lambda\rightarrow 0,\label{3.1.78}\\
&\tau U_{n+1}^{(1),\langle 0\rangle}+O(\lambda^{-1})=\tau q_n U_{n}^{(2),\langle -1\rangle}+O(\lambda^{-1}) ,&&\lambda\rightarrow
\infty,\label{3.1.79}\\
&\tau U_{n+1}^{(2),\langle -1\rangle}\lambda+O(1)=\tau U_{n}^{(2),\langle -1\rangle}\lambda+O(1),&&\lambda\rightarrow \infty.\label{3.1.80}
\end{alignat}
Then taking account the asymptotic behavior~(\ref{3.1.23})-(\ref{3.1.24}) one derive that
$\tilde{W}_{n}^{(2),\langle 0\rangle}=-\tau \chi_n^{-1}$, $\tilde{W}_{n}^{(1),\langle 0\rangle}=q_n \chi_n^{-1}$, $U_{n}^{(2),\langle
-1\rangle}=1$, and $U_{n}^{(1),\langle 0\rangle}=q_{n-1}$. Therefore,
\begin{align}\label{3.1.81}
\tilde{W}_n(\lambda)\sim \dfrac{1}{\chi_n} \left(\begin{array}{c}
                                  q_n\\
                                  -\tau
                                \end{array}\right),  \quad  \quad \text{as} \ \lambda\rightarrow0;\quad\quad\quad
 U_n(\lambda)\sim \left(\begin{array}{c}
                                  q_{n-1} \\
                                  \lambda
                                \end{array}\right),    \quad  \quad \text{as} \ \lambda\rightarrow\infty.
\end{align}

 From Eqs.~(\ref{3.1.42}), we get that $z, \  \lambda\rightarrow0$ as $\lambda\rightarrow\tau$ and   $z, \  \lambda\rightarrow\infty$ as
 $\lambda\rightarrow\tau^{-1}$, which suggests that $\lambda=\tau, \tau^{-1}$ also play a special role.

The WKB expansion of $\tilde{U}_n(\lambda)$ and $\tilde{W}_n(\lambda)$ at $\lambda=\tau$, and $U_n(\lambda)$ and $W_n(\lambda)$ at
$\lambda=\tau^{-1}$ can be written as
\begin{equation}\label{3.1.82}
\begin{alignedat}{2}
\tilde{U}_n^{(1)}(\lambda)&=\sum_{k=0}^{\infty}\tilde{U}_n^{(1),\langle k\rangle}(\lambda-\tau)^k,&\quad \quad&
\tilde{U}_n^{(2)}(\lambda)=\sum_{k=0}^{\infty}\tilde{U}_n^{(2),\langle k\rangle}(\lambda-\tau)^k,\\
\tilde{W}_n^{(1)}(\lambda)&=\sum_{k=0}^{\infty}\tilde{W}_n^{(1),\langle k\rangle}(\lambda-\tau)^k,&&
\tilde{W}_n^{(2)}(\lambda)=\sum_{k=0}^{\infty}\tilde{W}_n^{(2),\langle k\rangle}(\lambda-\tau)^k,\\
U_n^{(1)}(\lambda)&=\sum_{k=0}^{\infty}U_n^{(1),\langle k\rangle}(\lambda-\frac{1}{\tau})^{k},&&
 U_n^{(2)}(\lambda)=\sum_{k=0}^{\infty}U_n^{(2),\langle k\rangle}(\lambda-\frac{1}{\tau})^{k},\\
W_n^{(1)}(\lambda)&=\sum_{k=0}^{\infty}W_n^{(1),\langle k\rangle}(\lambda-\frac{1}{\tau})^{k},&&
 W_n^{(2)}(\lambda)=\sum_{k=0}^{\infty}W_n^{(2),\langle k\rangle}(\lambda-\frac{1}{\tau})^{k},
\end{alignedat}
\end{equation}
and  the expansions for the elements in~(\ref{3.1.25})-(\ref{3.1.26}) at $\lambda=\tau$ and $ \tau^{-1}$ can be expressed as
\begin{equation}\label{3.1.83}
\begin{alignedat}{3}
\gamma^2&=-\frac{\tau(\lambda-\tau)}{ Q^2}+O((\lambda-\tau)^2), &\quad & z\gamma= -\frac{\lambda-\tau}{ Q^2} +O((\lambda-\tau)^2), &\quad &
\lambda\rightarrow \tau,\\ \frac{1}{\gamma^2}&=-\frac{Q^2}{ \tau(\lambda-\tau)} +O(1), &&\frac{1}{z\gamma}=  -\frac{ Q^2}{\lambda-\tau} +O(1),
&&\lambda\rightarrow \tau,\\
\gamma^2&=\frac{Q^2}{\tau^3(\lambda-\frac{1}{\tau})}+O(1),  &&z\gamma= \frac{Q^2}{\tau^2(\lambda-\frac{1}{\tau})} +O(1),&&\lambda\rightarrow
\frac{1}{\tau},\\  \frac{1}{\gamma^2}&=\frac{\tau^3(\lambda-\frac{1}{\tau})}{Q^2} +O((\lambda-\frac{1}{\tau})^2),  &&\frac{1}{z\gamma}=
\frac{\tau^2(\lambda-\frac{1}{\tau})}{Q^2} +O((\lambda-\frac{1}{\tau})^2),  &&\lambda\rightarrow \frac{1}{\tau}.
\end{alignedat}
\end{equation}
Combining the expansions~(\ref{3.1.82})-(\ref{3.1.83}) with the scattering problems~(\ref{3.1.25})-(\ref{3.1.26}) yields
\begin{alignat*}{2}
&\tau \tilde{U}_{n+1}^{(1),\langle 0\rangle}+O(\lambda-\tau)=q_n \tilde{U}_{n}^{(2),\langle 0\rangle}+O(\lambda-\tau),&\quad \ &
\lambda\rightarrow \tau,\\
&\tau \tilde{U}_{n+1}^{(2),\langle 0\rangle}+O(\lambda-\tau)=\tau \tilde{U}_{n}^{(2),\langle 0\rangle}+O(\lambda-\tau ), &&\lambda\rightarrow
\tau,\\
&\tau\tilde{W}_{n}^{(1),\langle 0\rangle}+O(\lambda-\tau)=-\frac{q_n Q^2}{\tau(\lambda-\tau)}\tilde{W}_{n}^{(2),\langle 0\rangle}-\frac{q_n
Q^2}{\tau}\tilde{W}_{n}^{(2),\langle 1\rangle}+\frac{1}{\tau}\tilde{W}_{n}^{(1),\langle 0\rangle}+O(\lambda-\tau),&& \lambda\rightarrow \tau,
\\
&\tau\tilde{W}_{n}^{(2),\langle 0\rangle}+O(\lambda-\tau)=-\frac{ Q^2}{\lambda-\tau}\tilde{W}_{n}^{(2),\langle 0\rangle}-
Q^2\tilde{W}_{n}^{(2),\langle 1\rangle}+\bar{q}_n\tilde{W}_{n}^{(1),\langle 0\rangle}+O(\lambda-\tau),&& \lambda\rightarrow \tau, \\
&\tau U_{n+1}^{(1),\langle 0\rangle}+O(\lambda-\frac{1}{\tau})=\tau U_{n}^{(1),\langle 0\rangle}+O(\lambda-\frac{1}{\tau})
,&&\lambda\rightarrow \frac{1}{\tau},\\
&\tau U_{n+1}^{(2),\langle 0\rangle}+O(\lambda-\frac{1}{\tau})=\bar{q}_n U_{n}^{(1),\langle
0\rangle}+O(\lambda-\frac{1}{\tau}),&&\lambda\rightarrow \frac{1}{\tau},\\
&\tau W_{n+1}^{(1),\langle 0\rangle}+O(\lambda-\frac{1}{\tau})=\frac{Q^2}{\tau^2(\lambda-\frac{1}{\tau})}W_n^{(1),\langle
0\rangle}+\frac{Q^2}{\tau^2}W_n^{(1),\langle 1\rangle}+q_n W_n^{(2),\langle 0\rangle}+O(\lambda-\frac{1}{\tau}),&&\lambda\rightarrow
\frac{1}{\tau},\\
&\tau W_{n+1}^{(2),\langle 0\rangle}+O(\lambda-\frac{1}{\tau})=\frac{\bar{q}_n Q^2}{\tau^3(\lambda-\frac{1}{\tau})} W_{n}^{(1),\langle
0\rangle}+\frac{\bar{q}_n Q^2}{\tau^3} W_{n}^{(1),\langle 1\rangle}+\frac{1}{\tau} W_n^{(2),\langle
0\rangle}+O(\lambda-\frac{1}{\tau}),&&\lambda\rightarrow \frac{1}{\tau} ,
\end{alignat*}
and then
\begin{equation}\label{3.1.84}
\begin{alignedat}{3}
\tilde{U}_n(\lambda)\sim & -\bar{q}_l\left(\begin{array}{c}
                                  \dfrac{q_{n-1}}{\tau} \\
                                  1
                                \end{array}\right),  &\quad  \quad\quad&
\tilde{W}_n(\lambda)\sim \dfrac{q_r}{\chi_n} \left(\begin{array}{c}
                                  1\\
                                  \dfrac{\bar{q}_n(\lambda-\tau)}{Q^2}
                                \end{array}\right),  &\quad \quad \quad& \text{as} \ \lambda\rightarrow\tau,\\
 U_n(\lambda)\sim & \ q_l\left(\begin{array}{c}
                                  1 \\
                                 \dfrac{\bar{q}_{n-1}}{\tau}
                                \end{array}\right),
&&W_n(\lambda)\sim \dfrac{\bar{q}_r}{\chi_n}\left(\begin{array}{c}
                                  \dfrac{\tau^2 q_n(\lambda-\frac{1}{\tau})}{Q^2} \\
                                  -1
                                \end{array}\right),   && \text{as} \ \lambda\rightarrow\frac{1}{\tau},
\end{alignedat}
\end{equation}
which are compatible with the asymptotic behavior~(\ref{3.1.23})-(\ref{3.1.24}).

Accordingly, using Eqs.~(\ref{3.1.42}) one can obtain that
\begin{equation}\label{3.1.85}
\begin{alignedat}{4}
s_4(\lambda)&\sim \frac{q_{n} \bar{q}_{n-1} +\tau\lambda^{-1}}{\tau(\lambda+\lambda^{-1}-2 \tau)}\sim1,&\quad \quad&\lambda\rightarrow0,&\quad
\quad \quad&
s_1(\lambda)\sim \frac{q_{n-1} \bar{q}_n +\tau\lambda}{\tau(\lambda+\lambda^{-1}-2 \tau)}\sim1,& \quad \quad & \lambda\rightarrow\infty,\\
s_4(\lambda)&\sim \frac{q_r \bar{q}_l }{1-\tau^2}=e^{i(\theta_r-\theta_l)},&&\lambda\rightarrow \tau,&&
s_1(\lambda)\sim \frac{q_l \bar{q}_r }{1-\tau^2}=e^{i(\theta_l-\theta_r)},&&\lambda\rightarrow \frac{1}{\tau}.
\end{alignedat}
\end{equation}

\subsubsection{Symmetries}

\paragraph{Proposition 3.3} The eigenfunctions $\phi_{n}(z)$ and $\psi_{n}(z)$ have the following symmetry:
\begin{gather*}
\Phi_{n}(\lambda)=-\sigma_1\bar{\Phi}_{n}(\bar{\lambda}^{-1})\sigma_1,\quad
\Psi_{n}(\lambda)=-\sigma_1\bar{\Psi}_{n}(\bar{\lambda}^{-1})\sigma_1,
\end{gather*}
\begin{proof} It is easily to derive that $M_{n}(z)=\sigma_1\bar{M}_{n}(\bar{z}^{-1})\sigma_1$. Then from
$\bar{\Phi}_{n+1}(\bar{z}^{-1})=\bar{M}_n(\bar{z}^{-1}) \bar{\Phi}_{n}(\bar{z}^{-1})$
, we get
\begin{align*}
-\sigma_1\bar{\Phi}_{n+1}(\bar{z}^{-1})\sigma_1=\big(\sigma_1\bar{M}_n(\bar{z}^{-1})\sigma_1\big)\big(-
\sigma_1\bar{\Phi}_{n}(\bar{z}^{-1})\sigma_1\big)
={M}_n(z)\big(-\sigma_1\bar{\Phi}_{n}(\bar{z}^{-1})\sigma_1\big),
\end{align*}
that is to say, ${\Phi}_{n}(z)$ and $-\sigma_1\bar{\Phi}_{n}(\bar{z}^{-1})\sigma_1$ satisfy the same differential equation~(\ref{1.2}).
Likewise, ${\Phi}_{n}(z)$ and $-\sigma_1\bar{\Psi}_{n}(\bar{z}^{-1})\sigma_1$ also satisfy the same differential equation~(\ref{1.2}).
\\\indent  Moreover, according to the relation~(\ref{3.1.4}) and the BCs~(\ref{3.1.10})-(\ref{3.1.11}), one can derive that
$\Phi_{n}(z,\gamma)$, $-\sigma_1\bar{\Phi}_{n}(\bar{z}^{-1},\bar{\gamma}^{-1})\sigma_1\sim \tau^n \, V_{l} \, \Gamma^n$ as $n\rightarrow
-\infty$ and $\Psi_{n}(z,\gamma)$, $-\sigma_1\bar{\Psi}_{n}(\bar{z}^{-1},\bar{\gamma}^{-1})\sigma_1\sim \tau^n \, V_{r} \, \Gamma^n$ as
$n\rightarrow +\infty$ .
\\\indent In conclusion, there are $\Phi_{n}(z,\gamma)=-\sigma_1\bar{\Phi}_{n}(\bar{z}^{-1},\bar{\gamma}^{-1})\sigma_1$ and
$\Psi_{n}(z,\gamma)=-\sigma_1\bar{\Psi}_{n}(\bar{z}^{-1},\bar{\gamma}^{-1})\sigma_1$. Then we obtain
$\Phi_{n}(\lambda)=-\sigma_1\bar{\Phi}_{n}(\bar{\lambda}^{-1})\sigma_1$ and $
\Psi_{n}(\lambda)=-\sigma_1\bar{\Psi}_{n}(\bar{\lambda}^{-1})\sigma_1$ by the mapping~(\ref{3.1.12}) .
\end{proof}

\paragraph{Proposition 3.4} The scattering matrix $S(z)$ and $T(\lambda)$ have the following symmetry:
\begin{gather*}
S(z)=\sigma_1\bar{S}_{n}(\bar{z}^{-1})\sigma_1,\quad T(\lambda)=\sigma_1 T(\bar{\lambda}^{-1})\sigma_1.
\end{gather*}

\begin{proof}
From  Eq.~(\ref{3.1.32}) one obtain $S(z)=\Psi_n^{-1}(z)\Phi_n(z)$. Thus by the Proposition 3.3,
$\sigma_1\bar{S}(\bar{z}^{-1})\sigma_1=\sigma_1\Psi_n^{-1}(\bar{z}^{-1})\sigma_1
\sigma_1\Phi_n(\bar{z}^{-1})\sigma_1=\Psi_n^{-1}(z)\Phi_n(z)=S(z)$.
\\\indent   Since $\sigma_1\bar{\Lambda}(\bar{\gamma}^{-1})\sigma_1 = \gamma^{-1}\Lambda(\gamma)$ and
$\sigma_1\bar{\Lambda}^{-1}(\bar{\gamma}^{-1})\sigma_1 = \gamma\Lambda^{-1}(\gamma)$, according to Eq.~(\ref{3.1.38}) one has
$\sigma_1\bar{T}(\bar{\lambda}^{-1})\sigma_1=\sigma_1\bar{\Lambda}(\bar{\gamma}^{-1})\sigma_1\sigma_1 \bar{S}(\bar{z}^{-1})\sigma_1 \sigma_1
\Lambda^{-1}(\bar{\gamma}^{-1})\sigma_1=\Lambda(\gamma) S(z) \Lambda^{-1}(\gamma)=T(\lambda)$.
\end{proof}

It follows that $s_2(z)=\bar{s}_3(\bar{z}^{-1})$, $\tilde{s}_2(\lambda)=\bar{\tilde{s}}_3(\bar{\lambda}^{-1})  $ and
$s_1(\lambda)=\bar{s}_4(\bar{\lambda}^{-1})$ by the Proposition 3.4, and further $s_4'(\lambda)=-\lambda^{-2}\bar{s}_1'(\bar{\lambda}^{-1})$
can be derived.

Since $|\lambda_l|>1$ $\Leftrightarrow$ $|\tilde{\lambda}_l|<1$, where $ \tilde{\lambda}_l=\bar{\lambda}_l^{-1}$, then $\lambda_l$ is a zero of $s_1(\lambda)$ outside $|\lambda|=1$ if and
only if $\tilde{\lambda}_l$ is a zero of $s_4(\lambda)$ inside $|\lambda|=1$. Moreover, it can be checked that
$|\lambda_l-1/\tau|=Q/\tau$ $\Leftrightarrow$ $|\tilde{\lambda}_l-1/\tau|=Q/\tau$.
Then since the discrete  eigenvalues are the zeros of the scattering coefficients $s_1(\lambda)$ and $s_4(\lambda)$, according to Eq.~(\ref{3.1.19}), we obtain that there are $2N$ discrete eigenvalues, the set of which can be written as $\Omega=\{\lambda_l,\
\tilde{\lambda}_l,\  l=1, 2, \cdots,  N :|\lambda_l-1/\tau|=Q/\tau,\ |\lambda_l|>1,\ \tilde{\lambda}_l=\bar{\lambda}^{-1}_l , \ \text{and} \
\lambda_l \neq \lambda_0, \ \bar{\lambda}_0\}$, where $2N=J$.

\subsection{  Inverse scattering problem}
\renewcommand{\theequation}{\arabic{section}.\arabic{subsection}.\arabic{equation}}\setcounter{equation}{0}

\subsubsection{RH problem and reconstruction formula }

\hspace{1.5em}It follows from Eq.~(\ref{3.1.37}) that
\begin{align*}
\frac{U_n(\lambda)}{s_1(\lambda)}&=\tilde{W}_n(\lambda)+\gamma^{-2n}(\lambda)W_n(\lambda)\kappa(\lambda),\\
\frac{\tilde{U}_n(\lambda)}{s_4(\lambda)}&=\gamma^{2n}(\lambda)\tilde{W}_n(\lambda)\tilde{\kappa}(\lambda)+W_n(\lambda),
\end{align*}
where the reflection coefficients are introduced as
\begin{equation}\label{3.2.1}
\kappa(\lambda)=\frac{\tilde{s}_3(\lambda)}{s_1(\lambda)},\quad\quad\quad\tilde{\kappa}(\lambda)=\frac{\tilde{s}_2(\lambda)}{s_4(\lambda)}.
\end{equation}
That is,
\begin{align}\label{3.2.2}
\Big(W_n(\lambda),\ \frac{U_n(\lambda)}{s_1(\lambda)}\Big)=\Big(\frac{\tilde{U}_n(\lambda)}{s_4(\lambda)},\
\tilde{W}_n(\lambda)\Big)H_n(\lambda),
\end{align}
where
\begin{align*}
H_n(\lambda)=\left(\begin{matrix}
                                                       1 & \gamma^{-2n}\kappa(\lambda)\\-\gamma^{2n}\tilde{\kappa}(\lambda)&
                                                       1-\kappa(\lambda)\tilde{\kappa}(\lambda)
  \end{matrix}\right).
\end{align*}

Besides, from Proposition 3.2 and the asymptotic behavior~(\ref{3.1.76}), and (\ref{3.1.81}), we get that $W_{n}(\lambda)$ and $\tilde{W}_{n}(\lambda)$ are
analytic in $|z|>1$ and $|\lambda|<1$, respectively. Meanwhile, combining with~(\ref{3.1.85}) yields that$\dfrac{U_n(\lambda)}{s_1(\lambda)}$ are analytic in
$|\lambda|>1$ except not at $\lambda=\infty$ and the zeros of $s_1(\lambda)$, and that $\dfrac{\tilde{U}_n(\lambda)}{s_4(\lambda)}$ are analytic in
$|\lambda|>1$ except not at $\lambda=0$ and the zeros of $s_4(\lambda)$.

Define a counterclockwise directed curve $\Sigma=\{ |\lambda|=1\}  $ and the piecewise function
\begin{gather}\label{3.2.3}
R_n\left(\lambda\right)= \begin{cases} \, R^{+}_n\left(\lambda\right)=
	A_n\Big(\dfrac{\tilde{U}_n(\lambda)}{s_4(\lambda)},\ \tilde{W}_n(\lambda)\Big) ,\quad \quad \quad \quad\
\quad\ \quad\, \ |\lambda|<1,\\
\, R^{-}_n\left(\lambda\right)=A_n\Big(W_n(\lambda),\ \dfrac{U_n(\lambda)}{s_1(\lambda)(\lambda-\tau)}\Big)
,\quad\ \quad \quad \quad   |\lambda|>1,\\
\end{cases}
\end{gather}
where $ A_n=\tau^{-1}\left( \begin{matrix}
                              \chi_n & 0 \\
                             \bar{ q}_n & \tau
                            \end{matrix}\right)$.

Therefore, by Eq.~(\ref{3.2.2}), one has
\begin{align}\label{3.2.4}
R^{-}_n\left(\lambda\right)=R^{+}_n\left(\lambda\right)H_n(\lambda)\tilde{A}_n(\lambda),
\end{align}
where $ \tilde{A}_n(\lambda)=\text{diag}\big(1,\, (\lambda-\tau)^{-1}\big)$.

Then we obtain the following generalized RH problem:

\vspace{5mm}{\bf RHP 3.1}\quad Find a $2\times2$ matrix $R_n\left(\lambda\right)$ such that\\
(1)\ $R_n\left(\lambda\right)$ is meromorphic on $\mathbb{C}\setminus\Sigma$,\\
(2)\ $R^{-}_n\left(\lambda\right)=R^{+}_n\left(\lambda\right)H_n(\lambda)\tilde{A}_n(\lambda)$ on $\Sigma$,\\
(3)\ $R_n\left(\lambda\right)\rightarrow I$ as $\left|\lambda \right|\rightarrow \infty$.

Subtracting the behavior of $\dfrac{U_n(\lambda)}{s_1(\lambda)(\lambda-\tau)}$ at $\infty$, the jump condition Eq.~(\ref{3.2.4}) can be
rewritten as
\begin{align}
&\dfrac{\tilde{U}_n(\lambda)}{s_4(\lambda)(\lambda-\tau^{-1})}-\dfrac{W_n(\lambda)}{\lambda-\tau^{-1}}=\gamma^{2n}\tilde{\kappa}(\lambda)
\frac{\tilde{W}_n(\lambda)}{\lambda-\tau^{-1}}.\label{3.2.5}\\
&\dfrac{U_n(\lambda)}{s_1(\lambda)(\lambda-\tau)}-\left(\begin{array}{c}
                                                    0 \\
                                                    1
                                                  \end{array}\right)-\frac{\tilde{W}_n(\lambda)}{\lambda-\tau}=\gamma^{-2n}\kappa(\lambda)
                                                  {W}_n(\lambda)-\left(\begin{array}{c}
                                                    0 \\
                                                    1
                                                  \end{array}\right).\label{3.2.6}
\end{align}

According to Ref.~\cite{ck29}, the scattering coefficients $s_1(\lambda)$ and $s_4(\lambda)$ has only simple zeros, then  removing the singularity of
$\dfrac{U_n(\lambda)}{s_1(\lambda)(\lambda-\tau)}$ at the discrete eigenvalues $\lambda_l $, $l=1,2, \cdots N$,  the singularity of
$\dfrac{\tilde{U}_n(\lambda)}{s_4(\lambda)(\lambda-\tau^{-1})}$ at 0, $\tilde{\lambda}_l $, and the singularity of
$\dfrac{\tilde{W}_n(\lambda)}{\lambda-\tau} $, $\dfrac{W_n(\lambda)}{\lambda-\tau^{-1}}$ at $\tau $, $\tau^{-1}$ respectively yields:
\begin{align}
&\Bigg(\dfrac{\tilde{U}_n\left( \lambda \right)}{s_4\left( \lambda \right) ( \lambda -\tau ^{-1} )}-\sum_{l=1}^N{\frac{\text{Res}\Big(
\frac{\tilde{U}_n\left( \lambda \right)}{s_4\left( \lambda \right)};\tilde{\lambda}_l \Big)}{( \tilde{\lambda}_l-\tau ^{-1} ) ( \lambda
-\tilde{\lambda}_l )}}-\frac{\tau}{\lambda}\left(\begin{array}{c}
	1\\
	0\\
\end{array} \right)\Bigg)-\left( \dfrac{W_n\left( \lambda \right)}{\lambda -\tau ^{-1}}+\frac{\bar{q}_r}{\chi _n(\lambda -\tau ^{-1})}\left(
\begin{array}{c}
	0\\
	1\\
\end{array} \right)\right)\notag \\
=&\tilde{\kappa}\left( \lambda \right) \frac{\gamma ^{2n}\tilde{W}_n\left( \lambda \right)}{\lambda -\tau
^{-1}}-\sum_{l=1}^N{\frac{\text{Res}\left( \frac{\tilde{U}_n\left( \lambda \right)}{s_4\left( \lambda \right)};\tilde{\lambda}_l \right)}{(
\tilde{\lambda}_l-\tau ^{-1} ) ( \lambda -\tilde{\lambda}_l )}}-\frac{\tau}{\lambda}\left(\begin{array}{c}
	1\\
	0\\
\end{array} \right)-\frac{\bar{q}_r}{\chi _n(\lambda -\tau ^{-1})}\left( \begin{array}{c}
	0\\
	1\\
\end{array} \right) ,\label{3.2.7}\\
&\Bigg(\dfrac{U_n(\lambda)}{s_1(\lambda)(\lambda-\tau)}-\sum_{l=1}^N{\frac{\text{Res}\left( \frac{U_n\left( \lambda \right)}{s_1\left( \lambda
\right)};\lambda _l \right)}{\left( \lambda _l-\tau \right) \left( \lambda -\lambda _l \right)}}
-\left(\begin{array}{c}
                                                    0 \\
                                                    1
                                                  \end{array}\right)\Bigg)-\Bigg(\frac{\tilde{W}_n(\lambda)}{\lambda-\tau}-\frac{q_r}{\chi
                                                  _n(\lambda -\tau )}\left( \begin{array}{c}
	1\\
	0\\
\end{array} \right)\Bigg)\notag\\
=&\kappa(\lambda) \dfrac{\gamma^{-2n}W_n(\lambda)}{\lambda-\tau}-\sum_{l=1}^N{\frac{\text{Res}\left( \frac{U_n\left( \lambda
\right)}{s_1\left( \lambda \right)};\lambda _l \right)}{\left( \lambda _l-\tau \right) \left( \lambda -\lambda _l \right)}}
-\left(\begin{array}{c}
                                                    0 \\
                                                    1
                                                  \end{array}\right)+\frac{q_r}{\chi _n(\lambda -\tau )}\left( \begin{array}{c}
	1\\
	0\\
\end{array} \right),\label{3.2.8}
\end{align}
in which Eqs.~(\ref{3.1.76}) and~(\ref{3.1.84}) are used.

Define integral operators $P$ and $\tilde{P}$ as follows:
\begin{gather*}
P[f](\lambda)=\frac{1}{2\pi i}\lim_{\substack{\lambda'\to \lambda\\ |\lambda'|>1}}\oint_{|\rho |=1}{\frac{f\left( \rho \right)}{\rho
-\lambda'}}\text{d}\rho,  \quad\quad |\lambda|\geq1,\\
\tilde{P}[f](\lambda)=\frac{1}{2\pi i}\lim_{\substack{\lambda'\to \lambda\\ |\lambda'|<1}}\oint_{|\rho |=1}{\frac{f\left( \rho \right)}{\rho
-\lambda'}}\text{d}\rho,\quad\quad |\lambda|\leq1.
\end{gather*}
Then by applying the integral operators $P$ and $\tilde{P}$ to both sides of Eqs.~(\ref{3.2.7}) and (\ref{3.2.8}), respectively, and using the
Cauchy’s integral formula and the Cauchy’s theorem, one has
\begin{align}
 \dfrac{W_n\left( \lambda \right)}{\lambda -\tau ^{-1}}+\frac{\bar{q}_r}{\chi _n(\lambda -\tau ^{-1})}\left( \begin{array}{c}
	0\\
	1\\
\end{array} \right)&=\sum_{l=1}^N{\frac{\text{Res}\Big( \frac{\tilde{U}_n\left( \lambda \right)}{s_4\left( \lambda \right)};\tilde{\lambda}_l
\Big)}{( \tilde{\lambda}_l-\tau ^{-1} ) ( \lambda -\tilde{\lambda}_l )}}+\left(\begin{array}{c}
\dfrac{\tau}{\lambda}\\
	0\\
\end{array} \right)+\frac{1}{2\pi i}\lim_{\substack{\lambda'\to \lambda\\ |\lambda'|>1}}\oint_{|\rho |=1}{\frac{\gamma
^{2n}(\rho)\tilde{\kappa}\left(\rho \right) \tilde{W}_n\left( \rho \right)}{(\rho -\tau ^{-1})(\rho -\lambda')}}\text{d}\rho,\label{3.2.9}\\
\dfrac{\tilde{W}_n\left( \lambda \right)}{\lambda -\tau }-\frac{q_r}{\chi _n(\lambda -\tau )}\left( \begin{array}{c}
	1\\
	0\\
\end{array} \right)&=\sum_{l=1}^N{\frac{\text{Res}\Big( \frac{U_n\left( \lambda \right)}{s_1\left( \lambda \right)};\lambda_l \Big)}{(
\lambda_l-\tau ) ( \lambda -\lambda_l )}}+\left( \begin{array}{c}
	0\\
	1\\
\end{array} \right)-\frac{1}{2\pi i}\lim_{\substack{\lambda'\to \lambda\\ |\lambda'|>1}}\oint_{|\rho |=1}{\frac{\gamma
^{-2n}(\rho)\tilde{\kappa}\left(\rho \right)W_n\left( \rho \right)}{(\rho -\tau)(\rho -\lambda')}}\text{d}\rho.\label{3.2.10}
\end{align}

\paragraph{Proposition 3.5} The residues of $\dfrac{\tilde{U}_n\left( \lambda \right)}{s_4\left( \lambda \right)}$ and $ \dfrac{U_n\left(
\lambda \right)}{s_1\left( \lambda \right)}$ at discrete eigenvalues can be expressed as
\begin{align}
\text{Res}\Big( \frac{\tilde{U}_n\left( \lambda \right)}{s_4\left( \lambda \right)};\tilde{\lambda}_l \Big)&= \gamma^{2n}(\tilde{\lambda}_l
)\, \tilde{C}_l\, \tilde{W}_n( \tilde{\lambda}_l )  ,\label{3.2.11}\\
\text{Res}\Big( \frac{U_n\left( \lambda \right)}{s_1\left( \lambda \right)};\lambda_l \Big)&= \gamma^{-2n}(\lambda_l )
\, C_l\, W_n( \lambda_l )     ,\label{3.2.12}
\end{align}
where the norming constants
\begin{equation}\label{3.2.13}
\ \ C_l=\frac{\tilde{s}_3(\lambda_l)}{s_1'(\lambda_l)},\quad \quad \quad
\tilde{C}_l=\frac{\tilde{s}_2(\tilde{\lambda}_l)}{s_4'(\tilde{\lambda}_l)}=-\tilde{\lambda}^2_l\bar{C}_l.
\end{equation}
\begin{proof}
According to Eq.~(\ref{3.1.37}), one has
\begin{align*}
\text{Res}\Big( \frac{\tilde{U}_n( \lambda )}{s_4\left( \lambda \right)};\tilde{\lambda}_l \Big)&=\frac{\tilde{U}_n( \tilde{\lambda}_l)}{s'_4(
\tilde{\lambda}_l )}=\gamma^{2n}(\tilde{\lambda}_l)\frac{\tilde{s}_2(\tilde{\lambda}_l)}{s'_4( \tilde{\lambda}_l )}\tilde{W}_n(
\tilde{\lambda}_l)=\gamma^{2n}(\tilde{\lambda}_l )\, \tilde{C}_l\, \tilde{W}_n( \tilde{\lambda}_l ),\\
\text{Res}\Big( \frac{U_n( \lambda )}{s_1\left( \lambda \right)};\lambda_l \Big)&=\frac{U_n( \lambda_l)}{s'_1\left( \lambda_l
\right)}=\gamma^{-2n}(\lambda_l)\frac{\tilde{s}_3(\lambda_l)}{s'_1\left( \lambda_l  \right)}W_n( \lambda_l)=\gamma^{-2n}(\lambda_l )
\, C_l\, W_n( \lambda_l ) .\end{align*}
Meanwhile, using the Proposition 3.4 can yield
\begin{align*}
\tilde{C}_l=\frac{\tilde{s}_2(\tilde{\lambda}_l)}{s_4'(\tilde{\lambda}_l)}=\frac{\bar{\tilde{s}}_3(\lambda_l)}{-\tilde{\lambda}_l^{-2}\bar{s}_1'
(\lambda_l)}=-\tilde{\lambda}_l^{2}\frac{\bar{\tilde{s}}_3(\lambda_l)}{\bar{s}_1' (\lambda_l)}=-\tilde{\lambda}_l^{2}\bar{C}_l.
\end{align*}
\end{proof}

Combining Eqs.~(\ref{3.2.9})-(\ref{3.2.10}) with the Proposition 3.5, one has
\begin{align}
 W_n\left( \lambda \right)&=\left( \begin{array}{c}
	\tau-\frac{1}{\lambda}\\
	-\frac{\bar{q}_r}{\chi _n}\\
\end{array} \right)+\sum_{l=1}^N{\frac{(\lambda -\tau ^{-1})\gamma^{2n}(\tilde{\lambda}_l )\, \tilde{C}_l\, \tilde{W}_n( \tilde{\lambda}_l
)}{( \tilde{\lambda}_l-\tau ^{-1} ) ( \lambda -\tilde{\lambda}_l )}}
+\frac{1}{2\pi i}\lim_{\substack{\lambda'\to \lambda\\ |\lambda'|>1}}\oint_{|\rho |=1}{\frac{(\lambda -\tau^{-1} )\gamma
^{2n}(\rho)\tilde{\kappa}\left(\rho \right) \tilde{W}_n\left( \rho \right)}{(\rho -\tau ^{-1})(\rho -\lambda')}}\text{d}\rho,\label{3.2.14}\\
\tilde{W}_n\left( \lambda \right)&=\left( \begin{array}{c}
	\frac{q_r}{\chi _n}\\
	\lambda-\tau\\
\end{array} \right)+\sum_{l=1}^N{\frac{(\lambda -\tau )\gamma^{-2n}(\lambda_l )
\, C_l\, W_n( \lambda_l )}{( \lambda_l-\tau ) ( \lambda -\lambda_l )}}-\frac{1}{2\pi i}\lim_{\substack{\lambda'\to \lambda\\
|\lambda'|>1}}\oint_{|\rho |=1}{\frac{(\lambda -\tau )\gamma ^{-2n}(\rho)\tilde{\kappa}\left(\rho \right) W_n\left( \rho \right)}{(\rho
-\tau)(\rho -\lambda')}}\text{d}\rho.\label{3.2.15}
\end{align}

According to Eq.~(\ref{3.1.76}), if we take the limit that $\lambda \to \infty$ on both sides of Eq.~(\ref{3.2.14}), then
\begin{align}
\frac{1}{\chi_n}&=1+\sum_{l=1}^N{\frac{\gamma^{2n}(\tilde{\lambda}_l )\, \tilde{C}_l\, \tilde{W}^{(1)}_n( \tilde{\lambda}_l )}{
\tau\tilde{\lambda}_l-1 }}-\frac{1}{2\pi i}\lim_{\substack{\lambda'\to \lambda\\ |\lambda'|>1}}\oint_{|\rho |=1}{\frac{ \gamma
^{2n}(\rho)\tilde{\kappa}\left(\rho \right) \tilde{W}^{(1)}_n\left( \rho \right)}{\tau\rho -1}}\text{d}\rho,\label{3.2.16}\\
\bar{q}_n&=\bar{q}_r-\chi_n\sum_{l=1}^N{\frac{\gamma^{2n}(\tilde{\lambda}_l )\, \tilde{C}_l\, \tilde{W}^{(2)}_n( \tilde{\lambda}_l )}{
\tilde{\lambda}_l-\tau^{-1} }}+\frac{\chi_n}{2\pi i}\lim_{\substack{\lambda'\to \lambda\\ |\lambda'|>1}}\oint_{|\rho |=1}{\frac{ \gamma
^{2n}(\rho)\tilde{\kappa}\left(\rho \right) \tilde{W}^{(2)}_n\left( \rho \right)}{\rho -\tau ^{-1}}}\text{d}\rho.\label{3.2.17}
\end{align}

Therefore, Eq.~(\ref{3.2.17}) can be used to reconstruct the potential $\bar{q}_n$ combining with  Eqs.~(\ref{3.2.14})-(\ref{3.2.16}).

\subsubsection{ Trace formula }

\hspace{1.5em}Considering the Proposition 3.2 and the  asymptotic behavior~(\ref{3.1.85}), one can deduce that the scattering coefficient $s_1 ( \lambda )$ can be analytic for $|\lambda|>1$ , and $s_4 ( \lambda )$ for $|\lambda|<1$ . Since that their zeros are $\{ \lambda_l, | \ l=1, 2, \cdots,  N\}$ and $\{
\tilde{\lambda}_l |\ l=1, 2, \cdots,  N\}$, respectively,
 we define
\begin{gather}\label{3.2.18}
\hat{s}_1(\lambda)=\prod_{l=1}^N{\frac{\lambda-\tilde{\lambda}_{l}}{\lambda-\lambda_{l}}}s_1(\lambda),\quad\quad
\hat{s}_4(\lambda)=\prod_{l=1}^N{\frac{\lambda-\lambda_{l}}{\lambda-\tilde{\lambda}_{l}}}s_4(\lambda),
\end{gather}
to remove the zeros.

According to Eqs.~(\ref{3.1.35}),~(\ref{3.1.39}),~(\ref{3.2.1}) and the Proposition 3.4, one has
\begin{gather}
\hat{s}_1(\lambda)\hat{s}_4(\lambda)=\chi_{-\infty}(1-|\kappa(\lambda)|^2)^{-1},\quad\quad\quad|\lambda|=1.
\end{gather}

Define
\begin{gather*}
\varsigma(\lambda)=\left\{ \begin{array}{l}
	\varsigma^+(\lambda)=\hat{s}_4(\lambda), \quad\quad|\lambda|<1,\\
	\varsigma^-(\lambda)=\hat{s}_1(\lambda), \quad\quad|\lambda|>1.\\
\end{array} \right.
\end{gather*}
From Eqs.~(\ref{3.1.85}), $\varsigma(\lambda) \to 1$ as $\left|z  \right|\rightarrow \infty$ can be derived. Then a RH problem about $\varsigma(\lambda)$ can be given as below:

\vspace{5mm}{\bf RHP 3.2}\quad Find a scalar function $\varsigma(\lambda)$ such that\\
(1)\ $\varsigma(\lambda)$ is analytic on $\mathbb{C}\setminus\Sigma$,
\begin{flalign*}
&\text{(2)}\ \log\big(\varsigma^+(z)\big)+\log\big(\varsigma^-(z)\big)=\log(\chi_{-\infty})-\log\big(1-|\kappa(\lambda)|^2\big)\ \text{on}\ \Sigma,&
\end{flalign*}
(3)\ $\varsigma(\lambda)\rightarrow 1$ as $\left|z  \right|\rightarrow \infty$.

 Applying the integral operators $P$ and $\tilde{P}$ to the jump condition respectively yields
 \begin{align*}
\log\big(\hat{s}_1(\lambda)\big)&= \log\big(\varsigma^-(\lambda)\big)=\lim_{\substack{\lambda'\to \lambda\\ |\lambda'|>1}}\frac{1}{2\pi i}\oint_{|\rho
|=1}{\frac{\log\big(1-|\kappa(\rho)|^2\big)}{\rho-\lambda}}\text{d}\rho,\\
\log\big(\hat{s}_4(\lambda)\big)&= \log\big(\varsigma^+(\lambda)\big)=-\lim_{\substack{\lambda'\to \lambda\\ |\lambda'|>1}}\frac{1}{2\pi i}\oint_{|\rho
|=1}{\frac{\log\big(1-|\kappa(\rho)|^2\big)}{\rho-\lambda}}\text{d}\rho.
\end{align*}

Therefore, using the relation~(\ref{3.2.18}) can yield the trace formulas:
 \begin{alignat*}{2}
s_1(\lambda)&= \prod_{l=1}^N\frac{\lambda-\lambda_{l}}{\lambda-\tilde{\lambda}_{l}}\exp\Big[\frac{1}{2\pi i}\oint_{|\rho
|=1}{\frac{\log\big(1-|\kappa(\rho)|^2\big)}{\rho-\lambda}}\text{d}\rho\Big],&\quad \quad\ &|\lambda|>1.\\
s_4(\lambda)&= \prod_{l=1}^N\frac{\lambda-\tilde{\lambda}_{l}}{\lambda-\lambda_{l}}\exp\Big[\frac{-1}{2\pi i}\oint_{|\rho
|=1}{\frac{\log\big(1-|\kappa(\rho)|^2\big)}{\rho-\lambda}}\text{d}\rho\Big],&&|\lambda|<1.
\end{alignat*}

Moreover, the $\theta$-condition can be derived according to the trace formulas and the asymptotic behavior~(\ref{3.1.85}), which can be expressed as
 \begin{gather*}
 e^{i(\theta_l-\theta_r)}=\prod_{l=1}^N\frac{1-\tau\lambda_{l}}{1-\tau\tilde{\lambda}_{l}}\exp\Big[\frac{1}{2\pi i}\oint_{|\rho
|=1}{\frac{\tau\log\big(1-|\kappa(\rho)|^2\big)}{\tau\rho-1}}\text{d}\rho\Big].
 \end{gather*}

\subsection{ Time evolution}
\renewcommand{\theequation}{\arabic{section}.\arabic{subsection}.\arabic{equation}}\setcounter{equation}{0}

\hspace{1.5em}In this section, we will consider the time-dependence of the eigenfunctions, scattering coefficients and  norming constants, which can be
determined by the time-dependence equation~(\ref{1.3}) whose asymptotic form at large $n$ can be expressed as
 \begin{gather}\label{3.3.1}
\frac{\text{d}v_n}{\text{d}t}\sim \left( \begin{matrix}
	\xi_1&		\xi_2\\
	\xi_3&		\xi_4\\
\end{matrix} \right) v_n,\quad \quad\ n\rightarrow\pm\infty,
\end{gather}
where
\begin{align*}
\xi_1&=-\frac{3}{4}z^{-4}+(2+Q^2(t))z^{-2}+3Q^4-\frac{3}{2}-Q^2(t)z^2+\frac{1}{4}z^4,\\
\xi_2&=q_{r/l}(z^{-3}+z^3)-(1+2Q^2(t))q_{r/l}(z^{-1}+z),\\
\xi_3&=q_{r/l}(z^{-3}+z^3)-(1+2Q^2(t))q_{r/l}(z^{-1}+z),\\
\xi_4&=-\frac{3}{4}z^{4}+(2+Q^2(t))z^{2}+3Q^4(t)-\frac{3}{2}-Q^2(t)z^{-2}+\frac{1}{4}z^{-4}.
\end{align*}
That is,
\begin{align}
\frac{\text{d}v^{(1)}_n}{\text{d}t}&\sim \xi_1 v^{(1)}_n +\xi_2v^{(2)}_n,\quad \quad\ n\rightarrow\pm\infty,\label{3.3.2}\\
\frac{\text{d}v^{(2)}_n}{\text{d}t}&\sim \xi_3 v^{(1)}_n +\xi_4v^{(2)}_n, \quad \quad\ n\rightarrow\pm\infty.\label{3.3.3}
\end{align}

We define the  simultaneous solutions of the Lax pair as
\begin{align}\label{3.3.4}
\tilde{\Phi}_{n}(z,t)=\Phi_{n}(z,t)\Theta(z,t),\quad\quad\quad \tilde{\Psi}_{n}(z,t)=\Psi_{n}(z,t)\Theta(z,t),
\end{align}
where $\Theta(z,t)=\text{diag}( e^{\omega_1(z) t},\, e^{\omega_2(z) t})$ . Meanwhile, $\Phi_{n}(z,t)$ and $\Psi_{n}(z,t)$ are eigenfunctions
of the scattering problem~(\ref{1.2}) which satisfy the BCs~(\ref{3.1.10})-(\ref{3.1.11}).
Therefore,
\begin{align}
\dfrac{\text{d}\tilde{\Phi}_{n}(z,t)}{\text{d}t}
=\Phi_{n}(z,t)\dfrac{\text{d}\Theta(z,t)}{\text{d}t}+\dfrac{\text{d}\Phi_{n}(z,t)}{\text{d}t}\Theta(z,t)=\tilde{\Phi}_{n}(z,t)\tilde{\Theta}(z)+\dfrac{\text{d}\Phi_{n}(z,t)}{\text{d}t}\Theta(z,t),\label{3.3.5}\\
\dfrac{\text{d}\tilde{\Psi}_{n}(z,t)}{\text{d}t}=\Psi_{n}(z,t)\dfrac{\text{d}\Theta(z,t)}{\text{d}t}+\dfrac{\text{d}\Phi_{n}(z,t)}{\text{d}t}\Theta(z,t)
=\tilde{\Psi}_{n}(z,t)\tilde{\Theta}(z)+\dfrac{\text{d}\Psi_{n}(z,t)}{\text{d}t}\Theta(z,t),\label{3.3.6}
\end{align}
where $\tilde{\Theta}(z)=\text{diag}( \omega_1(z),\, \omega_2(z) )$.

Taking the limits that $n \to \pm \infty$ on  both sides of Eq.~(\ref{1.1}),
one can obtain that $\dfrac{\text{d}q_{r/l}}{\text{d}t}=0$, i.e. $q_{r/l}$ are $t$-independent, and then $Q(t)=Q(0)$, $\theta_{r/l}(t)=\theta_{r/l}(0)$. Therefore it follows from Eqs.~(\ref{3.1.10})-(\ref{3.1.11}) that $\dfrac{\text{d}\Phi_{n}
(z,t)}{\text{d}t} \to 0 $ as $n \to -\infty$ and $\dfrac{\text{d}\Psi_{n} (z,t)}{\text{d}t} \to 0 $ as $n \to +\infty$. As a result, from
Eqs.~(\ref{3.3.5})-(\ref{3.3.6}), one has
\begin{align}\label{3.3.7}
\dfrac{\text{d}\tilde{\Phi}_{n}(z,t)}{\text{d}t} =\tilde{\Phi}_{n}(z,t)\tilde{\Theta}(z),\quad\quad n \to -\infty;\quad\quad\quad
\dfrac{\text{d}\tilde{\Psi}_{n}(z,t)}{\text{d}t}=\tilde{\Psi}_{n}(z,t)\tilde{\Theta}(z),\quad\quad n \to +\infty.
\end{align}

In addition, from Eqs.~(\ref{1.2}),~(\ref{3.1.10})-(\ref{3.1.11}), and (\ref{3.3.4}), we can obtain that
\begin{equation}\label{3.3.8}
\begin{alignedat}{3}
q_{r/l}v^{(2)}_n&\sim v^{(1)}_{n+1}-zv^{(1)}_n,&\quad\quad& q_{r/l}v^{(1)}_n\sim v^{(2)}_{n+1}-z^{-1}v^{(2)}_n,&\quad\quad \ &
n\rightarrow\mp\infty,\\
\tilde{\Phi}_{n+1,1}&\sim \tau\gamma \tilde{\Phi}_{n,1},&& \tilde{\Phi}_{n+1,2}\sim \tau\gamma^{-1} \tilde{\Phi}_{n,1},&&
n\rightarrow-\infty,\\
\tilde{\Psi}_{n+1,1}&\sim \tau\gamma \tilde{\Psi}_{n,1},&& \tilde{\Psi}_{n+1,2}\sim \tau\gamma^{-1} \tilde{\Phi}_{n,1},&&n\rightarrow+\infty.
\end{alignedat}
\end{equation}
Consequently, combining Eqs.~(\ref{3.3.2})-(\ref{3.3.3}), (\ref{3.3.7}) with (\ref{3.3.8}) to derive that
\begin{align}
\omega_1(z)& =\tau\gamma
\big((z^{-3}+z^3)-(1+2Q^2)(z^{-1}+z)\big)-\big(\frac{3}{4}(z^{-4}+z^{4})-(1+Q^2)(z^{-2}+z^{2})-3Q^4+\frac{1}{2}\big),\label{3.3.9}\\
\omega_2(z) &=\frac{\tau}{\gamma}
\big((z^{-3}+z^3)-(1+2Q^2)(z^{-1}+z)\big)-\big(\frac{3}{4}(z^{-4}+z^{4})-(1+Q^2)(z^{-2}+z^{2})-3Q^4+\frac{1}{2}\big).\label{3.3.10}
\end{align}

According to Eqs.~(\ref{3.3.4}) and (\ref{3.1.29})-(\ref{3.1.30}),  we can derive that
\begin{align}
\det\tilde{\Phi}_n(t)&=-e^{(\omega_1(z)+\omega_2(z)) t}\tau^{2n}\big((\gamma \tau-z)^2+Q^2\big)\frac{\chi_{-\infty}}{\chi_{n}}\neq0 ,
\quad\quad z\neq \pm z_0,\ \pm\bar{z}_0,\label{3.3.11}\\
\det\tilde{\Psi}_n(t)&=-e^{(\omega_1(z)+\omega_2(z)) t}\tau^{2n}\big((\gamma \tau-z)^2+Q^2\big)\frac{1}{\chi_{n}}\neq0,\quad\quad\quad z\neq
\pm z_0,\ \pm\bar{z}_0 .\label{3.3.12}
\end{align}
Consequently,  there is the scattering matrix $\tilde{S}(z)$ that is independent of $n$ and $t$ and satisfies
\begin{gather}\label{3.3.13}
\tilde{\Phi}_n(z,t)=\tilde{\Psi}_n(z,t)\tilde{S}(z) ,\quad\quad\  z\neq \pm z_0,\, \pm\bar{z}_0.
\end{gather}

Combining Eqs.~(\ref{3.1.32}) with (\ref{3.3.4}) and (\ref{3.3.13}), one has the $n$-independent scattering matrix
$S(z,t)=\Psi_{n}^{-1}(z,t)\Phi_n(z,t)=\Theta(z)\tilde{\Psi}_{n}^{-1}(z,t)\tilde{\Phi}_n(z,t)\Theta^{-1}(z)=\Theta(z)\tilde{S}(z)\Theta^{-1}(z)$.
That is,
\begin{align}
\tilde{S}(z)=\Theta^{-1}(z)S(z,t)\Theta(z)=\left(\begin{matrix}
                                                   s_1(z,t) &  s_2(z,t)e^{(\omega_2(z)-\omega_1(z)) t} \\
                                                    s_3(z,t)e^{(\omega_1(z)-\omega_2(z)) t}  &  s_4(z,t)
                                                 \end{matrix}\right).
\end{align}

Due to  $\tilde{S}(z)$ is independent of $t$, then $\dfrac{\text{d}\tilde{S}(z)}{\text{d}t}=0$, which suggests that
\begin{equation}
\begin{aligned}
 s_1(z,t)&= s_1(z,0),\quad\quad \ s_3(z,t)=s_3(z,0)e^{\tilde{\omega}(z) t},\\s_4(z,t)&= s_4(z,0),\quad\quad \
 s_2(z,t)=s_2(z,0)e^{-\tilde{\omega}(z) t},
\end{aligned}
\end{equation}
where $\tilde{\omega}(z) =\omega_2(z)-\omega_1(z)=(\frac{\tau}{\gamma} -\tau\gamma) \big((z^{-3}+z^3)-(1+2Q^2)(z^{-1}+z)\big)$.

Note that, using the relations~(\ref{3.1.13}) can yield
\begin{equation}
\tilde{\omega}(\lambda)=\frac{\tau^2 (-1 +
   2 \tau \lambda -\lambda^2) (-1 + \lambda^2) (2 (2 +
      Q^2) (\lambda^2 -  \tau \lambda (1 + \lambda^2) )+
   \tau^2 (1 +
      2 (1 + Q^2) \lambda^2 + \lambda^4)}{\lambda^2(\tau - \lambda)^2
(-1 + \tau \lambda)^2}.
\end{equation}
Then it follows from Eq.~(\ref{3.1.38}) that
\begin{equation}
\begin{aligned}
 s_1(\lambda,t)&= s_1(\lambda,0),\quad\quad \ \tilde{s}_3(\lambda,t)=\tilde{s}_3(\lambda,0)e^{\tilde{\omega}(\lambda) t},\\s_4(\lambda,t)&=
 s_4(\lambda,0),\quad\quad \ \tilde{s}_2(\lambda,t)=\tilde{s}_2(\lambda,0)e^{-\tilde{\omega}(\lambda) t}.
\end{aligned}
\end{equation}
That is, $s_1(\lambda)$ and $s_4(\lambda)$ are $t$-independent, and accordingly the eigenvalues are constant with the time evolution of
solutions.

It follows from  Eqs.~(\ref{3.2.1}) and  (\ref{3.1.13}) that the time evolution of the reflection coefficients and the norming constants are
\begin{align*}
&\kappa(\lambda,t)=\kappa(\lambda,0)e^{\tilde{\omega}(\lambda) t},\quad\
\tilde{\kappa}(\lambda,t)=\tilde{\kappa}(\lambda,0)e^{-\tilde{\omega}(\lambda) t},\\
&C_l(t)=C_l(0)e^{\tilde{\omega}(\lambda_l) t},\quad\  \ \ \,
  \tilde{C}_l(t)=\tilde{C}_l(0)e^{-\tilde{\omega}(\tilde{\lambda}_l) t}=-\tilde{\lambda}^2_l\bar{C}_l(0)e^{\bar{\tilde{\omega}}(\lambda_l )
t}.
\end{align*}

\subsection{  $N$-soliton solution}
\renewcommand{\theequation}{\arabic{section}.\arabic{subsection}.\arabic{equation}}\setcounter{equation}{0}

\hspace{1.5em} We consider the reflectionless case, i.e.,  both the reflection coefficients $\kappa(\lambda)$ and $\tilde{\kappa}(\lambda)$ vanish on $|\lambda|=1$. In this
 situation the integrals vanish in the algebraic-integral system~(\ref{3.2.14})-(\ref{3.2.16}), and then an algebraic system can be derived as
\begin{align}
W_n^{(1)}\left( \lambda_k \right)&=
	\tau-\frac{1}{ \lambda_k}+\sum_{s=1}^N{\frac{( \lambda_k -\tau ^{-1})\gamma^{2n}(\tilde{\lambda}_s )\, \tilde{C}_s(t)\, \tilde{W}^{(1)}_n( \tilde{\lambda}_s
)}{( \tilde{\lambda}_s-\tau ^{-1} ) (  \lambda_k -\tilde{\lambda}_s )}},\label{3.4.1}\\
W_n^{(2)}\left( \lambda_k \right)&=
	-\bar{q}_r-\bar{q}_r\sum_{s=1}^N{\frac{\gamma^{2n}(\tilde{\lambda}_s )\, \tilde{C}_s(t)\, \tilde{W}^{(1)}_n( \tilde{\lambda}_s )}{
\tau\tilde{\lambda}_s-1 }}
+\sum_{s=1}^N{\frac{(\lambda_k -\tau ^{-1})\gamma^{2n}(\tilde{\lambda}_s )\, \tilde{C}_s(t)\, \tilde{W}^{(2)}_n( \tilde{\lambda}_s
)}{( \tilde{\lambda}_s-\tau ^{-1} ) ( \lambda -\tilde{\lambda}_s )}},\label{3.4.2}\\
\tilde{W}^{(1)}_n( \tilde{\lambda}_k )&=
	q_r+q_r\sum_{s=1}^N{\frac{\gamma^{2n}(\tilde{\lambda}_s )\, \tilde{C}_s(t)\, \tilde{W}^{(1)}_n( \tilde{\lambda}_s )}{
\tau\tilde{\lambda}_s-1 }}
	+\sum_{s=1}^N{\frac{(\tilde{\lambda}_k -\tau )\gamma^{-2n}(\lambda_s )
\, C_s(t)\, W^{(1)}_n( \lambda_s )}{( \lambda_s-\tau ) ( \tilde{\lambda}_k -\lambda_s )}},\label{3.4.3}\\
\tilde{W}_n^{(2)}( \tilde{\lambda}_k )&=
	\tilde{\lambda}_k-\tau+\sum_{s=1}^N{\frac{(\tilde{\lambda}_k-\tau )\gamma^{-2n}(\lambda_s )
\, C_s(t)\, W_n^{(2)}( \lambda_s )}{( \lambda_s-\tau ) (\tilde{\lambda}_k -\lambda_s)}}.\label{3.4.4}
\end{align}
Accordingly, the reconstruction formula becomes
\begin{gather}
\bar{q}_n=\bar{q}_r-\frac{\sum_{s=1}^N{\dfrac{\gamma^{2n}(\tilde{\lambda}_s )\, \tilde{C}_s(t)\, \tilde{W}^{(2)}_n( \tilde{\lambda}_s )}{
\tilde{\lambda}_s-\tau^{-1} }}}{1+\sum_{s=1}^N{\dfrac{\gamma^{2n}(\tilde{\lambda}_s )\, \tilde{C}_s(t)\, \tilde{W}^{(1)}_n( \tilde{\lambda}_s
)}{ \tau\tilde{\lambda}_s-1 }}}.
\end{gather}

If we set
\begin{align*}
X&=\big(X_1,\cdots,\
X_N,\ X_{N+1}, \cdots,\
X_{2N},\ X_{2N+1}, \cdots,\
X_{3N},\ X_{3N+1}, \cdots,\
X_{4N}\big)^T\\
Y&=\big(Y_1,\cdots,\
Y_N,\ Y_{N+1}, \cdots,\
Y_{2N},\ Y_{2N+1}, \cdots,\
Y_{3N},\ Y_{3N+1}, \cdots,\
Y_{4N}\big)^T,
\end{align*}
where
\begin{alignat*}{5}
X_{j}&=W_n^{(1)}( \lambda_j ),&\quad\ &  X_{N+j}=W_n^{(2)}( \lambda_j ),&\quad\ & X_{2N+j}=\tilde{W}_n^{(1)}( \tilde{\lambda}_j ),&\quad\  & X_{3N+j}=\tilde{W}_n^{(2)}( \tilde{\lambda}_j ),&\quad\  & j=1,2,\cdots,N, \\
Y_{j}&=\tau-\lambda_j^{-1},&&Y_{N+j}=-\bar{q}_r, &&Y_{2N+j}=q_r, && Y_{3N+j}= \tilde{\lambda}_j-\tau, && j=1,2,\cdots,N,
\end{alignat*}
and set $\Upsilon_{4N\times4N}=I_{4N\times4N}+\Xi_{4N\times4N}$, in which
\begin{align*}
 \Xi_{4N\times4N}=\left( \begin{matrix}
	0&		0&		\varXi _{1}&		0\\
	0&		0&		\bar{q}_r\varXi _{2}&		\varXi _{1}\\
	\varXi _{3}&		0&		-q_r\varXi _{2}&		0\\
	0&		\varXi _{3}&		0&		0\\
\end{matrix} \right)_{4N\times4N},
\end{align*}
where
\begin{gather*}
 \varXi _{j}=\big(\varXi_{j}^{(k,s)}\big)_{N\times N},\quad\quad\quad j=1,2,3,
 \end{gather*}
 \begin{gather*}
\varXi_{1}^{(k,s)}=-\frac{( \lambda_k -\tau ^{-1})\gamma^{2n}(\tilde{\lambda}_s )\, \tilde{C}_s(t)\, }{( \tilde{\lambda}_s-\tau ^{-1} ) (  \lambda_k -\tilde{\lambda}_s )},\quad\quad \ \varXi_{2}^{(k,s)}=\frac{\gamma^{2n}(\tilde{\lambda}_s )\, \tilde{C}_s(t)}{\tau\tilde{\lambda}_s-1 },\quad\quad \ \varXi_{3}^{(k,s)}=-\frac{(\tilde{\lambda}_k-\tau )\gamma^{-2n}(\lambda_s )
\, C_s(t)}{( \lambda_s-\tau ) (\tilde{\lambda}_k -\lambda_s )},
\end{gather*}
then there is
$\Upsilon X=Y$  according to Eqs.~(\ref{3.4.1})-(\ref{3.4.4}).

 By using the Cramer's rule, we consequently derive the $N$-soliton solution:
\begin{gather}\label{3.4.6}
\bar{q}_n=\bar{q}_r+\frac{\text{det}\hat{\Upsilon}_1}{\text{det}\Upsilon- \text{det}\hat{\Upsilon}_2 },
\end{gather}
 in which
 \begin{equation*}
 \hat{\Upsilon}_1=\left( \begin{matrix}
	0&	F	\\
	Y&	\Upsilon	\\
\end{matrix} \right),\quad\quad\quad\hat{\Upsilon}_2=\left( \begin{matrix}
	0&	\tilde{F}	\\
	Y&	\Upsilon	\\
\end{matrix} \right),
  \end{equation*}
and
\begin{align*}
F&=\big(F_1,\cdots,\
F_N,\ F_{N+1}, \cdots,\
F_{2N},\ F_{2N+1}, \cdots,\
F_{3N},\ F_{3N+1}, \cdots,\
F_{4N}\big)^T\\
\tilde{F}&=\big(\tilde{F}_1,\cdots,\
\tilde{F}_N,\ \tilde{F}_{N+1}, \cdots,\
\tilde{F}_{2N},\ \tilde{F}_{2N+1}, \cdots,\
\tilde{F}_{3N},\ \tilde{F}_{3N+1}, \cdots,\
\tilde{F}_{4N}\big)^T,
\end{align*}
where
\begin{gather*}
F_{j}=F_{N+j}=F_{2N+j}=\tilde{F}_{j}=\tilde{F}_{N+j}=\tilde{F}_{3N+j}=0,\\ F_{3N+j}=\frac{\gamma^{2n}(\tilde{\lambda}_j )\, \tilde{C}_j(t)}{\tilde{\lambda}_j-\tau^{-1} },\quad \tilde{F}_{2N+j}=\dfrac{\gamma^{2n}(\tilde{\lambda}_j )\, \tilde{C}_j(t)}{ \tau\tilde{\lambda}_j-1 }=\frac{F_{3N+j}}{\tau}, \quad j=1,2,\cdots,N.
\end{gather*}

\subsubsection{ 1-soliton }
\hspace{1.5em}When $N=1$, i.e., the set of discrete eigenvalues is $\Omega=\{ \lambda_1,\tilde{\lambda}_1\}$ , the solution~(\ref{3.4.6}) reduces to
\begin{align}
\bar{q}^{[1]}_n=\bar{q}_r+\frac{a_4 (q_r \lambda_1 + \lambda_1 (
      a_2 q_r-1 ) (\tilde{\lambda}_1 - \tau) +
   a_3^2 \bar{q}_r (a_1 \lambda_1 + a_2 ( \tau \lambda_1-1)) -
   a_3 ( \lambda_1 (\bar{q}_r + \tau +
         a_1 (\tau-\tilde{\lambda}_1 ))-1))}{\lambda_1(a_1 a_3-1) ( a_1 a_3 + a_2 q_r-1)+(\tau^{-1} a_4 (a_1 a_3-1) (
    a_3 (\tau\lambda_1-1)-q_r \lambda_1))},
\end{align}
where
\begin{align*}
a_1=\frac{( \lambda_1 -\tau ^{-1})\gamma^{2n}(\tilde{\lambda}_1 )\, \tilde{C}_1(t)\, }{( \tilde{\lambda}_1-\tau ^{-1} ) ( \tilde{\lambda}_1 - \lambda_1 )},\ \, \, a_2=\frac{\gamma^{2n}(\tilde{\lambda}_1 )\, \tilde{C}_1(t)}{\tau\tilde{\lambda}_1-1 },\ \, \, a_3=\frac{(\tilde{\lambda}_1-\tau )\gamma^{-2n}(\lambda_1 )
\, C_1(t)}{( \lambda_1-\tau ) (\lambda_1 -\tilde{\lambda}_1 )},\ \, \, a_4=\frac{\gamma^{2n}(\tilde{\lambda}_1 )\, \tilde{C}_1(t)}{\tilde{\lambda}_1-\tau^{-1} },
\end{align*}
and
\begin{align*}
\gamma^{2n}(\tilde{\lambda}_1 )=\Bigg(\frac{\tilde{\lambda}_1(\tilde{\lambda}_1-\tau)}{\tau \tilde{\lambda}_1-1}\Bigg)^n,\quad \quad \gamma^{-2n}(\lambda_1 )=\Bigg(\frac{\tau \lambda_1-1}{\lambda_1(\lambda_1-\tau)}\Bigg)^n,\\
\tilde{C}_1(t)=\tilde{C}_1(0)e^{-\tilde{\omega}(\tilde{\lambda}_l) t},\quad \quad C_1(t)=C_1(0)e^{\tilde{\omega}(\lambda_1) t}=-\lambda_1^{2}\bar{\tilde{C}}_1(t).
\end{align*}

By taking appropriate parameters, the black dark 1-soliton and grey dark 1-soliton solutions can be derived respectively by taking different parameter values, whose shapes are plainly presented in Figs. 3.1.

\begin{figure}[H]
\setcounter{subfigure}{0}
\centering
\subfigure[]{\includegraphics[width=6cm]{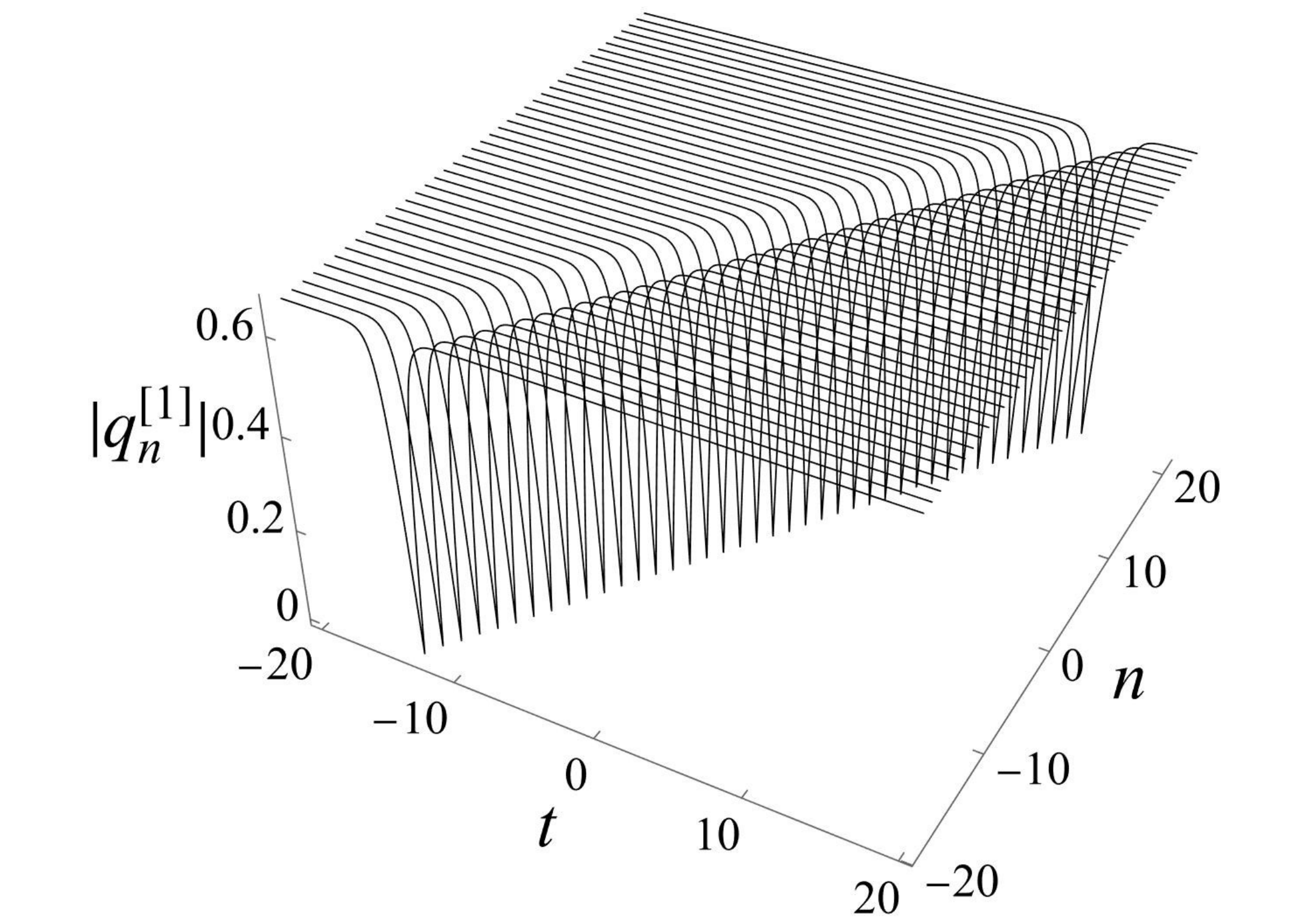}}
 \quad \quad\quad \
\subfigure[]{\includegraphics[width=6cm]{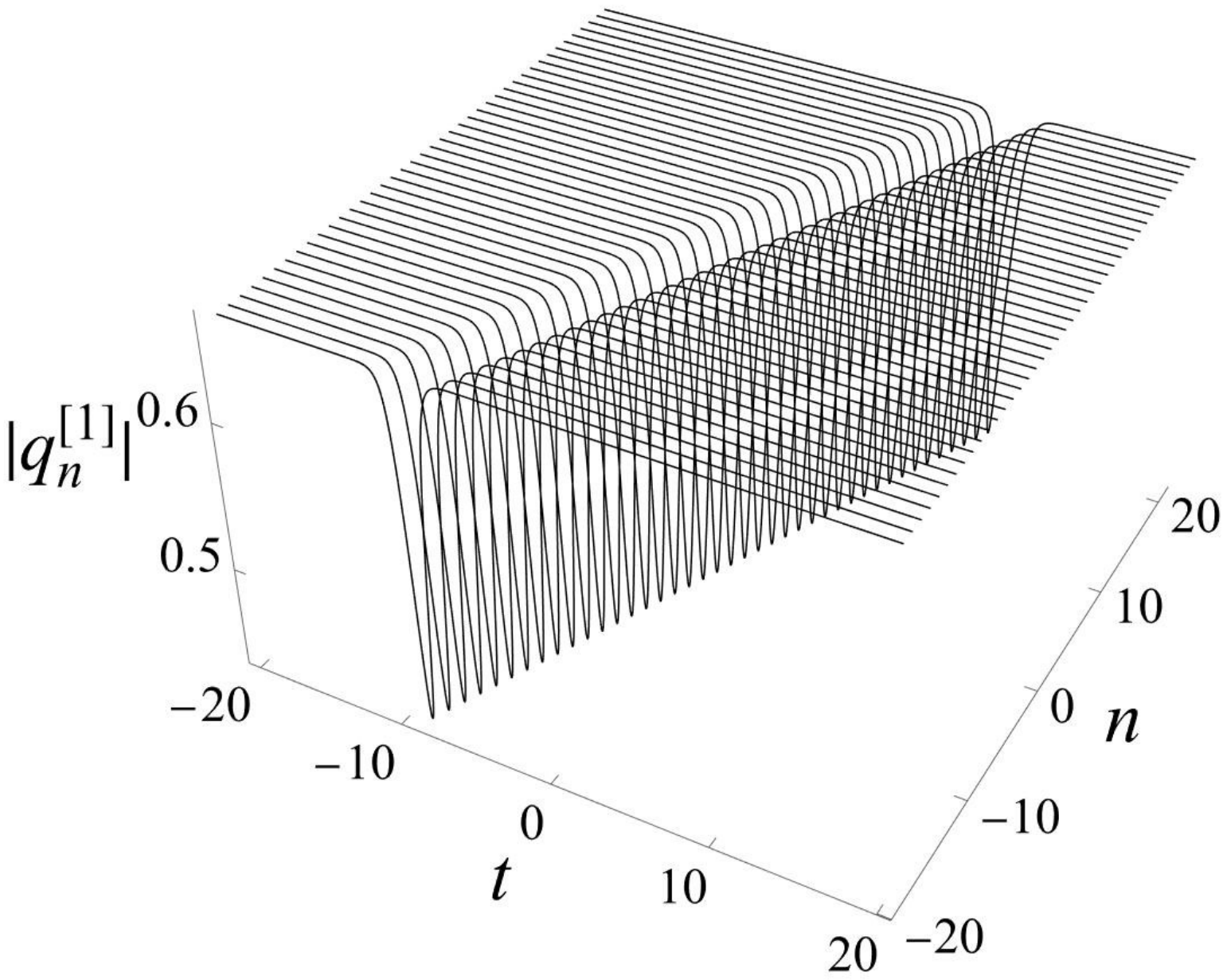}}
\\
 \subfigure[]{\includegraphics[width=5.5cm]{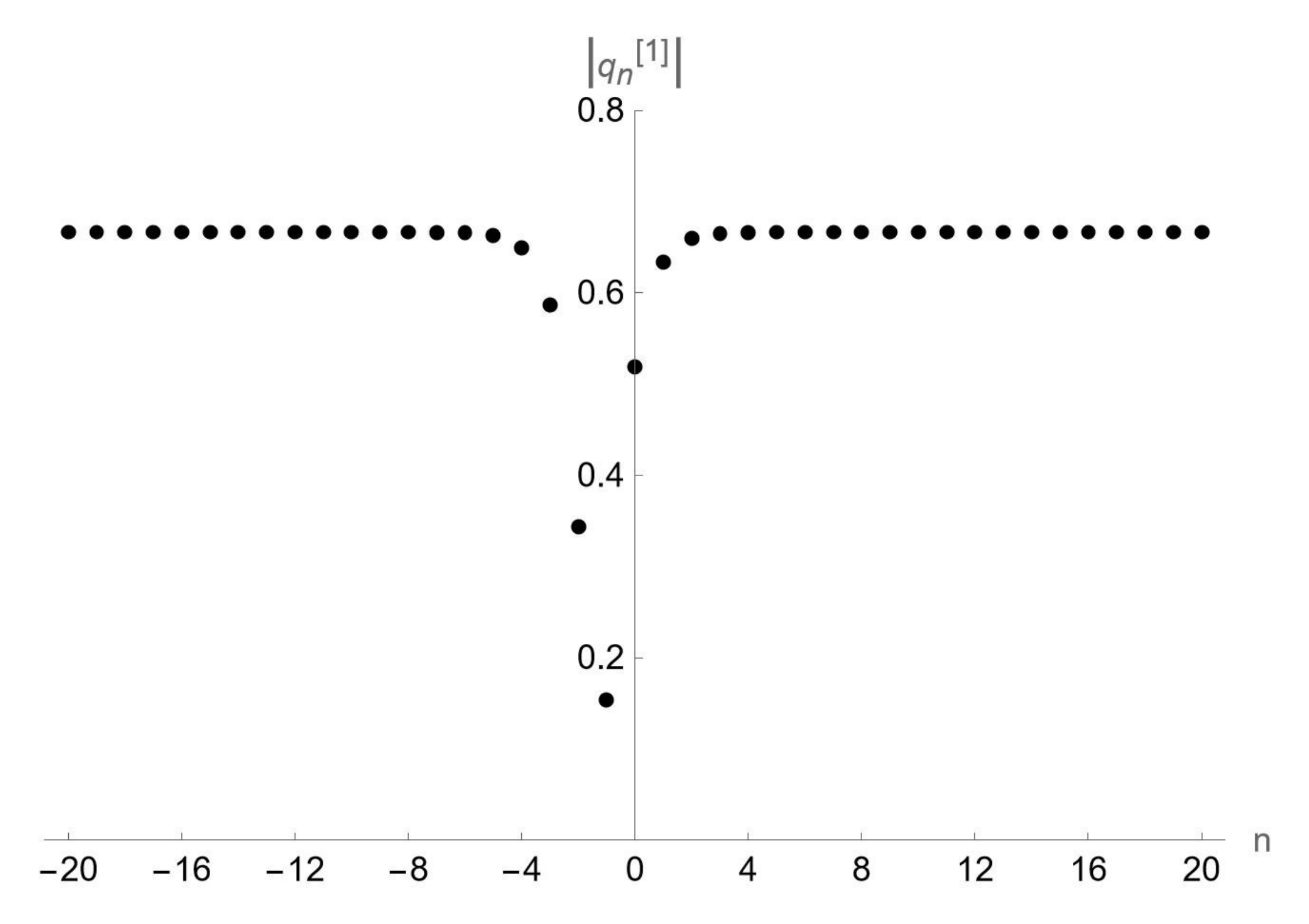}}
 \quad \quad\quad \quad
\subfigure[]{\includegraphics[width=5.5cm]{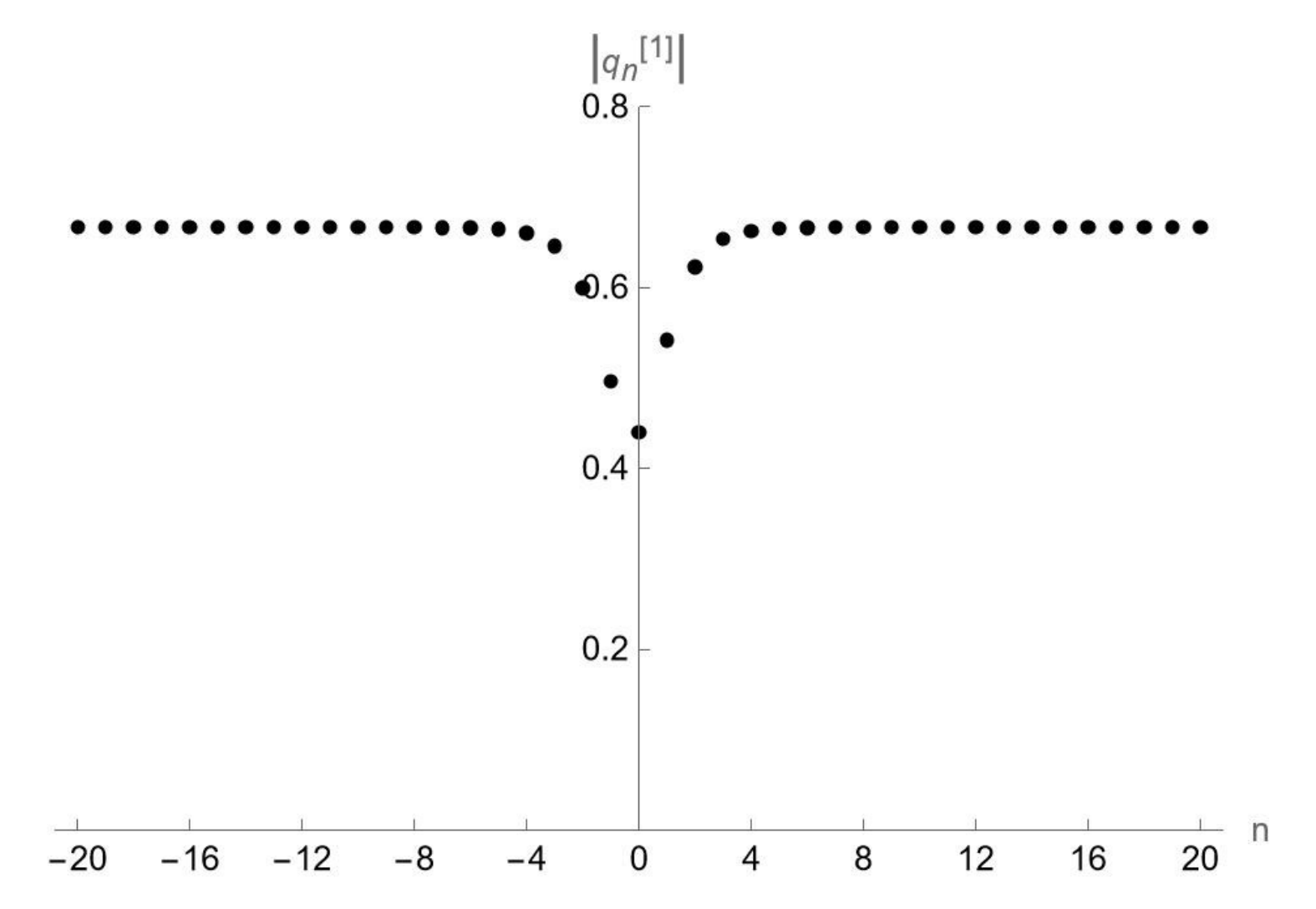}}

\flushleft{\footnotesize
\textbf{Figs.~$3.1$.} (a) The black dark 1-soliton of (\ref{1.1}) with $ Q=\frac{2}{3}$, $\theta_r=0$, $\tilde{C}_1(0)=\frac{1}{2}$, and $ \tilde{\lambda}_1=\frac{1-Q}{\tau}$. (b) The grey dark 1-soliton of (\ref{1.1}) with $ Q=\frac{2}{3}$, $\theta_r=0$, $\tilde{C}_1(0)=2 e^{\frac{i \pi}{12}}$, and $ \tilde{\lambda}_1=\frac{1-Q e^{\frac{i \pi}{12}}}{\tau}$ . (c) and (d) respectively show the sectional views of (a) and (b) at $t =0$.}
\end{figure}

\subsubsection{ 2-soliton}

\hspace{1.5em}When $N=2$, i.e., the set of discrete eigenvalues is $\Omega=\{\lambda_1,\tilde{\lambda}_1,\, \lambda_2, \tilde{\lambda}_2\}$,  the corresponding dark 2-soliton solution can be derived from the solution~(\ref{3.4.6}) by taking the appropriate parameter values and shown in
Fig. 3.2.

\begin{figure}[H]
\setcounter{subfigure}{0}
\centering
{\includegraphics[width=8cm]{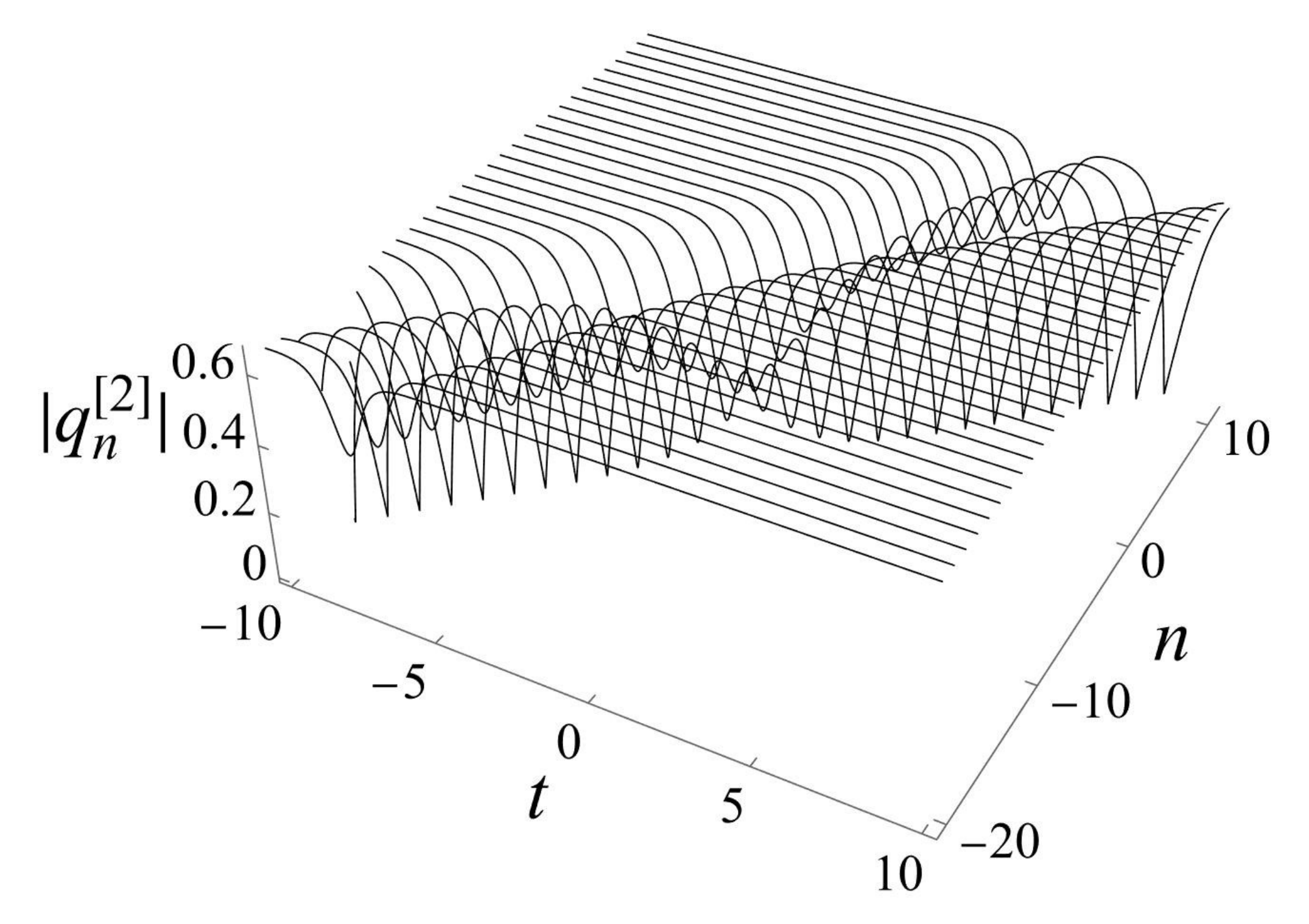}}
\flushleft{\footnotesize
\textbf{Figs.~$3.2$.} The dark 2-soliton of Eq.~(\ref{1.1}) with $ Q=\frac{2}{3}$, $\theta_r=0$ , and $\tilde{C}_1(0)=\frac{1}{2}, \tilde{\lambda}_1=\frac{1-Q}{\tau}$,  $\tilde{C}_2(0)=2 e^{\frac{i \pi}{12}}, \tilde{\lambda}_2=\frac{1-Q e^{\frac{i \pi}{12}}}{\tau} $. }
\end{figure}

\section{Conclusions}

\hspace{1.5em}In this paper, the DcmKdV equation has been investigated under the zero and non-zero BCs by the RH approach. The eigenfunctions and scattering matrix have been introduced according to the spatial spectrum problem, and then the properties including analyticity, asymptotics and symmetries of them have been studied detailedly. On the basis of the research results in direct problems, the inverse problems have been formulated as the RH problems, from which the reconstruction potential formulas have been derived. Meanwhile, the trace formulas have been provided by establishing and solving the corresponding RH problems. Since the dependence on time $t$  was omitted for simplicity in the part of direct and inverse scattering problems, the temporal spectrum problem has been considered to complete the time evolution. Moreover, taking into account the reflectionless case, the $N$-soliton solutions on the zero and nonzero backgrounds have been constructed respectively.

In particular, under the zero BCs, i.e. the potential $q_n$ vanishes as $n \to \pm\infty$, the case of $\sigma=-1$ has been studied. The eigenfunctions which are the solutions of the spatial spectrum and the modified eigenfunctions whose BCs are identity matrix have been defined, and then the scattering matrix has been introduced owing to the linear independence of the two eigenfunctions. In order to prove the properties of the modified eigenfunctions and the scattering matrix, the summation equations that the modified eigenfunctions satisfy have been derived by the Green's functions method. Accordingly, the analyticity and asymptotics have been analyzed respectively by introducing the Neumann series and researching the Laurent series expansions about 0 or infinity. Meanwhile,  the two symmetries have been obtained from the spatial spectrum problem and the Laurent series expansions, which has implied that the number of discrete eigenvalues was multiple of 4. Based on these properties, we have formulated the inverse problem as a generalized RH problem for a defined piecewise function with simple poles at discrete eigenvalues. Further, a new RH problem has been established by removing the singularity at discrete eigenvalues, and has been solved to present the reconstruction formula and a algebraic-integral system, from which the $N$-soliton solutions have been derived in the case of reflectionless. If $N=1$ and $N=2$, the bright 1-soliton and 2-soliton on the zero background have been obtained by taking the appropriate parameters and have been visually demonstrated in images.

For the zero BCs, i.e. the potential $q_n\in\mathcal{H}_k$, we have considered the case of $\sigma=1$. The eigenfunctions have been defined, and the two-sheeted Rieman surface has been introduced since the correspondence between the spectral parameter $z$ and the parameter $\gamma$ in the BCs of eigenfunctions was 1-to-2. For fear of the complexity of the Rieman surface,  the uniformization variable $\lambda$ has been introduced by establishing a conformal mapping. The mapping relations between variables $z$, $\gamma$ and $\lambda$ have been discussed clearly, which has contributed to clarify the mappings between the complex $z$-plane and the complex $\lambda$-plane for discrete spectrum and continuous spectrum, and has promoted the investigation of the scattering problem on the $\lambda$-plane. Unlike the case of zero BCs, a modified scattering problem has been introduced to research the analytic properties of the eigenfunctions and scattering coefficients except for the points $\lambda=0,\, \infty$ and the branch points, while the asymptotics obtained by WKB expansion and the summation equations for eigenfunctions have given an indication of whether the eigenfunctions were well defined at these points, and then the behavior of scattering coefficients at these points has been analyzed as well. In the part of inverse problem, a generalized RH problem has been constructed on the $\lambda$-plane, which has yielded the reconstruction formula and further the $N$-soliton solution in the reflectionless case. Specifically, when $N=1$ and $N=2$, the black dark 1-soliton and black grey 1-soliton, 2-soliton on the small norm non-zero background have been obtained and illustrated graphically. In addition, the $\theta$-condition has been derived according to the presented trace formulas and the asymptotic behavior.

\section*{Acknowledgments}

\hspace{1.5em}We express our sincere thanks to each member of our discussion group for their suggestions. This work has been supported by the National Natural Science Foundation of China under Grant No. 11905155, and the Fund Program for the Scientific Activities of Selected Returned Overseas Scholars in Shanxi Province under Grant No. 20220008.

\end{document}